\documentclass[letterpaper,11pt]{article}
\usepackage[utf8]{inputenc}
\usepackage{graphicx,fullpage,paralist}
\usepackage{amsmath, amssymb, amsthm,dsfont}

\usepackage{xr-hyper}
\makeatletter
\newcommand*{\addFileDependency}[1]{
  \typeout{(#1)}
  \@addtofilelist{#1}
  \IfFileExists{#1}{}{\typeout{No file #1.}}
}
\makeatother

\newcommand*{\myexternaldocument}[1]{%
    \externaldocument{#1}%
    \addFileDependency{#1.tex}%
    \addFileDependency{#1.aux}%
}

\myexternaldocument{appendix}

\usepackage{comment,hyperref}
\usepackage{thmtools}
\usepackage{tikz}
\usepackage{pgfplots}
\usepackage{varwidth}
\usepackage{graphicx}
\usepackage{csquotes}
\usepackage[font=small,labelfont=bf]{caption}

\usepackage[percent]{overpic}

\usepackage[square,numbers]{natbib}
\usepackage{booktabs}
\usepackage{longtable}
\usepackage{array}
\usepackage{multirow}
\usepackage{wrapfig}
\usepackage{float}
\usepackage{colortbl}
\usepackage{pdflscape}
\usepackage{tabu}
\usepackage{threeparttable}
\usepackage{threeparttablex}
\usepackage[normalem]{ulem}
\usepackage{makecell}
\usepackage{xcolor}

\pgfplotsset{compat=1.10}
\usepgfplotslibrary{fillbetween}
\usetikzlibrary{backgrounds}
\usetikzlibrary{patterns}
\newcommand{\PP}{\mathbb{P}}

\newcommand{\EE}{\mathbb{E}}

\usepackage{bbm}

\usepackage{amsbsy}

\newlength{\algofontsize}
\usepackage[textsize=tiny,textwidth=2cm]{todonotes}

\newcommand{\newmo}[1]{\noindent \textcolor{black}{#1}}

\newcommand{\colsm}[1]{\noindent \textcolor{black}{#1}}

\usepackage{xcolor}
\hypersetup{
	colorlinks,
	linkcolor={red!50!black},
	citecolor={blue!50!black},
	urlcolor={blue!80!black}
}

\usepackage[left=2cm,top=2cm,right=2cm, bottom=2cm]{geometry}
\usepackage{multirow}
\usepackage[noend]{algpseudocode}
\usepackage{algorithm,algorithmicx}

\begin{document}
	
	\algrenewcommand\algorithmicrequire{\textbf{Input:}}
	\algrenewcommand\algorithmicensure{\textbf{Output:}}

{\large
\textbf{Replacing Quarantine of COVID-19 Contacts with Periodic Testing is also Effective in Mitigating the Risk of Transmission}}

\bigskip

Patricio Foncea\textsuperscript{1} \hspace{3mm}
Susana Mondschein \textsuperscript{2,3} \hspace{3mm}
Marcelo Olivares\textsuperscript{2,3}

\medskip


\bigskip

\begin{flushleft}
$^1$ Operations Research Center, MIT.\\
$^2$ Department of Industrial Engineering, University of Chile.\\
$^3$ Instituto Sistemas Complejos de Ingenier\'ia.

\bigskip

Corresponding author: Susana Mondschein, susana.mondschein@uchile.cl
\end{flushleft}

\begin{abstract}

The quarantine of identified close contacts has been vital to reducing transmission rates and averting
secondary infection risk before symptom onset and by asymptomatic cases. The effectiveness of this
contact tracing strategy to mitigate transmission is sensitive to the adherence to quarantines, which
may be lower for longer quarantine periods or in vaccinated populations (where perceptions of risk
are reduced). This study develops a simulation model to evaluate contact tracing strategies based
on the sequential testing of identified contacts after exposure as an alternative to quarantines, in
which contacts are isolated only after confirmation by a positive test. The analysis considers different
number and types of tests (PCR and lateral flow antigen tests (LFA)) to identify the cost-effective
testing policies that minimize the expected infecting days post-exposure considering different levels
 of testing capacity. This analysis suggests that even a limited number of tests can be effective at
 reducing secondary infection risk: \colsm{two LFA tests (with optimal timing) avert infectiousness at a level
 that is comparable to 14-day quarantine with 80-90\% adherence, or equivalently, 7-9 day quarantine
 with full adherence (depending on the sensitivity of the LFA test). Adding a third test (PCR or LFA)
 reaches the efficiency of a 14-day quarantine with 90-100\% adherence. These results are robust to the
 exposure dates of the contact, test sensitivity of LFA and alternative models of viral load evolution,} which suggests that simple testing rules can be effective for improving contact tracing in settings where
strict quarantine adherence is difficult to implement.

\end{abstract}

\thispagestyle{empty}

\section{Introduction}

The COVID-19 pandemic has imposed many challenges on societies around the world. The virulence of the outbreak has required strict nonpharmaceutical interventions, such as massive lockdowns, curfews, contact quarantines, sanitary measures, travel restrictions, and testing surveillance. Although many of these policies have been useful for containing outbreaks (\cite{chinazzi2020effect}, \cite{tang2020effectiveness}), they have also imposed a significant social and economic burden on most countries (\cite{jin2021economic}).

Since the first outbreak of COVID-19 in early 2020, new scientific knowledge has been rapidly developed regarding the characteristics of this virus, such as the viral load evolution of an infected individual (\cite{larremore2021test}), infectiousness profile (\cite{he2020temporal}), transmission patterns (\cite{meyerowitz2020transmission}) and cardinal symptoms (\cite{zoabi2021machine}). A significant challenge in containing transmission is to halt infections generated before symptom onset and by asymptomatic cases, thus making symptom monitoring insufficient to contain the spread of the virus (\cite{ferretti2020quantifying}, \cite{li2020asymptomatic}), even with close monitoring of close contacts (\cite{peak2020individual}). Therefore, preventive quarantines of potentially exposed individuals have been a fundamental mitigation measure to reduce transmission in the community. These quarantine policies vary across countries, both in terms of the target population and the quarantine protocol. Most countries require preventive quarantine of traced contact between 10 and 14 days (\cite{NHSprotocol},\cite{CDCq}). Restrictions to incoming international travelers also vary across countries, ranging from no quarantine when a recent negative test result is provided to others requiring strict quarantines ranging from 10 to 14 days. Some countries even use dedicated facilities to quarantine incoming travelers. These traveling restrictions have led many traveling website hubs to provide detailed information on quarantine and testing protocols by country,  (Wego: \url{https://blog.wego.com/covid19-travel-restrictions-by-destination-country/}; Kayak: \url{https://www.kayak.com/travel-restrictions}).

\colsm{The design of targeted and temporally restricted quarantine 
protocols for traced contacts and higher-risk individuals should account for the associated risk reduction of the policy as well as the costs imposed on the target population. Quarantines have been associated with economic cost and adverse mental health effects, both when targeted to specific individuals (e.g. travellers and close contacts) (\cite{brooks2020psychological}), and massive lockdowns and mobility restrictions(\cite{bonaccorsi2020economic}).These studies report negative consequences of quarantines such as  post-traumatic stress symptoms, confusion, and anger, which are worsened by longer quarantine duration, infection fears, frustration, boredom,  inadequate information, and financial loss among others. }
Quarantine measures that are too strict may reduce compliance and the incentives to report close contacts, thereby reducing the effectiveness of contact tracing strategies (\cite{WEBSTER2020163}). Approximately 75\% of U.S. subjects who were surveyed indicated that they would adhere with quarantine for 14 days when mandated by a health official; however, compliance can be as low as 60\% in specific demographic groups (\cite{mcclain2020challenges}). Of those who declare their lack of willingness to comply, 44\% indicate that they do not think that quarantining is necessary.

Improvements in testing technologies have helped to shorten quarantine periods while maintaining a low risk of secondary infections by exposed contacts (\cite{xu2020covid}). For example, the WHO quarantine recommendations for contacts of individuals with a
confirmed or probable case of COVID-19 have been made more flexible and evolved from 14 days from their last exposure (\cite{world2020considerations}) to more discretionary measures, such as advising local public health authorities to account for local conditions and needs to determine the length of quarantine. These options include stopping quarantine for contacts that have not presented symptoms after day 10 or after day 7 with a negative diagnostic specimen test (\cite{CDCq}, \cite{CDCshorten}).

As vaccination campaigns continue to advance, transmission rates are expected to fall, thereby reducing the risk of infection of contacts exposed to a confirmed case. Nevertheless, some risk of transmission is still present due to the 
lower effectiveness of some vaccines in preventing infection and uncertainty associated with virus variants (\cite{whowho}); therefore, contact tracing will continue to be relevant. However, vaccination is likely to reduce the perception of risk of exposed contacts, which could lower compliance with strict quarantine measures (\cite{WEBSTER2020163}). Hence, the focus of this study is to analyze alternatives to quarantine of traced contacts to reduce the risk of secondary infections.

Access to low-cost PCR and lateral flow antigen (LFA) tests has become widespread (\cite{mercer2021testing}), and this massive availability of detection tests enables the close monitoring of traced contacts without the need to confine exposed individuals (unless a positive test result), which lowers the quarantine costs without increasing the secondary transmission risks. Thus, we analyze the optimal timing of different types of tests to reduce the risk of exposure of active (not quarantined) unconfirmed contacts to susceptible individuals, thereby helping to reduce both infection risk and the costs of quarantine through a cost-efficient use of testing resources. This finding is particularly important for minimizing disruptions in essential activities, such as highly specialized workers, teachers, students and healthcare workers, where quarantines may require major re-organization of the operations. Similar strategies could be used to ease quarantine requirements on foreign travel.

Our study contributes to the literature on the analysis of quarantine strategies of traced contact in different settings. Several modeling studies suggest that quarantine periods can be shortened to 7 days with a negative PCR test at the end of this period because it has a residual risk equivalent to a quarantine period of 14 days with no testing (\cite{van2021intra}, \cite{wells2020optimal}). The recent modeling study by \cite{quilty2021quarantine} also suggests that daily LFA testing of traced contacts over 5 days \textit{without quarantine} if all tests are negative can actually reduce the risk of secondary infections relative to a mitigation strategy of 14 quarantine days with moderate levels of adherence. Following that idea, we evaluate alternative sequential testing schemes when different numbers and types of tests are available to monitor traced contacts that are not under quarantine, with isolation only triggered when the case is confirmed through a positive test.

This study was motivated through the design of testing and quarantine policies for schools in Chile, where in-person teaching has been prohibited during most of the pandemic. In planning a safe return to in-person schooling, Chilean health authorities have developed protocols on how to handle confirmed cases and require quarantines of the complete classroom of an infected student with flexibility on the quarantine strategies for teachers, who received priority in the immunization campaign and whose quarantine may induce severe disruptions in the school operation. An alternative to quarantine is to allow teachers to continue face-to-face teaching but closely monitor them through an optimal design of PCR and LFA tests to reduce the risk of secondary infections. A similar strategy can be used to ease the quarantine requirements of the classroom of infected students, where the risk of transmission has been shown to be relatively low for younger students (\cite{CHILDRENSUSC}) along with the adoption of masks and other mitigation measures (\cite{chernozhukov2021association}, \cite{lessler2021household}). \newmo{This \textit{test-to-stay} approach has been implemented in the US and UK, reporting significant increase in in-person teaching (\cite{lanier2021covid,nemoto2021evaluation,harris2021evaluation,couzin2021schools}}).

Our modeling approach is similar to that of \cite{larremore2021test} and \cite{wells2020optimal} and used simulation methods to generate scenarios of viral loads of infected contacts that may or may not present symptoms. These simulated viral load paths relate the infectiousness of the contact with test sensitivity during post-exposure time, enabling us to model the reduction of secondary infections under alternative sequential testing schemes. Our modeling analysis confirms the findings of \cite{larremore2021test} that despite the lower sensitivity of LFA tests relative to PCR, they are more efficient in averting infections when PCR tests take more than one day to confirm the results. We also corroborate the result of \cite{wells2020optimal} that daily LFA testing during 5 days postexposure, with isolation required after a positive test result, essentially averts all the risk of secondary infections and is equivalent to a 14-day quarantine policy for high adherence scenarios. \newmo{We show that these results are robust to the days of exposure of the traced contact with the index case, test sensitivity scenarios and to alternative models of viral load evolution}. When testing resources are scarce, our modeling analysis suggests that using three LFA tests with an appropriate timing during the postexposure period can also achieve a very low risk of secondary infections, which is superior to that of a 14-day quarantine policy with 90\% adherence. Our analysis shows that the timing of these sequential tests is important because suboptimal testing schedules may substantially increase the risk of secondary infections.

Another important difference of our work compared to that of \cite{larremore2021test} and \cite{wells2020optimal} is that we analyze settings with uncertainty on the exact day of exposure of the contact (\newmo{\cite{van2021intra} follows a similar approach to model exposure uncertainty of incoming travelers based on the incidence of country of origin}). This difference is important for studying settings with structured contact networks that meet recurrently, such as workplaces, schools, healthcare facilities and households. We show that modeling this uncertainty is relevant for the design of an optimal testing schedule and should also account for different types of index cases: we cover scenarios where the index case is identified at symptom onset or by surveillance testing, among others.

Our modeling analysis suggests that an optimal design of testing strategies of traced contacts after exposure can be effective for gradually easing quarantine requirements for essential activities where the costs of quarantines are high or have low adherence rates. Nevertheless, the implications of the proposed quarantine/testing strategies need to be evaluated with caution because they might impact the behavior of confirmed cases and their contacts in multiple dimensions. On the positive side, easing quarantine requirements may lead to higher adherence of these policies by the traced contacts and a higher proportion of contacts reported by an index case. On the negative side, relaxing quarantine policies may reduce the adoption of other mitigation measures in the community and work environments (such as the use of personal protective equipment and physical distancing). Further research is needed to empirically evaluate the overall impact of the proposed contact monitoring schemes on community transmission.

\section{Overview of the Modeling Approach}
\label{model}
To relate test sensitivity with infectiousness, we model the evolution of viral load of infected individuals by replicating the methodology used in \cite{larremore2021test}. Test sensitivity is affected by three complementary factors: (i) the timing of the test (relative to exposure time); (ii) the level of detection (LOD) of the type of test used; and (iii) the quality of the swab sample. We incorporate all three into our modeling approach.

\begin{figure}
    \centering
    \includegraphics[scale=0.6]{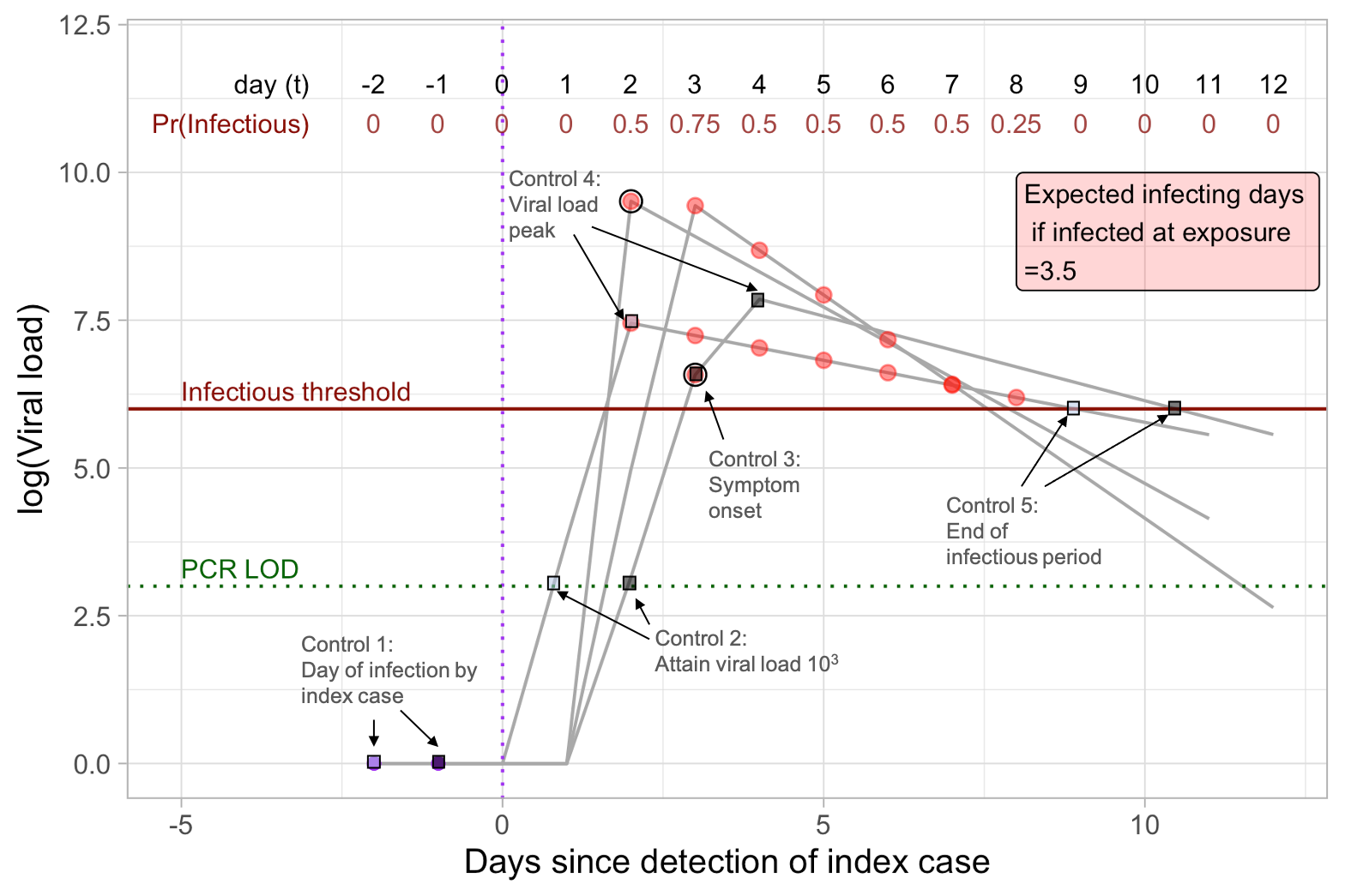}
\caption{\textbf{Description of the simulation of viral load paths.}
The horizontal axis represents a timeline, with $t=0$ representing the date of detection of the index case and its contact. Each gray line indicates one simulated viral load path of the infected contact, which is generated randomly using 5 control points shown with squares for 2 independent paths (with light and dark colors). Control 1 is the day of infection, which in the example includes days -1 and -2 for each respective path. Control point 2 is the day on which a viral load is detectable by PCR. Control 3 is generated only for symptomatic cases and corresponds to the day of symptom onset (represented with a dark circle). Control 4 is the peak viral load and the day it is attained. Control 5 is the day at which the infectious period ends and indicates the slope of the viral load decline. Red dots indicate the infectious days on each viral path; individuals self-isolate the day after presenting symptoms; therefore infecting days post-symptoms are averted. The top part of the figure shows the probability that the infected contact is contagious on that day (excluding days where infection is averted). Expected infecting days, which are conditional on the contact being infected, are equal to the sum of these probabilities. }
    \label{fig:sim_path}
\end{figure}

Given a set of days of exposure, we generated a sample of random paths describing potential scenarios of viral load evolution over time. Individuals become infectious when their viral load exceeds $10^6$ cp/ml (\cite{larremore2021test,scheiblauer2021comparative}). Each viral load path is simulated using five control points generated as random variables: (1) the day of infection; (2) the time (since the infection date) at which the minimum level of detection (LOD) with PCR test is reached; (3) the peak level of viral load and the time it is reached; (4) the time of symptom onset for symptomatic cases; and (5) the time at which the infectious period ends. This simulation procedure is illustrated in Figure \ref{fig:sim_path}, where the horizontal axis is a timeline, with $t=0$ representing the time at which the index case is confirmed and the individual is identified as contact. Exposure dates of the contact occur during or before the confirmation date ($t\leq 0$). Further details on the simulation, including the probability distributions used to simulate the control points, are described in the Appendix \ref{sec:app_math}.

The red points in Figure \ref{fig:sim_path} show the days on each path in which the individual was infectious, i.e., when the viral load exceeds the level of infectiousness ($10^6$ cp/ml). Symptomatic cases are assumed to self-isolate after symptom onset, whereas asymptomatic cases are not isolated and therefore continue to infect throughout the infectious period. Conditional on being infected at exposure, the probability that the individual is infecting others on a given day is the fraction of sample paths that are above the infectiousness threshold on that day. The expected number of infecting days is the sum of these probabilities across all days after the first exposure date. An example of these calculations is provided in Figure \ref{fig:sim_path} for the illustrative sample paths that were simulated. In the actual simulation, we consider 200,000 sample paths for each exposure date. The probability distribution of the exposure data is described next.

\subsection{Modeling uncertainty in the exposure time}

Our methodology incorporates uncertainty on the day in which the contact has been infected, considering a range of possible exposure days of index case with the traced contact. This modeling approach is more realistic in settings with structured contact networks that interact frequently (e.g. school and workplace).

The uncertainty in the exposure time is modeled using a probabilistic approach, deriving the probability distribution for the days in which the transmission from the index case to the contact may have occurred; this probability distribution is used to simulate the contact's viral load. Specifically, let $t\in \{0,-1,\ldots,-14\}$ represent the set of possible exposure days, where $t=0$ is the day of index case confirmation (we consider up to two weeks before confirmation as possible exposure dates). Infection occurs on day $t$ when: (i) the index case is during the infectious period on that day, which is presented by the probability $p_t$; and (ii) the contact was not previously infected and transmission from the infectious index to the susceptible contact. The latter is represented by the \textit{infectivity} parameter $\beta$, which represents the transmission probability, conditional on the index case been infectious.

\begin{figure}
    \centering
    \includegraphics[width=0.8\textwidth]{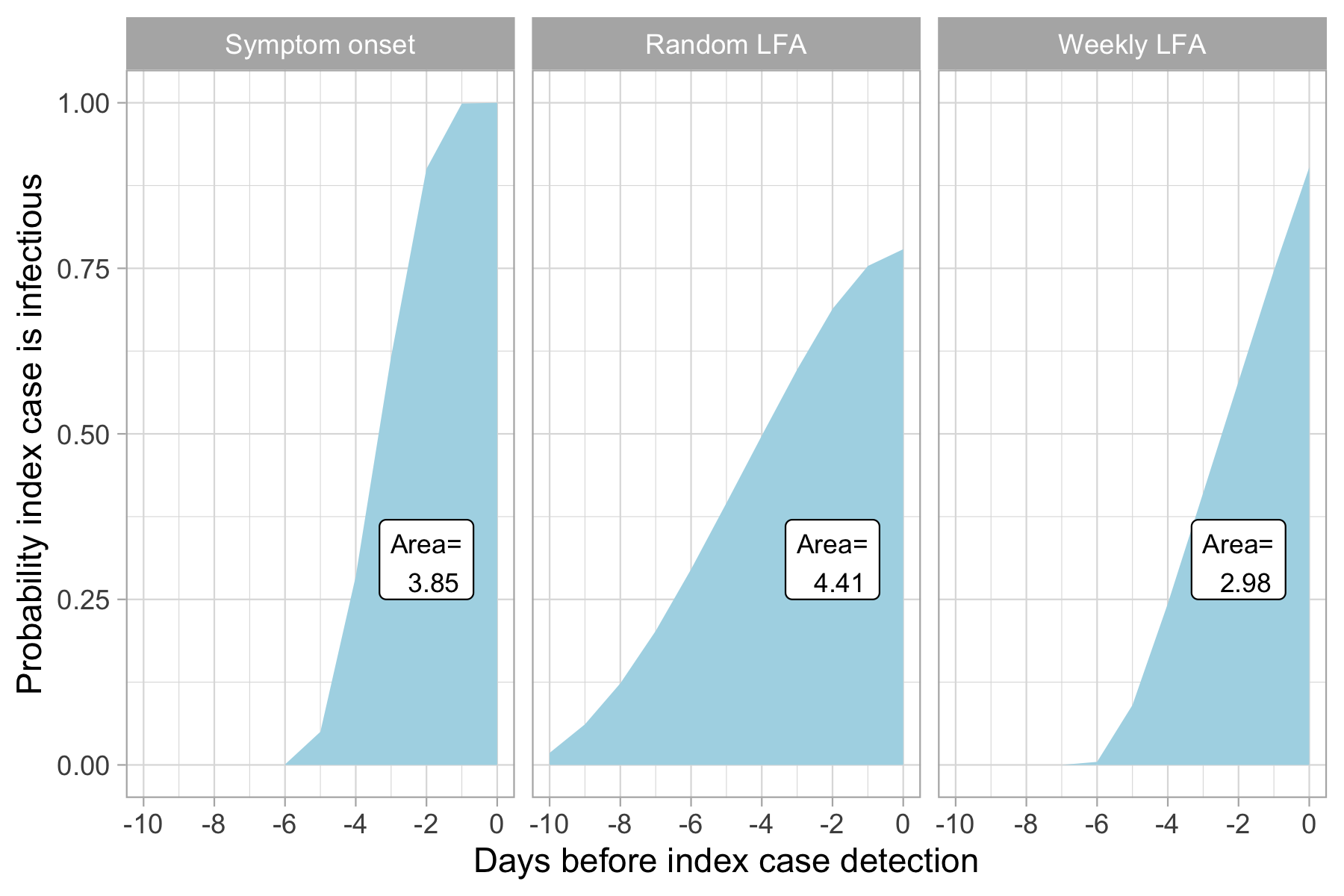}
    \includegraphics[width=0.8\textwidth]{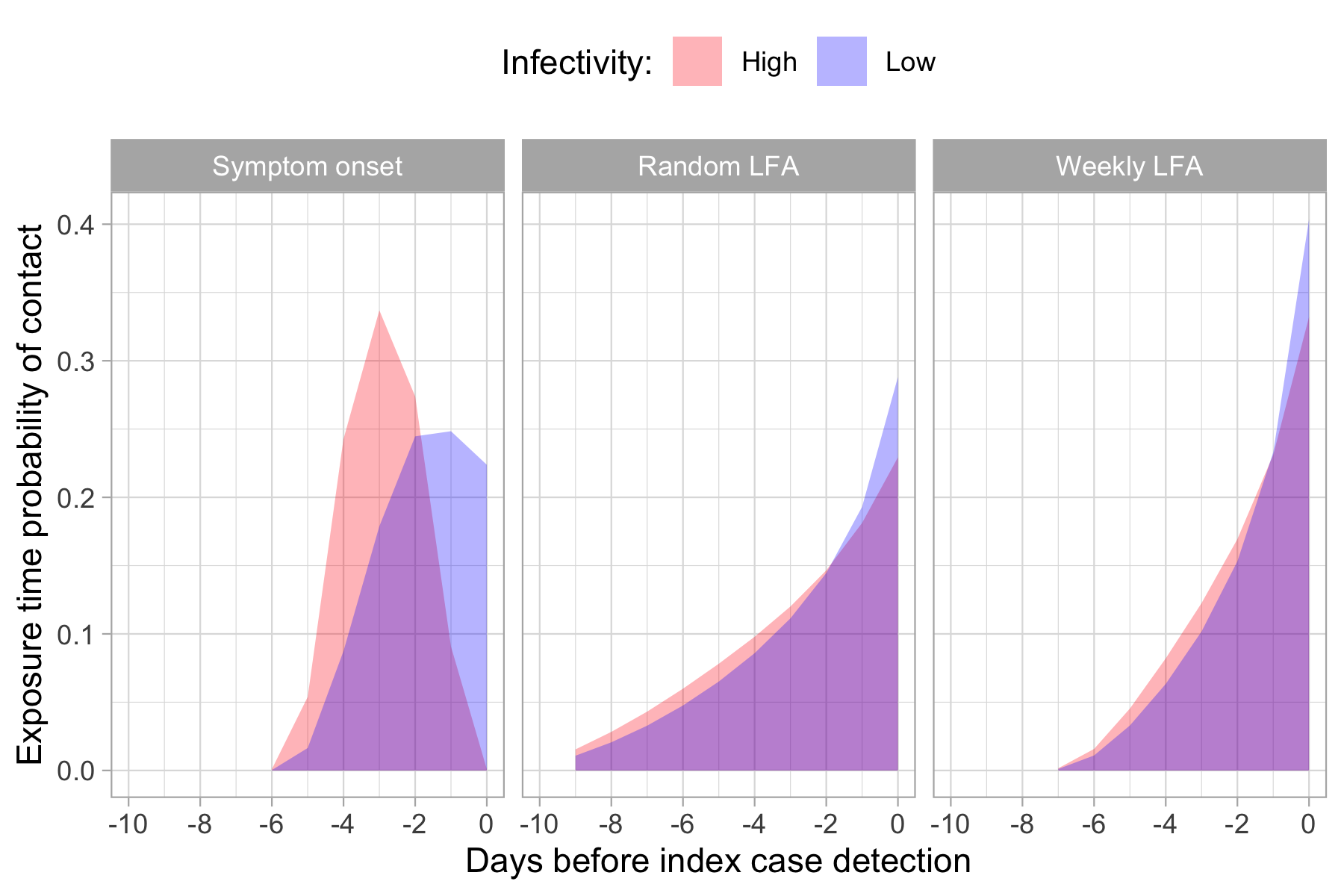}
    \caption{The top panel shows the probability of the index case been infectious on each day prior to the confirmation date ($t=0$). Each facet describes a different scenario on how the index case was detected: (i) at symptom onset; (ii) asymptomatic detected with a random LFA test; (iii) asymptomatic detected with a weekly surveillance LFA test. The bottom panel shows the distribution of the exposure time of a contact that was infected on or before the index confirmation date, for each scenario. The distribution is calculated using two infectivity parameter values, Low (0.1) and High (1.0). The overlap between these distributions is shown in purple color.}
    \label{fig:exposure}
\end{figure}

The probability distribution $p_t$ (index case is infectious on day $t$) depends on how the index case was detected at $t=0$. The model considers three types of index case detection: (1) symptomatic index case detected at symptom onset; (2) asymptomatic index case detected by a randomly performed LFA test; and (3) asymptomatic index case detected by a weekly surveillance screening with LFA test. To compute $p_t$ on each of these three scenarios, we simulate a large sample of viral load paths of the index case starting on each possible infection date $t\in[-14,-1]$. From this large sample, we select the paths that are feasible with the index case detection on $t$. For example, for the scenario where the index case is detected at symptom onset, only the simulated paths that present symptoms on day $t=0$ are selected. For the scenario detected by a random LFA, the selected paths include the simulations with viral load above the LOD ($10^5 cp/ml)$) on day $t=0$. Using this selected sample, $p_t$ is computed as the fraction of selected paths that exceed the infectious threshold ($10^{6}$) on day $t$. The top panel of Figure \ref{fig:exposure} shows the calculations of $p_t$ for the three scenarios considered in the model. The area under the curve represents the average number of days in which the index case was infectious previous to detection.

Conditional on been infectious, the probability that the index case infects the contact on a given day is given by the infectivity parameter $\beta$ (we assume that the infectivity is constant during the infectious period, that is, when the viral load is above $10^6$). Define the events: (i) $S_{t}$= the contact has not been infected up to time $t$; and (ii) $I_t$= index case is infectious at time $t$.  The probability that the contact is infected at day $t$ can be expressed as:
\[
r_t =\beta \Pr(I_t| S_{t}) \cdot \prod_{j\leq t-1} (1 - \beta \Pr(I_j|S_{j})),
\]
where the term in the product represents the probability that the contact was not infected up to time $t$ (i.e. $\Pr(S_t)$). Appendix \ref{sec:app_math} provides further details on how to compute $r_t$ using simulation methods. Conditioning on the event that the contact was infected, the probability that the exposure occured in day $t$ is obtained by normalization, $r_t \big / \sum_{j=0}^{-14}r_j$. Note that this exposure time distribution depends on the infectivity parameter $\beta$. The bottom panel of Figure \ref{fig:exposure} shows the (normalized) probability distribution of the exposure date for an infected contact for two values of $\beta$ equal to 0.1 and 1.0 (Low and High) under the different scenarios of index case confirmation. As the figure illustrates, increasing the infectivity parameter $\beta$ moves the distribution of the infection time to the left, because the the exposure time is more likely to occur during the first interactions of the index case with the contact. This effect is larger for the scenario where the index case is detected at symptom onset, which has a narrower range of possible exposure days. The figure suggests that for the other two scenarios (random LFA test and weekly LFA test), the exposure time distribution is not very sensitive to the infectivity parameter.

The simulations were generated using multiple values of $\beta$ (0.01, 0.1, 0.5 and 1.0) to assess whether the efficiency of the testing schedules are sensitive to the infectivity profile. This is important because infectivity may vary depending on the context, including the usage of personal protective equipment, indoor ventilation, vaccine adoption, type of contact (e.g. household) and potential risk factors (\cite{hu2021infectivity}).

\subsection{Modeling testing strategies}

Expected infecting days can be reduced with contact tracing and immediate quarantine. Note that quarantine at $t=1$ does not fully mitigate the contact's infecting days because the infectious period of the contact may start before the index case was detected. As an alternative to quarantine, identified contacts may continue with active circulation with a test schedule to detect a potential infection, thereby reducing the costs of unnecessary quarantines when the contact case has not been infected. \colsm{A test schedule is defined as a set of test interventions on specified dates, where each test performed has an associated LOD, sensitivity and delay to inform the test result. Two types of tests were considered for this analysis: (1) PCR test, with LOD=$10^{3}$ and a one-day delay to report results, and (2) LFA test, with LOD=$10^{4.5}$ and immediate reporting (zero delay).}

\begin{figure}
    \centering
    \includegraphics[scale=0.25]{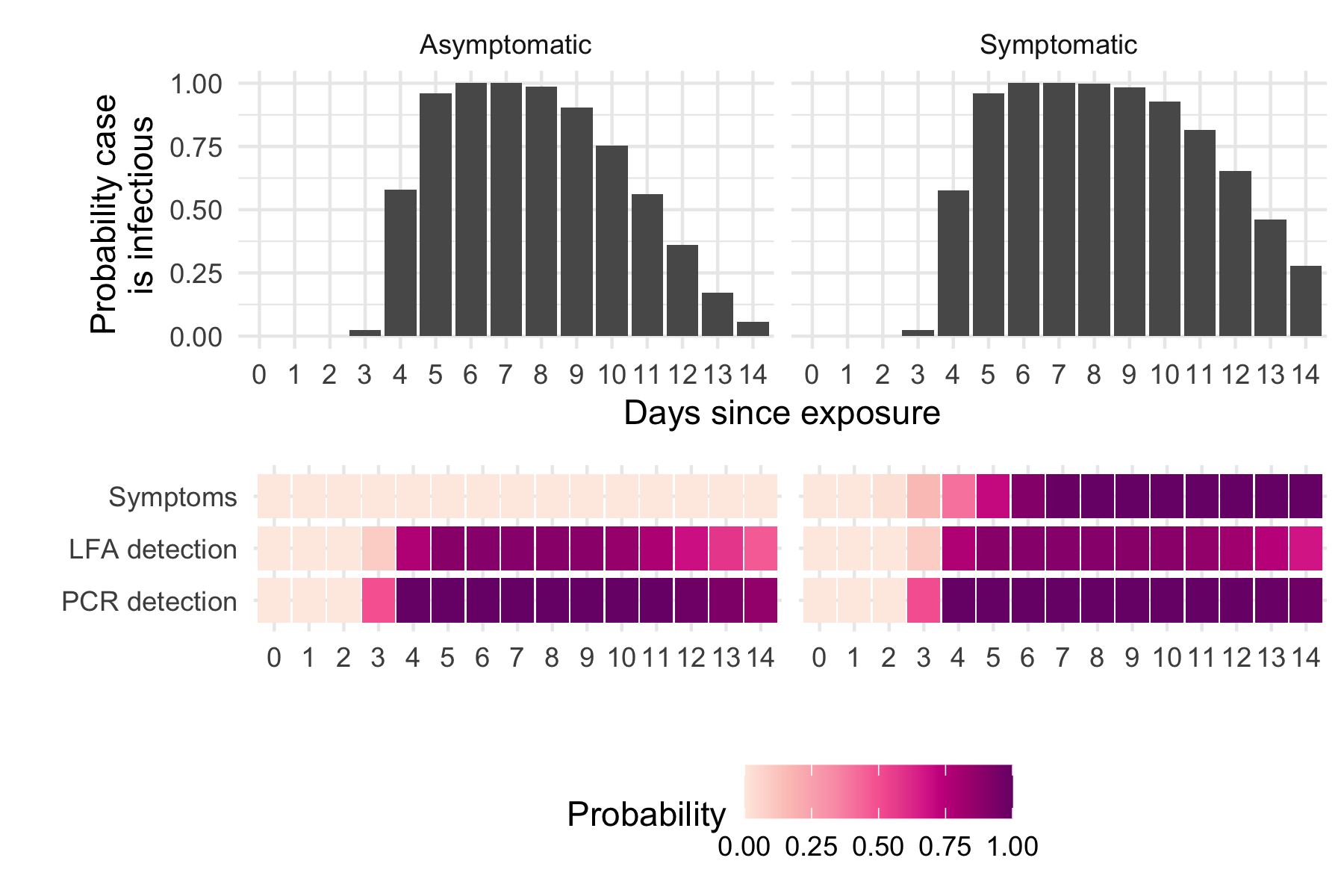}
    \caption{\textbf{Infectiousness of a Covid-19 case}, based on 10 thousand viral load simulations of asymptomatic and symptomatic cases with exposure time at $t=0$. The bottom panel of the figure shows the probability of developing symptoms on each day following exposure and the sensitivity of PCR tests and LFA tests (using the \textit{LFA Med} scenario from Table \ref{tab:sensitivity}.}
    \label{fig:inf_prob}
\end{figure}

\colsm{The sensitivity of the test depends on the viral load of the subject, the test's LOD and the quality of the sample swab (pre-analytic factors).
Table \ref{tab:sensitivity} describes the sensitivity of the different tests considered in our analysis. Sensitivity of PCR is set to 100\% above a viral load of $10^3$ (\cite{vogels2020analytical}). For LFA, we considered different scenarios which account for reported differences among manufacturers and the type of sample swab used. \cite{scheiblauer2021comparative} compare more than 100 LFA tests, reporting test sensitivity relative to PCR for different cycle threshold values. We converted cycle thresholds to viral load assuming an equivalence of Ct values of 35 and 25 to viral loads of $10^3$ and $10^6$, interpolating for intermediate values assuming that a Ct increase of one corresponds to a factor of two in the viral load concentration. We used the top decile of the reported tests to define the \textit{LFA High} scenario. Note that the sample swabs in that study where collected by professional health care personnel, which can be higher compared to self-collected samples (i.e. home testing). Studies by \cite{lindner2021head}, \cite{frediani2021multidisciplinary}, and \cite{brummer2021PlosMed} comparing self-collected versus professional-collected swabs show differences in test sensitivity in the order of 5-15\%. Hence, we considered an intermediate sensitivity parameter reducing the high performance LFA test by 10\%, generating the \textit{LFA Med} scenario shown in Table \ref{tab:sensitivity}. The \textit{LFA Med-Low} scenario was constructed using the bottom decile of the tests studied in \cite{scheiblauer2021comparative}, using professional-collected swabs; the LFA Low reduces that sensitivity by 10\% to account for a self-collected swab.
\textcolor{blue}{
This range of sensitivity considered for LFA tests is consistent with the values reported in recent studies (\cite{pilecky2022performance}, \cite{wagenhauser2021clinical}).}
PCR tests are reported with a one day delay, while LFA tests provide immediate results with no delay in all the scenarios.
Figure \ref{fig:inf_prob} shows the infectiousness profile and test sensitivity during post-exposure date ($t=0$) based on the simulated viral load paths of our model.
}

\begin{table}[hbp]
\centering
\begin{tabular}{lllclcccl}
\hline
Ct              & \multicolumn{1}{c}{log(Viral load)}     &  & PCR   &                      & \multicolumn{4}{c}{LFA}                 \\
                & \multicolumn{1}{c}{}                    &  &       &                      & High     & Med     & Med-Low     & Low \\ \cline{1-2} \cline{4-4} \cline{6-9} 
\textless 25    & \textgreater 6  &  & 1.0   &                      & 1.0      & 0.9     & 0.85    & 0.75     \\
{[} 25 , 30 {]} & {[} 4.5 , 6 {]} &  & 1.0   &                      & 0.85     & 0.75    & 0.15    & 0.05     \\
{[} 30 , 35 {]} & {[} 3 , 4.5 {]} &  & 1.0   &                      & 0        & 0       & 0       & 0        \\ \cline{1-2} \cline{4-4} \cline{6-9} 
\multicolumn{2}{c}{Test delay}                            &  & 1 day & \multicolumn{1}{c}{} & \multicolumn{4}{c}{No delay}          \\ \cline{1-2} \cline{4-4} \cline{6-9} 
\end{tabular}

\caption{Sensitivity of the different types of tests considered in the study. PCR test have 100\% sensitivity when viral load exceeds $10^6$ (cp/ml), and zero below that threshold. Four scenarios of sensitivities were considered for LFA tests (High, Med and Low), all with a LOD of $10^{4.5}$.}
\label{tab:sensitivity}
\end{table}

\begin{figure}
\centering

\centering
\includegraphics[scale=0.16,trim=0 80 0 0,clip]{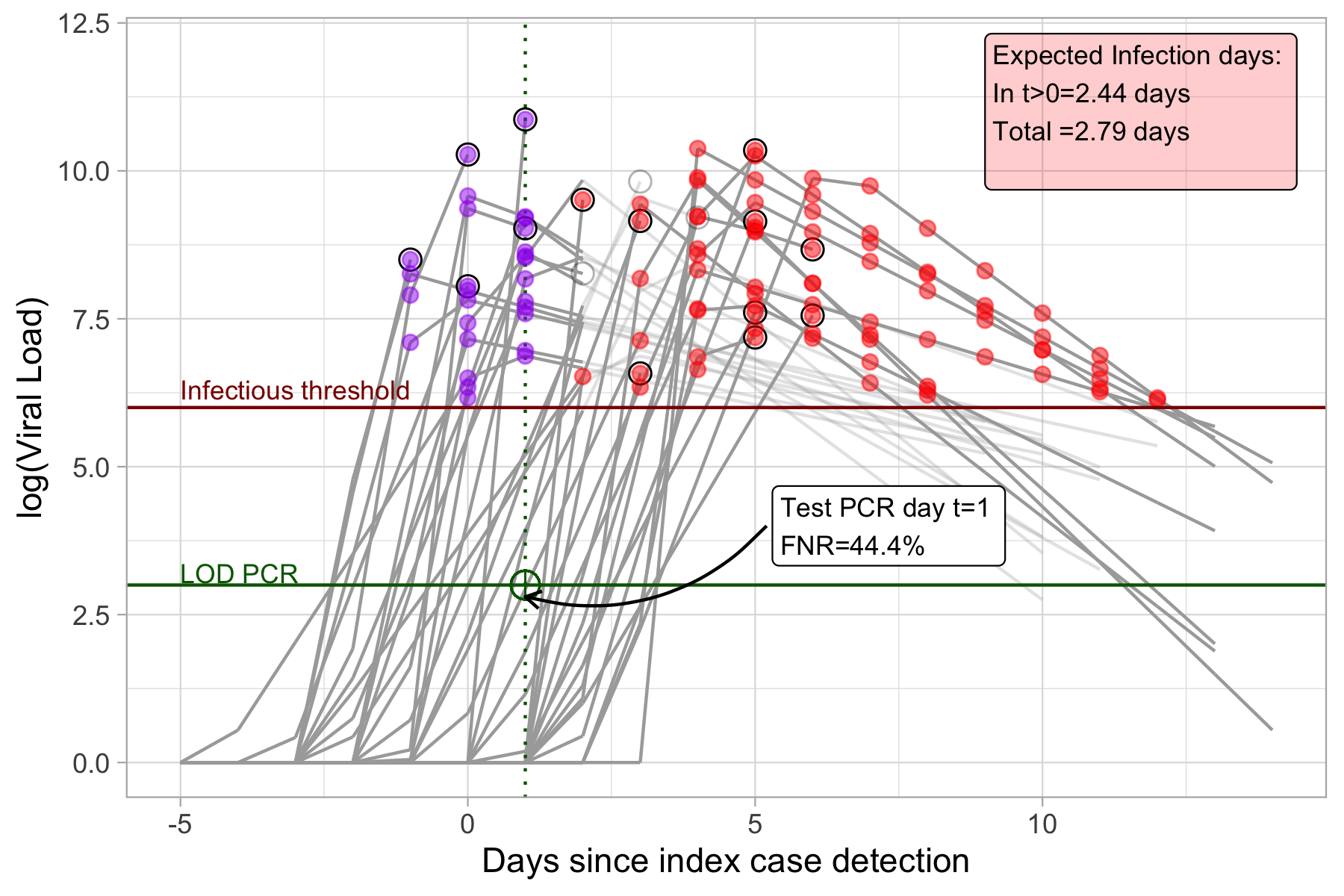}
\includegraphics[scale=0.16,trim=0 80 0 0,clip]{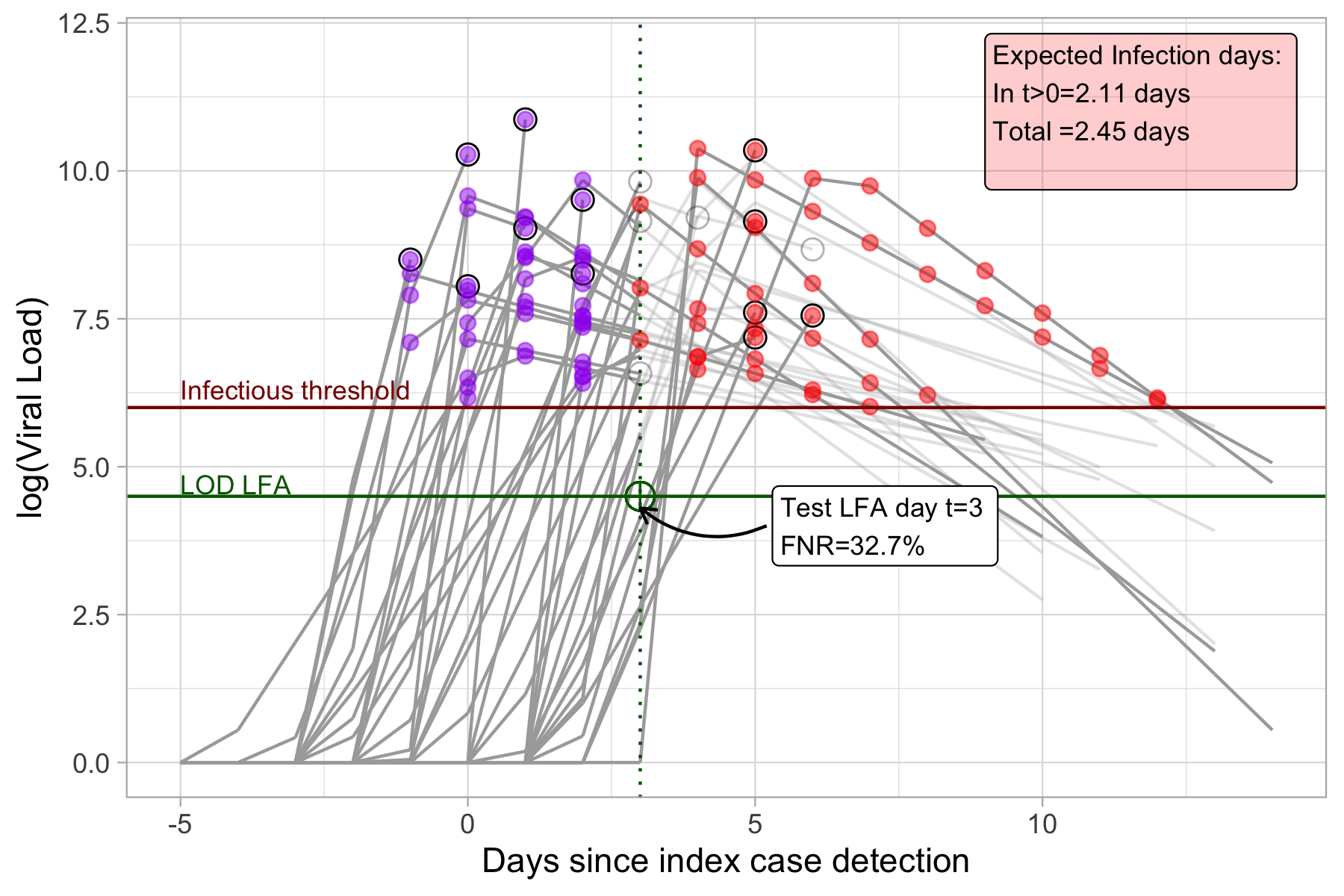}
\includegraphics[scale=0.16]{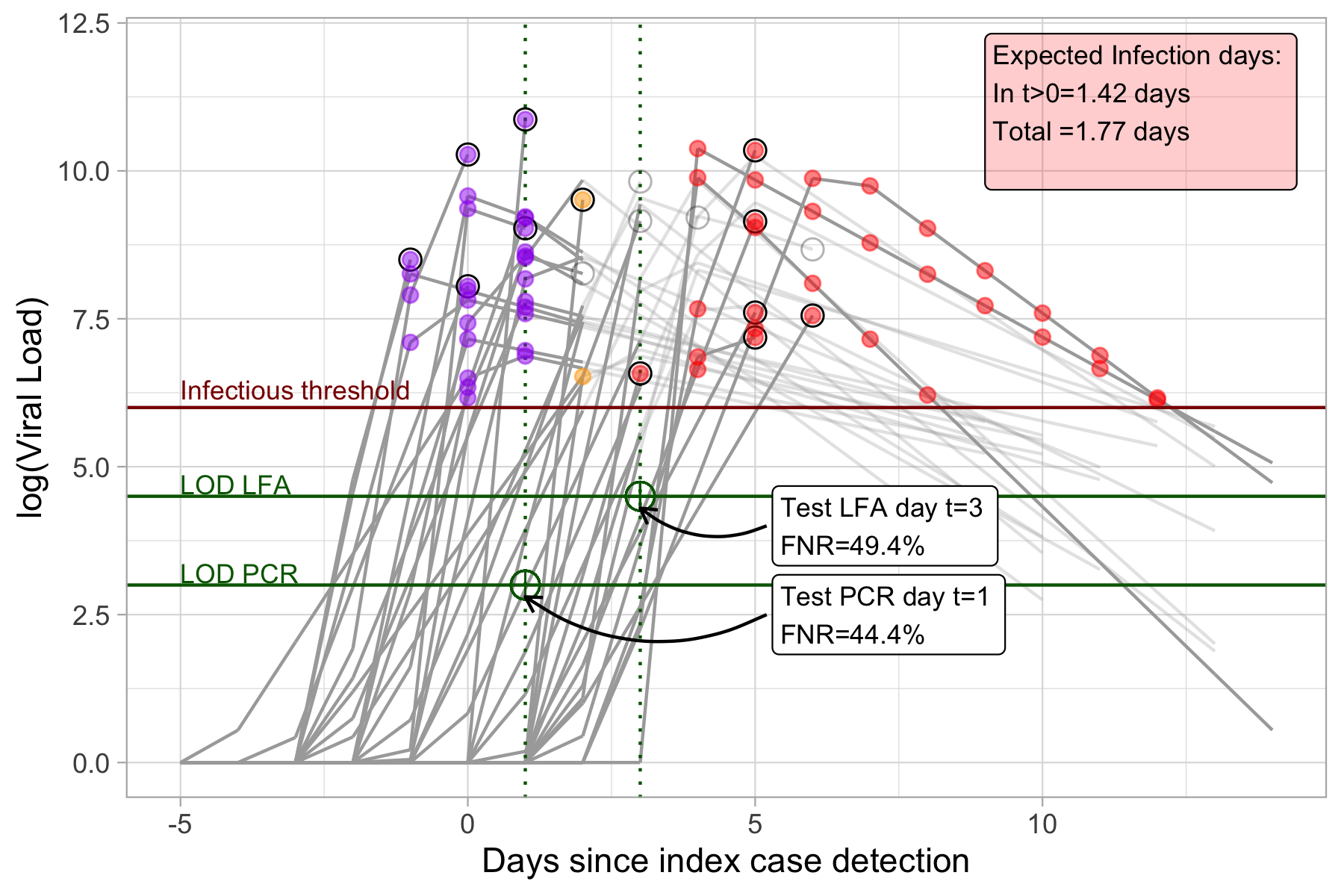}
\caption{\textbf{Examples of test schedules for an infected contact and their impacts on the infecting days.} \colsm{The top panel shows a schedule with the PCR test on day 1 after index case confirmation. Purple dots indicate infecting days for the contact when detected by the test and red dots show the infecting days for undetected cases. Viral paths are shown in light gray after they are detected by the corresponding test. The middle panel shows the performance of an LFA test on day 3. The bottom panel shows the performance of two sequential tests, PCR on day 1 and LFA on day 3, with the yellow dots representing the infection days for the scenarios that are detected with the second test after a false negative in the first test.}}
\label{fig:ag_t1}
\end{figure}

\colsm{The false negative rate (FNR) of a test is defined as the probability of obtaining a negative test result on an infected subject. In our simulation, the FNR of PCR tests can be calculated as the fraction of sample paths with viral load below the LOD of the test ($10^3$ cp/ml). The top panel of Figure \ref{fig:ag_t1} illustrates an example of a test schedule with one PCR test implemented one day after the index case detection ($t=1$). The contact is isolated when the test gives a positive results with one day of delay, and the infecting days for these cases correspond to the purple dots shown in the figure. Note that these detected case do infect during the day the PCR test was taken due to the one day delay in reporting. Negative results filter out all the sample paths with viral loads above LOD=$10^3$ on day $t=1$: all of these paths are discarded when test results are reported; therefore, an infected individual could evolve on only one of the remaining paths with viral loads below the LOD on the test date. The discarded paths are ``grayed-out`` in the figure, and their infection days are eliminated.
}

\colsm{The red dots in Figure \ref{fig:ag_t1} represent the possible infecting days when the infected contact remained active in the community after a false negative PCR test result. The fraction of paths above the infectious threshold that have not been isolated represents the probability that the individual is infectious on that day. These infecting days, which are referred to as the \textit{residual risk} (\cite{van2021intra}), are generated by the paths that were not filtered out by the PCR test on day 1. Considering both scenarios, namely, a true positive and false negative test result, the expected infecting days (conditional on infection at exposure) is equal to 2.79 in this example (shown by the label in the upper-right corner of the top panel). Of these total expected infecting days, 2.44 correspond to days after the index case detection date and could have been avoided by a strict quarantine of the contact.
}

\colsm{The middle panel of Figure \ref{fig:ag_t1} shows a test schedule with an LFA test performed at day $t=3$. In this example, we use the \textit{LFA Med} parameters from Table \ref{tab:sensitivity}, hence there could be false negatives even when viral load exceeds the LOD of the test. Despite its lower sensitivity relative to PCR testing, the FNR for this LFA test drops (relative to the PCR test on day 1) because a larger fraction of sample paths exceeds the LOD on day 3, thus implying a higher chance of detection. The expected infection days prior to the positive test result increase for this test schedule (i.e. there are more purple dots); however, this increase is compensated with a larger reduction in the residual risk of false negative results (fewer red dots). The overall effect is that delaying the PCR test from day 1 to a LFA test on day 3 reduces the overall expected infection days from 2.79 to 2.45.}

\colsm{The bottom panel of Figure \ref{fig:ag_t1} illustrates a test schedule that combines PCR and LFA tests taken on days 1 and 3 respectively. Note how the first PCR test on day 1 was capable of detecting infected contact in scenarios where infection occurred on earlier exposure dates, which reduces the infection days for the scenarios that are detected with the LFA test on day 3. This initial ``filtering`` of cases at day 1 also increases the FNR of the LFA test on day 3. Altogether, incorporating an additional PCR test on day 1 to a LFA test scheduled on day 3 reduces the expected infecting days from 2.45 to 1.77.}

\colsm{The above examples are provided to illustrate our modeling approach, which can be applied to different types of tests varying their LOD, sensitivity and reporting delay.} We applied this methodology to study all possible test combinations that can be generated with up to two PCR tests and five LFA tests within the 8 days following the index case detection date, considering different numbers of tests and testing dates.

\colsm{Although the proposed strategy of sequential testing increases the likelihood of false positives (thereby generating quarantines for non-infected contacts), we found that its impact is negligible relative to the benchmark policy of quarantining all close contacts. For example, for a LFA specificity of 99\% (\cite{frediani2021multidisciplinary} and \cite{scohy2020low}), a healthy close contact would have a probability of ($ 1-(0.99)^n $) of a false positive if $n$ tests are performed. Hence, a testing strategy with three LFA tests leads to a false positive rate of 0.03: for every 100 non-infected close contacts that would be subject to a strict quarantine, there would be, on average, only 3 people undergoing an unnecessary quarantine when the sequential testing strategy is used. This false positive rate can be further reduced by conducting repeated testing after a positive result, as suggested for surveillance testing (\cite{mercer2021testing}).
}

The results of the analysis are presented next. All the data used in this analysis are synthetic and generated via simulation using Python and R code, to be made publicly available.

\section{Results}
\label{results}

\begin{figure}
    \centering
    \includegraphics[scale=0.25]{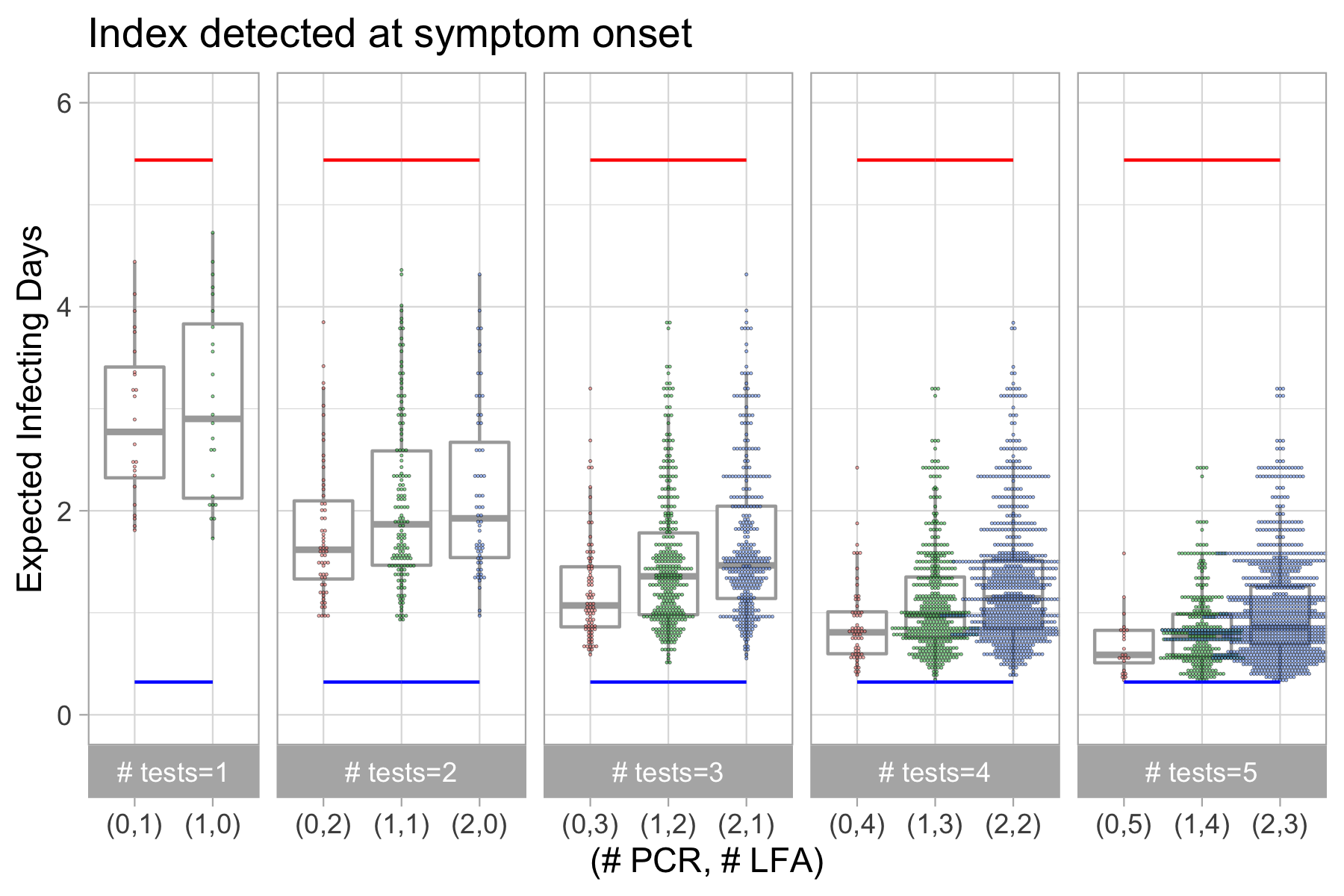}
    \begin{overpic}[scale=0.25,trim=0 0 0 75,clip]{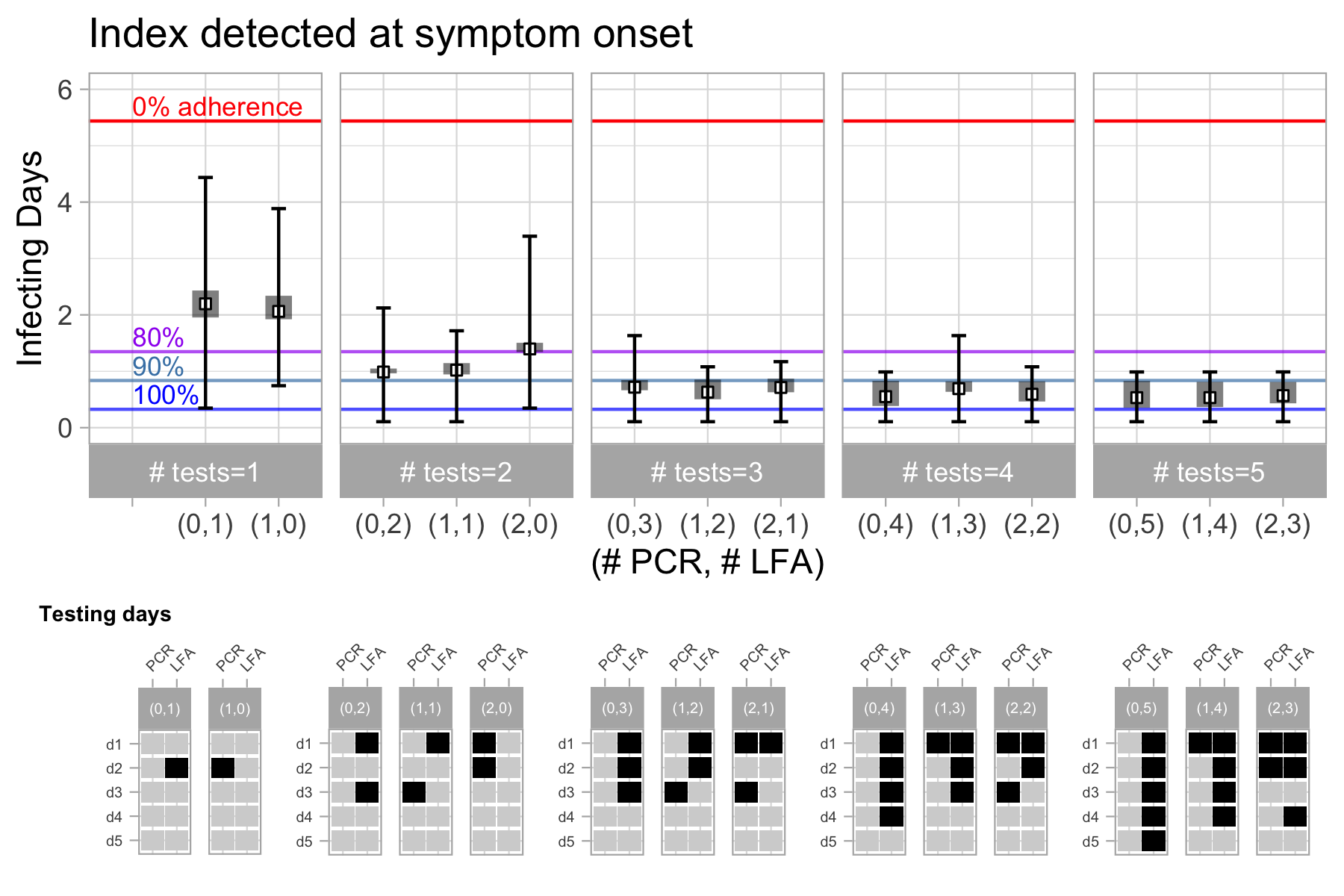}
        \put(85,46){\includegraphics[scale=0.32]{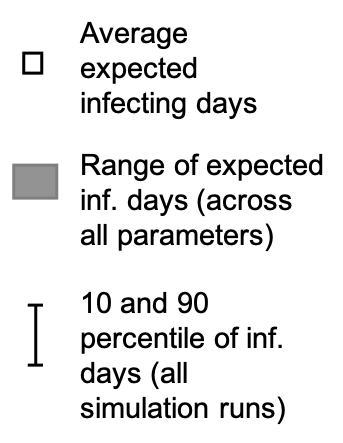}} 
    \end{overpic}

\caption{\textbf{Evaluation of testing policies for a traced contact exposed to an index case identified by symptom onset.} \small{In the upper and middle panels, the horizontal axis contains the number of PCR and LFA tests. Blue, green, purple, and red horizontal lines correspond to the average infecting days when traced contact is quarantined for 14 days with adherence of 100\%, 90\%, 80\%, and 0\%, respectively. In the upper panel,
each dot displays the performance of a testing schedule and infectivity parameter, and the lower and upper limits of boxes are the 25\% and 75\% quartiles. 
For the robust testing policies,  the middle panel displays
the average expected infecting days (small squares), the range of the expected infecting days across all parameters (gray rectangles)   and the 10\% and 90\% percentiles \newmo{across all the simulation runs analyzed}.
The lowest panel shows the schedule of the robust testing policy for each group of tests.
}}
    \label{fig:symp}
\end{figure}

We evaluated all testing policies considering a maximum of 2 PCR and 5 LFA tests. \colsm{For LFA tests, we considered the \textit{LFA Med} scenario described in Table \ref{tab:sensitivity} to model the sensitivity of the test. Later we show some results with other scenarios of the LFA sensitivity parameters.} Figure \ref{fig:symp} shows the results for the scenario where the contact was exposed to an index case detected at symptom onset. The top panel shows the performance of different numbers and combinations of tests, thus allowing two tests of different types on the same day, and different values of the infectivity parameter $\beta$ (0.01, 0.1, 0.5 and 1.0).
Each dot in the plot shows the expected infecting days of a feasible testing policy for a fixed infectivity parameter. The dispersion across testing policies is illustrated with dot plots and box plots, and the policies are grouped by the number of PCR and LFA tests used, with each pair (\#PCR,\#LFA) indicating the number of tests of each type. Dot plots with higher densities represent clusters of policies that achieve similar performance.
Testing policies are ordered from lower to higher costs on the horizontal axis; because PCR tests are typically more costly, policies within the same group are reported in increasing order of PCR tests. We notice that the costs of PCR testing can be lowered by pooling specimens from multiple samples; however, this cost reduction is less effective when prevalence is high, as would be expected with effective contact tracing (\cite{cherif2020simulation}).

The horizontal red line shows the expected number of infecting days of the traced contact when he/she remains active in the community until self-isolation only at symptom onset (for asymptomatic cases, there is no isolation), giving an upper bound of 5.44 expected infecting days when neither testing nor quarantine are used. The horizontal blue line shows the lowest expected number of infecting days of the traced contact if he/she is \emph{immediately} quarantined upon confirmation of the index case (with 100\% adherence), 
and it is equal to 0.26 expected infecting days, which represents a lower bound on the performance of all possible testing policies. The analysis suggests that with \newmo{5} tests, the averted risk reaches this lower bound; therefore, all reported results are limited to \newmo{5} tests or fewer (LFA and PCR combined).

For each pair (\#PCR,\#LFA), we identified the \textit{optimal policy} by selecting the testing schedule that minimizes the expected infecting days; we found that the optimal testing schedule was similar across all the parameter values of infectivity ($\beta$) that were used to simulate the exposure time distribution, with some exceptions. An example where the optimal policy changes with $\beta$ is when a single test is available: the simulations using a higher infectivity parameter suggest that earlier testing is more efficient to avert risk, because it is more likely that the contact was exposed earlier (see Figure \ref{fig:exposure}).
When the optimal testing schedule changes depending on the infectivity parameter, we also identify the policy that minimizes the worst-case scenario (i.e. highest expected infecting days) across all values of $\beta$, hereon referred to as the \textit{robust testing policy.}

The middle panel of Figure \ref{fig:symp} shows in further detail the performance of the robust testing policy for each pair (\#PCR,\#LFA). The small squares represents the average expected infecting days and the gray rectangles the range of expected infecting days, across all the values of the infectivity parameters used in the simulation. The error bars indicate the 10\% and 90\% percentiles of the number of infecting days across all the simulated sample paths for the selected policy. This graph also includes two additional benchmarks indicated by the light blue and purple horizontal lines, which correspond to a 14-day quarantine with 90\% and 80\% adherence (but imposing full isolation at symptom onset of the infected contact).

The bottom panel shows in further detail the days in which the tests are performed for the robust testing policies (black squares represent the days when the PCR/LFA tests should be performed). \colsm{In Appendix \ref{sec:add_results}, Tables \ref{TT1}, \ref{TT2}, and \ref{TT3}  show the detailed testing schedules for all robust policies, and the false negative rates at each epoch when a test is performed.}

Figure \ref{fig:symp} suggests that sequential testing strategies can be an effective alternative to quarantines to avert secondary infection risk of traced contacts. For example, two LFA tests can lead to a lower risk relative to a 14-day quarantine with \newmo{85\%} adherence; and three \newmo{LFA tests can be as effective as a quarantine with 90\% adherence.}

However, the results also suggest that the timing of these tests is highly relevant. The optimal schedule of the two LFA tests is on days 1 and 3, thus leading to 0.99 expected infection days. However, changing to a testing schedule on days 1 and 2 deteriorates the performance to 1.49 expected infecting days, a 50\% increase on the risk relative to the optimal strategy. Similarly, when using 2 LFAs and 1 PCR, the optimal schedule on days 1 and 2 for the LFA and day 3 for the PCR leads to an average of 0.63 infection days compared to 1.37 days when using a schedule of LFAs in days 1 and 2 and PCR at day 1 (a 117\% increase in the risk of secondary infection). The top panel of Figure \ref{fig:symp} shows significant dispersion on the performance across testing strategies using the same number of tests, suggesting that optimizing the dates of the tests matters.

\subsection{Analysis with other types of index case detection}

\begin{figure}
    \centering
        \includegraphics[scale=0.25]{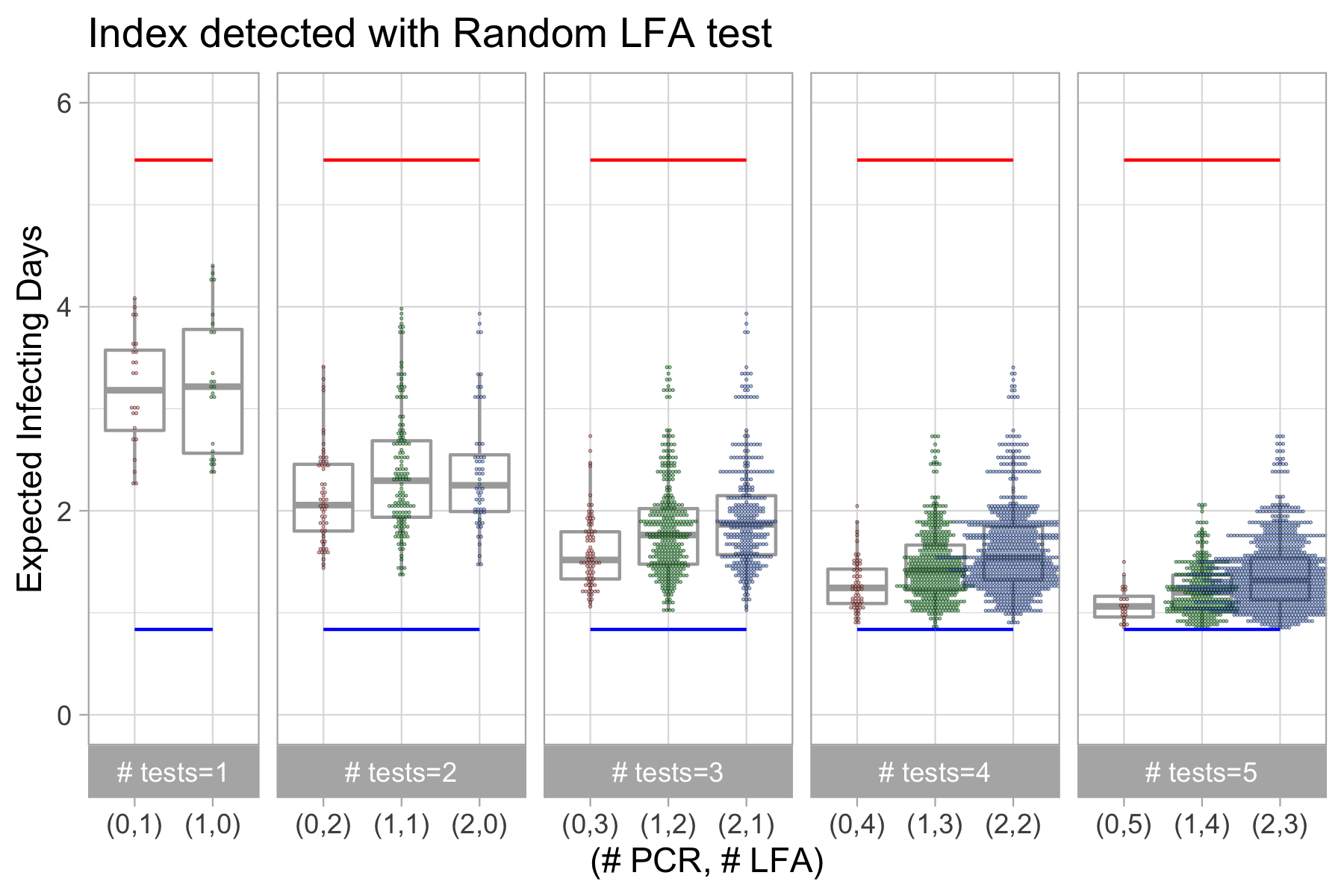}
    \begin{overpic}[scale=0.25,trim=0 0 0 75,clip]{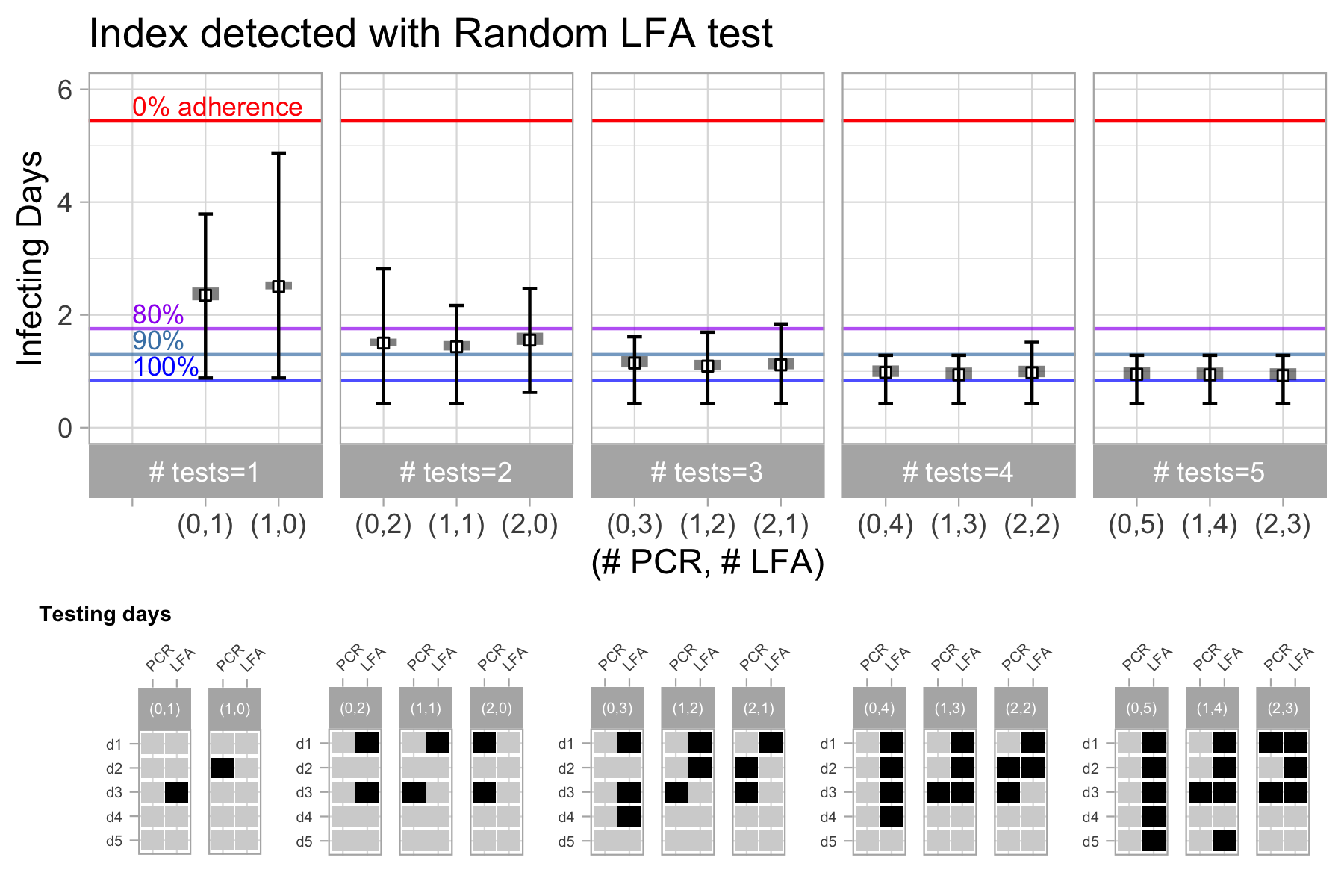}
        \put(85,46){\includegraphics[scale=0.32]{figures/legend_results.png}} 
    \end{overpic}
\caption{\textbf{Evaluation of testing policies for a traced contact exposed to an index case detected by random LFA test.}
\small{In the upper and middle panels, the horizontal axis contains the number of PCR and LFA tests. Blue, green, purple, and red horizontal lines correspond to the average infecting days when traced contact is quarantined for 14 days with adherence of 100\%, 90\%, 80\%, and 0\%, respectively. In the upper panel,
each dot displays the performance of a testing schedule and infectivity parameter, and the lower and upper limits of boxes are the 25\% and 75\% quartiles. 
For the robust testing policies,  the middle panel displays
the average expected infecting days (small squares), the range of the expected infecting days across all parameters (gray rectangles)   and the 10\% and 90\% percentiles.
The lowest panel shows the schedule of the robust testing policy for each group of tests.}
}
    \label{fig:LFA}
\end{figure}

Figure \ref{fig:LFA} shows the results for the scenario when the contact was exposed to an index case detected by a LFA test. In this scenario, the index case has no symptoms at the moment of detection and hence could be presymptomatic or asymptomatic, which in turn affects the possible dates of exposure. Specifically, since we model an environment where contacts are recurrent, the range of possible dates of infection is longer when the index case is asymptomatic (see Figure \ref{fig:exposure}). This longer time period of exposure increases the likelihood that the contact is already infectious at the time the index case is detected. Consequently, the lower bound represented by the blue horizontal line, which was attained with immediate quarantine of the traced contact at $t=1$ and 100\% adherence, leads to an expected infecting days of 0.78, which is significantly higher than the 0.26 bound attained when the index case is detected at symptom onset, see Figure \ref{fig:symp}. The upper bound, illustrated by the red line, is the expected infecting days without quarantine or testing, with isolation only at symptom onset. Hence, this upper bound does not depend on the exposure time of the contact.

In qualitative terms, the results of Figure \ref{fig:LFA} (i.e. index case detected by LFA) are similar to those obtained in Figure \ref{fig:symp}. Two LFA tests with optimal testing time reduce the secondary infection risk relative to a 14-day quarantine with 80\% adherence, and adding a third LFA test attains a lower risk relative to a quarantine with 90\% adherence. The optimal testing schedule for each PCR/LFA combination was similar across all the infectivity parameters applied in the simulation.

The two scenarios analyzed in Figures \ref{fig:symp} and \ref{fig:LFA} differ in the probability distribution of the exposure days (presented in Figure \ref{fig:exposure}). An intermediate scenario can be analyzed when the index case is detected by a weekly surveillance LFA test, with a range of 7 exposure days prior to index case detection. The results of this scenario, as reported in Figure \ref{fig:LFAweekly} in the Appendix, are qualitatively similar to those obtained in the previous two scenarios. The main difference is that the lower bound attained with immediate quarantine with 100\% adherence reaches 0.28, which represents a 64\% reduction relative to the bound attained when the index case is detected with a random LFA test. Hence, increasing the frequency of a surveillance testing program is useful for improving the case detection rate and simultaneously increasing the efficiency of contact tracing.

\subsection{\newmo{Alternative scenarios of LFA test sensitivity}}

Overall, the results suggest that sequential testing with LFA tests is a cost effective alternative to quarantines to mitigate infection risk of traced contacts, when considering tests of moderate sensitivity levels (90\% for viral loads above $10^6$ (cp/ml) and 75\% for loads between $10^{4.5}-10^6$ (cp/ml)). However, there is mixed evidence of the sensitivity of LFA in practice and it is of interest to analyze how this test sensitivity may impact the effectiveness of the sequential testing strategy relative to quarantines of close contacts.

\begin{figure}
    \centering
    \includegraphics[width=0.8\columnwidth]{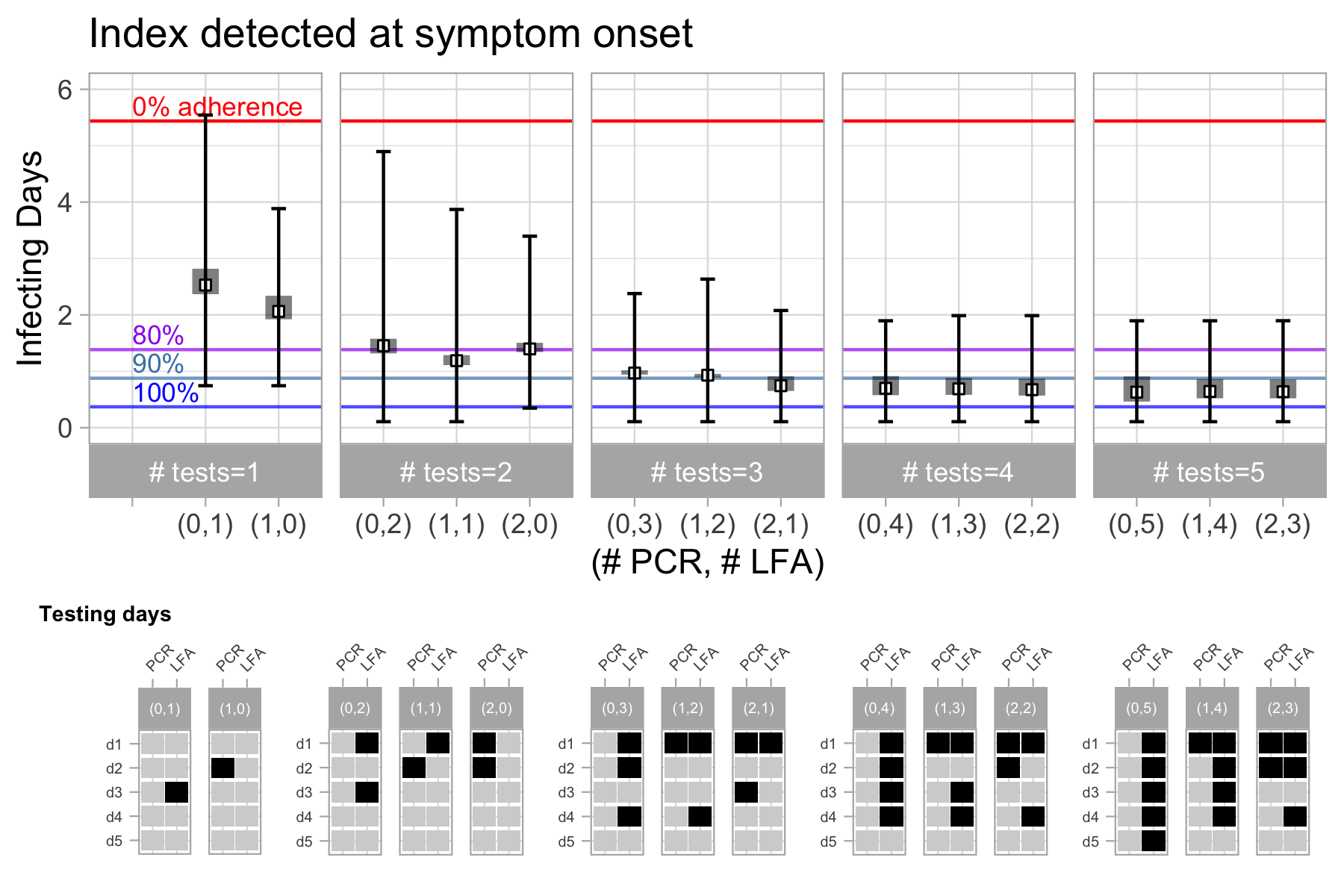}
    \includegraphics[width=0.8\columnwidth]{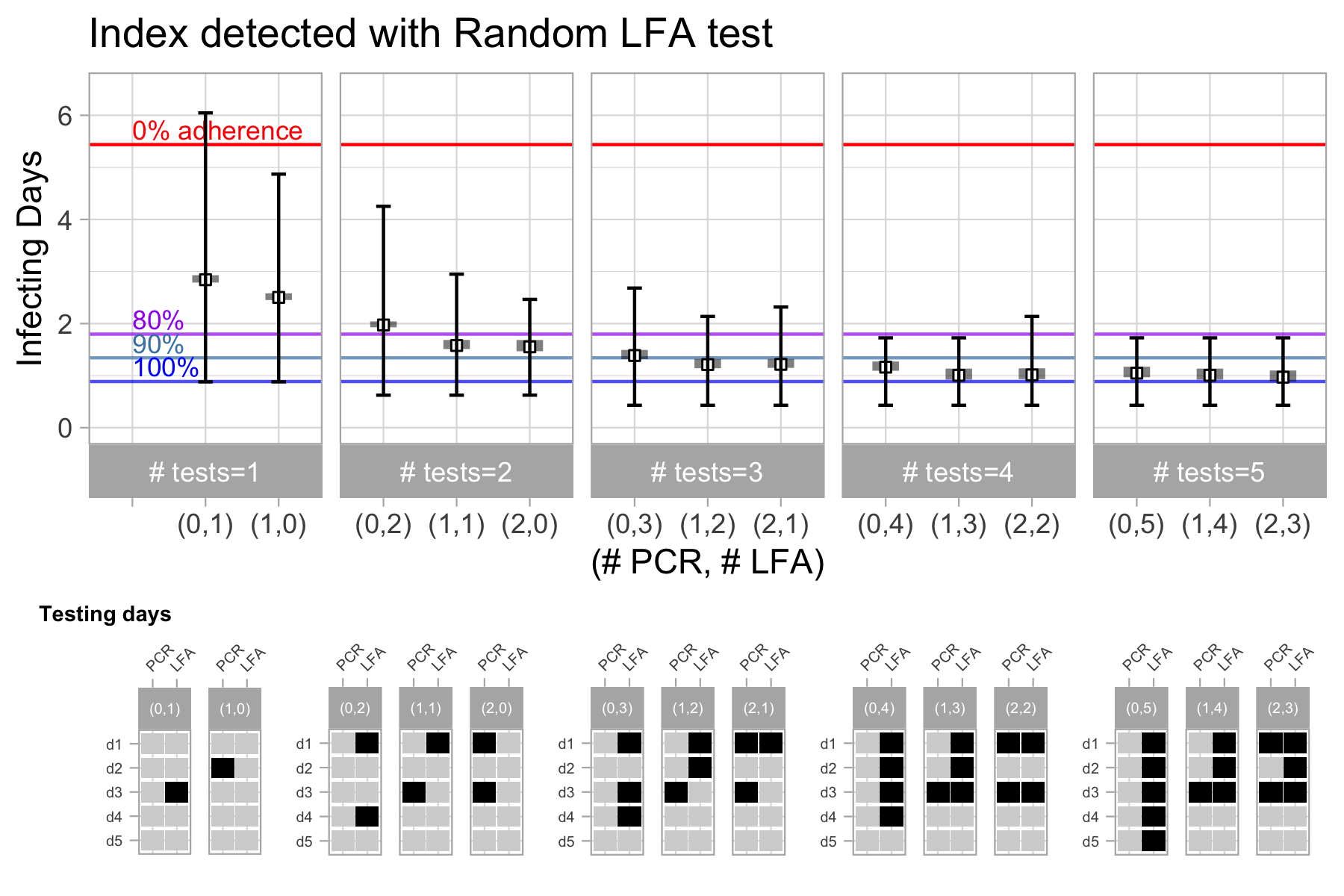}
    \caption{Robust testing policies for the \textit{LFA Med-Low} test sensitivity scenario.}
    \label{fig:results_test_MedLow}
\end{figure}

Figure \ref{fig:results_test_MedLow} shows the robust testing schedules generated based on the simulations with the \textit{LFA Med-Low} test sensitivity scenario (see Table \ref{tab:sensitivity}). As expected, there is an increase in the expected infecting days for the testing schemes using LFA tests alone; this drop is more pronounced when using one or two LFA tests, and is negligible when using three or more tests. This suggests that using sequential testing may compensate the lower sensitivity of the tests. Interestingly, a testing scheme with one LFA and one PCR achieves a similar performance relative to 2 sequential PCR tests, showing the value of combining tests with different LODs. The robust testing schedules for other levels of LFA test sensitivity are reported in Appendix \ref{app:sec:test_sensitivity}. 

\begin{figure}
    \centering
    \includegraphics[scale=0.25]{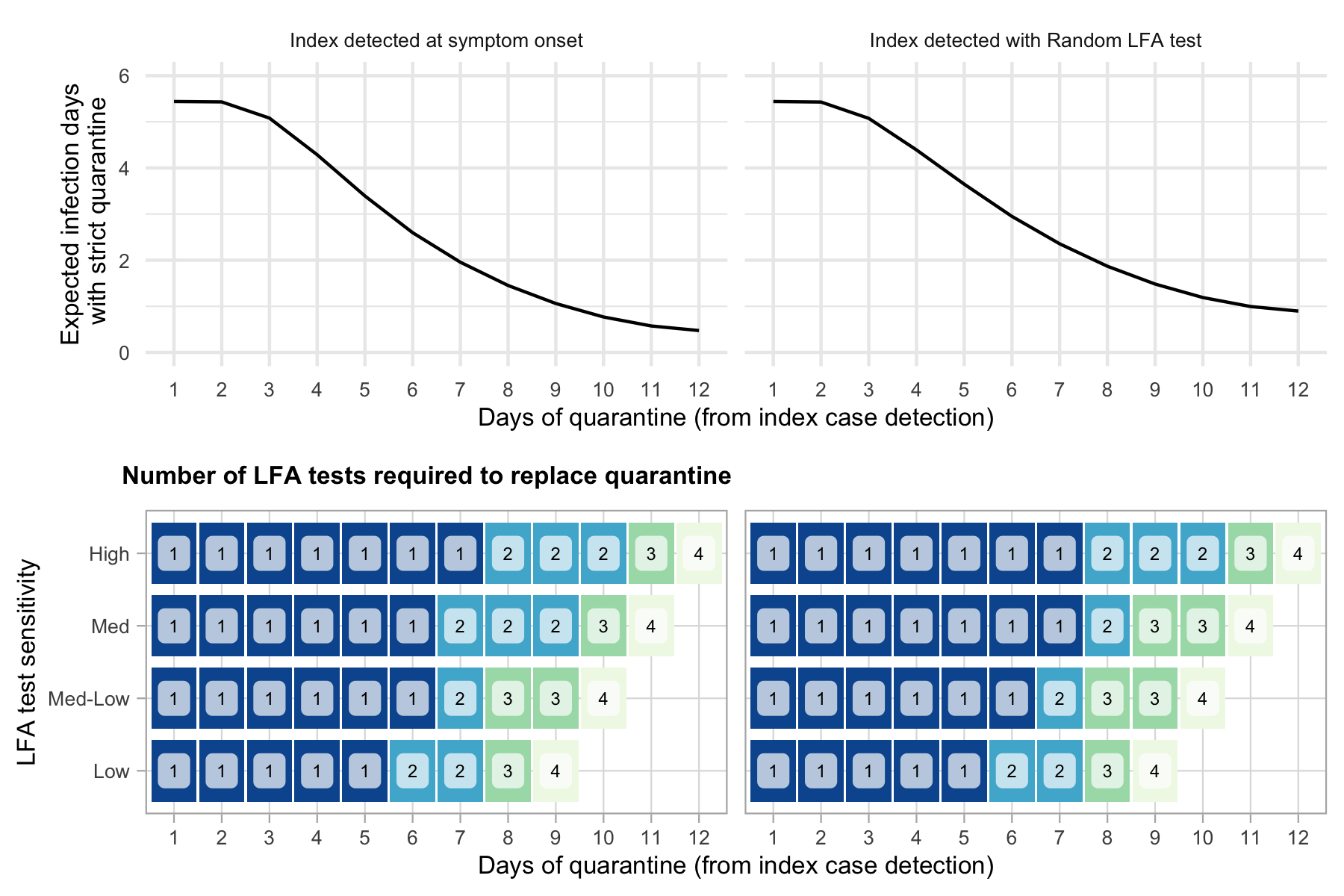}
    \caption{\textbf{Substituting quarantines with sequential testing without increasing infection risk.} The top panel shows the infection days of a pure quarantine strategy of different time extension (1 to 12 days) and 100\% adherence. Quarantine of the contact begins with the detection of the index case and each facet shows the calculations for alternative scenarios of index case detection. The bottom panel shows the number of LFA tests that would be needed to replace the quarantine with an optimal sequential testing strategy (with no quarantine when tests are negative) without increasing infection risk, considering different levels of test sensitivity (see Table \ref{tab:sensitivity}). Blank entries indicate that the corresponding quarantine time extension cannot be replaced with 5 or fewer tests.}
    \label{fig:sens_vs_quar}
\end{figure}

Our analysis has focused in comparing risk-reduction strategies of sequential testing with no quarantine relative to a 14-day quarantine with different levels of adherence. As an alternative, in what follows, we compare the performance of this sequential testing strategy with quarantines of different time extensions.  The top panel of Figure \ref{fig:sens_vs_quar} shows the expected infection days of using strict quarantines (with 100\% adherence and no testing) of different time length, between 1 and 12 days, starting at the time of detection of the index case. Expected infecting days is calculated taking an average across all the values of the infectivity parameter $\beta$ used in the simulations, and these calculations are presented separately for the different alternatives of index case detection. Quarantines of three or fewer days lead to more than 5 infection days on average, which can be significantly reduced by increasing the length of the quarantine, down to one or lower infection days for quarantines of more than 10 days. Note that the expected infecting days do not reach zero as we increase the length of the quarantine because the infection days of the contact prior to index case detection are unavoidable.

The bottom panel shows the number of sequential LFA tests without quarantine that reach the same effectiveness as a strict quarantine, for different time lengths. This number is calculated for the alternative scenarios of LFA test sensitivity reported in Table \ref{tab:sensitivity}. For example, using four LFA tests of High sensitivity (with the robust testing schedule shown in Figures \ref{fig:symp} and \ref{fig:LFA}) and no quarantine if results are negative, achieves a lower risk than 12 days of strict quarantine.

The results in Figure \ref{fig:sens_vs_quar} also suggest that test sensitivity is relevant to determine the effectiveness of sequential testing strategy. When using Low sensitivity LFA tests, it may not be possible to achieve the effectiveness of a 12 day quarantine: the simulations show that the best outcome that can be achieve with up to 5 LFA tests is equivalent to a 9-day strict quarantine, which is about 0.5 infection days higher than what could be achieved with the same number of tests with High sensitivity or, alternatively, 12 days of strict quarantine. Nevertheless, short quarantines of 7 or fewer days -- a policy that has been used by many countries to quarantine incoming travelers (e.g. Thailand, Italy, and Austria as for Dec 21st 2021, \url{https://www.kayak.com/travel-restrictions})
 -- could be replaced by 2 sequential LFA tests without increasing infection risk (the optimal schedule of these tests varies depending on the test sensitivity). In summary, these results reinforce the conclusion that sequential testing of close contacts can be an effective alternative to quarantines, even when using low sensitivity tests, if the timing of these tests is set appropriately.

\subsection{\newmo{Comparing alternative models of viral load trajectories}}

Our main results were generated based on the viral load trajectories developed in \cite{larremore2021test}, who combined empirical data from several sources published during 2020. Recent studies have shown that  some of the variants of concern may exhibit different viral load trajectories. \cite{jones2021estimating} shows that the Alpha variant tends to rise faster leading to an earlier infectious period. \cite{singanayagam2021community} shows differences in the rise and decline of viral loads among pre-Alpha, Alpha and Delta variants.

To check the robustness of our results, we replicated all the analysis using the more recent study by \cite{jones2021estimating}, who estimate viral load trajectories based on sequential PCR tests of more than 25 thousand confirmed Covid-19 cases in Germany. The detailed results of this analysis are reported in Appendix \ref{app:sec:jones}, leading to the following conclusions:

\begin{itemize}
    \item The performance of sequential testing relative to strict quarantines of traced contacts is similar to what we obtained in our main results. Based on the simulations with the \textit{LFA Med} test sensitivity scenario, we find that two tests with no quarantines is equivalent to a 14 day quarantine with 80-90\% adherence. The performance with 3 and 4 tests is equivalent to 90-100\% adherence. The performance with 5 tests is close to 100\% adherence of the 14-day strict quarantine.
    \item With the lower sensitivity scenario \textit{LFA Med-Low}, the results based on the viral load trajectories of \cite{jones2021estimating} have small drop in performance, similar to what we obtained in our main results.
    \item The performance for the cases where the index is detected at symptom onset are similar to those obtained when the detection is with a random LFA test, a pattern that is also consistent with our main results.
    \item The optimal schedules obtained with the \cite{jones2021estimating} viral loads are similar to those reported in our main results. In \cite{jones2021estimating}, some of the tests are scheduled one day later, because the viral load trajectories in this model tend to decline more slowly, leading to a more extended infectious period and thereby a higher residual risk (see Figure \ref{app:fig:comp_jones}).
\end{itemize}

Overall, our key results appear to be robust to adjustments in the viral load trajectories. Nevertheless, it is important to continuously revise the optimal testing strategies based on new empirical evidence of viral load evolution, and on this regard our simulation approach is relatively flexible to accommodate alternative models. Although the computational times of running these simulations was reasonably fast for the purpose of this study, evaluating testing strategies in real time could be achieved more efficiently by pre-computing infectious state transitions using a simplified Markovian model, adapting the approach implemented in \cite{van2021intra} and \cite{van2021covidstrategycalculator}.

\section{Discussion}
\label{disc}

Most countries use quarantines for traced contacts and isolation for confirmed cases of COVID-19, with the purpose of avoiding the further spread of the virus. These strategies are costly, and qualitative studies show that adherence to them is highly dependent on risk perception and the degree of monitoring by the health authority (\cite{reynolds2008understanding}, \cite{saurabh2020compliance}).

In this paper, we propose an alternative to quarantines for traced contacts based on sequential PCR and/or LFA tests (with isolation of confirmed cases) and show that by choosing the appropriate test mix and timing, it is possible to reach the same
risks of secondary infections compared to that of quarantines with high levels of adherence.\newmo{ For example, the use of 4 consecutive LFAs since notification is equivalent to a 14-day quarantine with 90 to 100\% adherence; this can be achieved even when using LFA tests with relatively low levels of sensitivity.}

When considering more realistic adherence to quarantines of 80-90\%, a testing approach that consists of two or three LFA tests can actually attain a \textit{lower} risk of secondary infections compared to those with quarantines. We show that the optimal timing of these tests is important to effectively avert infectiousness of the exposed contact. For example, in the case of an index case detected at symptom onset, conducting LFA tests \newmo{of high quality but with a self-collected sample} on the first and third days after contact is determined is more effective at averting secondary risk infections relative to a 14-day quarantine with 80\% adherence (assuming 100\% compliance in the isolation of the contact when confirmed by a positive test). \newmo{In the scenarios with lower LFA test sensitivity, combining one LFA test and one PCR test achieves a similar performance compared to a 14-day quarantine with 80-90\% adherence. These conclusions hold for alternative models of viral load trajectories (\cite{larremore2021test,jones2021estimating}}).

Our modeling analysis captures three important aspects that determine the effectiveness of sequential testing to reduce the infection risk of traced contacts.

First, for a number of available tests, not all feasible schedules lead to good results; therefore, among all possible test allocations during the contact tracing period, choosing the optimal one leads to significant differences in terms of effectiveness in reducing secondary infection risk.

Second, for a given number of available tests, \newmo{using only high sensitivity PCR tests does not necessarily result in the optimal testing plan, due to delays in the test results. }
This result extends the conclusions of \cite{larremore2021test} obtained when analyzing surveillance testing strategies. Using PCR is effective to confirm traced contact while maintaining strict quarantine; however, when compliance with quarantine is imperfect, the delay in reporting results increases the risk of secondary infection. This risk can be more effectively managed \newmo{ combining high sensitivity PCR tests with a lower sensitivity LFA test with immediate results; moreover, the cost of this strategy is usually lower}.

Third, our analysis suggests that in environments with structured contact networks with recurrent risk of exposure, the effectiveness of quarantines and post exposure testing of traced contacts depends on how the index case is detected. In this environment, asymptomatic index cases may lead to a wider range of possible exposure dates, thereby increasing the likelihood that the exposed contact is already infectious at the time of case notification. Increasing the frequency of surveillance testing is useful for reducing this risk, thereby improving the efficiency of the contact tracing strategies analyzed in this work. Interestingly, although the effectiveness of post exposure testing varies depending on the range and probability distribution of the exposure days, the optimal testing schedules that should be implemented to avert secondary infection risk are relatively similar across all the scenarios that were analyzed, and their performance relative to quarantines with different levels of adherence was also similar.

Our modeling approach is subject to limitations. First, we assume that confirmed cases fully adhere to strict isolation, which is plausible to implement in environments with stricter control, such as workplaces, healthcare facilities and schools, or where isolation in dedicated facilities is feasible. However, strict isolation may be difficult to implement in other environments, such as households or for social contact networks. Second, our analysis is based on simulated viral load trajectories that have been calibrated in previous work (\cite{larremore2021test, jones2021estimating}). However, recent work by \cite{li2021viral} and \cite{singanayagam2021community} suggests that the viral load of new variants (such as Delta) may exhibit differences from those reported for the original strains during the initial waves of the COVID-19 pandemic. \newmo{Interestingly, our results suggest that the effectiveness of sequential testing is similar for alternative models of viral load trajectories, some of which included variants of concern. Nevertheless, our results must be interpreted with caution and may require further analysis in the future with alternative models of viral load evolution.} Third, testing strategies may lead to changes in the behavior of the traced contacts on their adoption of complementary prevention measures, such as masking and personal hygiene, which are relevant when the individual is actively in contact with the susceptible population.

Our analysis is focused on improving contact tracing for essential workers, such as medical staff, teachers, and specialized workers, among others, where quarantines might heavily disturb the normal functioning of crucial activities. 
\newmo{ We implemented some of the insights obtained from the modeling analysis in   two schools in Chile, with the objective of maximizing in-person teaching during 2021. In Chile, health authorities mandate that upon a confirmed case of a student in the school, the whole class where the student attended two days prior to case confirmation or symptom onset should be quarantined for 14 days. The mandate leaves to the discretion of school management how to handle the teachers that followed the mandatory hygiene measures during teaching hours in the class (wearing mask, open windows and hand washing). A 14-day quarantine of the teacher -- a practice that was implemented by many schools in Chile -- would generate disruptions in in-person teaching for other classes at the school. Instead, these two pilot schools opted to maintain in-person teaching for the exposed teachers that were vaccinated, but conducting three LFA tests during the five days following the index case detection. These tests would complement the regular weekly surveillance testing that was implemented at the schools.}
    
\newmo{Between March 3rd and December 7th of 2021, 69 students were confirmed with a positive PCR tests, leading to 181 teachers who where suspect contacts and, therefore, followed the sequential LFA testing strategy (all of them were vaccinated). All of the 543 LFA tests gave negative results; 65 of the teachers also had a negative PCR test. Overall, this resulted in 1810 in-person teaching days that would have otherwise been lost under the strategy of strict quarantine of the teachers. }

Furthermore, as countries are working on finding ways to normalize certain economic activities, foreign travel has been at the center of discussion. Travel has been restricted, and testing at airports and quarantines upon arrival have been implemented in many countries. However, these strategies will become difficult to implement and enforce at a large scale as airport traffic approaches pre-pandemic levels. Therefore, the sequential testing strategies studied in this work might become an effective alternative to complement quarantines for travelers or other settings where adherence to quarantine mandates is low.

\bigskip

\noindent \textbf{Code availability}\\
 \textcolor{blue}{All codes used in the research are documented and available at \url{
https://datos.uchile.cl/dataset.xhtml?persistentId=doi:10.34691/FK2/GT1XHA}.}

\bigskip

\bibliography{main}

\newpage

\noindent \textbf{\Large {Appendix: Mathematical Formulation}}

\section{Disease Evolution and Infectiousness}
\label{evol}

Since the moment a susceptible person becomes infected with COVID-19, the viral load steadily increases until it reaches the limit of detection (LOD), which is the minimum viral load that can be detected with a PCR test. Then, the viral load keeps growing until reaching its peak value. After that, it continuously decreases until it becomes undetectable.
Following \cite{larremore2021test}, we use a LOD of $10^3$ cp/ml for PCR tests and $10^5$ cp/ml for LFA tests. This methodology adjusts the parameters to simulate viral load for symptomatic and asymptomatic patients.

For the viral load evolution, we use the model described in \cite{larremore2021test}, and use  $V_t$ to denote the viral load of an infected individual at time $t$ since exposure to the virus. The following parameters describe the control points used to generate sample paths of $V_t$ (see Figure 1): 

\begin{itemize}
	\item $t_0$= Time when an infected person reaches the LOD of $10^3$ cp/ml of viral load. $t_0\sim U[2.5, 3.5]$.
	\item $t_{peak}$ = Time when an infected individual reaches peak viral load. $t_{peak} = t_0 + \min(3, (0.5 + \gamma))$, where $\gamma$ is a random variable that follows a gamma distribution with a shape of 1.5 and scale of 1.
	\item $V_{peak}$ = Value of the logarithm of viral load at its peak. $V_{peak} = V_{t_{peak}} \sim U[7,11]$.
	\item $t_{sympt}$= Time when symptoms appear for a symptomatic individual. An individual is symptomatic with probability $1/2$. $t_{sympt} = t_{peak} + s$, where $s\sim U[0, 3]$.
	\item $t_f$: Ending time of infectious period. The probability distribution of $t_f$ depends on whether the individual is symptomatic or not. Thus, if it is symptomatic, then $t_f=t_{sympt} + f$; if it is asymptomatic, then $t_f= t_{peak} + f$, where $f\sim U[4,9]$. Note that this is the time when the viral load drops below $10^6$ and the individual is no longer infectious.
\end{itemize}

We assume that the logarithm of the viral load follows a linear function (constant growth) between the time it reaches the LOD for PCR and the time of peak load (from $3$ to $V_{peak})$) and another linear function (constant decrease) from the peak time to the end time of infection, from ${V_{peak}}$ to $6$.
These model parameters fully characterize the viral load evolution for both a symptomatic and an asymptomatic patient in the course of the disease.

\section{Mathematical Formulation}\label{sec:app_math}
Our goal is to determine the most effective testing policies for traced contact to minimize the secondary infection risk in the community,
without the need of an immediate isolation. Thus, we minimized the expected number of contagious days (i.e. days in which the contact is infecting, with viral load greater than or equal to $10^6$ cp/ml), and the traced contact was active in the community before detection and isolation.

Let $T$ be the time horizon in days in which we monitor the traced contact (set to $T=14$). Thus, a test can be scheduled on any day $t=0,1,\ldots, T$, where $t=0$ corresponds to the day on which the contact is identified. A testing schedule is defined by the specific days in which the LFA and/or PCR tests are taken: we denote a schedule as a tuple $(P, A)$, where $A$ is the set of days an LFA test is performed, and $P$ is the set of days a PCR test is performed. If we fix the number of LFA and PCR tests that can be used, then the set $\Pi(i,j)$ consists of all feasible schedules to perform exactly $j$ LFA tests and $i$ PCR tests. We notice that all policies within this set use exactly the same number of tests of each type. For each schedule, we compute the expected number of infectious days a traced contact is active in the community before being isolated. Additionally, we denote $\Pi=\cup_{i,j\in \mathbb{N}} \Pi(i, j)$ as the set of all possible policies.

The dynamics of the testing and isolation process are as follows. At the beginning of $t$, an LFA or PCR test is taken if they are scheduled at $t$. In the case of a LFA test, we assume that its result is observed immediately. In practice, this would take at most 30 minutes, but we assume that the individual is isolated until the LFA test result is back, which makes this is a realistic assumption; in the case of a PCR test, we observe the test result at the beginning of the next day ($t+1$). When positive test results are observed, the individual is immediately isolated and starts quarantine. If the test result is negative, then the individual remains active in the community until the next scheduled test or if symptoms develop.

We evaluated the performance of a test schedule based on the number of days a suspected infected individual was contagious before being identified as such and therefore imposed a risk to the community. For this, we take the perspective of a decision maker who has a budget that specifies the number of LFA and PCR tests that can be performed and needs to decide on which days to take these tests to minimize the number of days an infected individual was contagious in the workplace before being isolated. We remark on the difference between being infected and being contagious, and only the latter imposes an exposure risk to others.


There are several sources of randomness when measuring the number of infectious days of a traced contact before isolation. In what follows, we explain how we consider this randomness in our model and its effect on the computation of the expected number of infectious days.

\begin{itemize}
	
	\item \textbf{Uncertainty on whether or not a traced contact has been infected:} in our analysis, we assume the individual \emph{is} actually infected (i.e., we condition on the event that the contact was infected by the index case at some date of exposure). Although this may seem paradoxical at first since an infected individual should always be isolated, this is methodologically correct in our setting. Our optimization minimizes the number of infecting days subject to the individual not being isolated until confirmed by a positive test or self-isolated at symptom onset. In this optimization, a non infected individual will always contribute zero to the infecting days regardless of the selected policy; therefore, the expected value is conditional on the event of infection. Another reason is that given the applications we consider, we believe decision makers are more concerned with measuring the performance of a testing policy with respect to how good it is at isolating infecting individuals rather than focusing on cases in which the individual is actually not infected. Additionally, taking this approach makes us indifferent to the underlying probability of being infected, which is difficult to estimate and context dependent.
	
	Formally, the objective is to minimize is the expected number of infecting days, and our control is the schedule of the tests (given a number of tests). Denote $\pi$ the testing policy (days in which the tests are performed), $N_{inf}(\pi)$ a random variable representing the number of infecting days under that policy and $\EE[N_{inf}(\pi)]$ its expectation. Because $N_{inf}(\pi)$ equals to zero when the contact was not infected, conditioning on the event that the contact is infected yields:  
	\begin{align*}
	\EE[N_{inf}(\pi)] 
	&= \EE[N_{inf}(\pi)\mid \text{Contact \textbf{is} infected}]\times\Pr(\text{Contact \textbf{is} infected})
	\end{align*}
	Because the probability that the contact is infected  ($\Pr(\text{Contact \textbf{is} infected})$) is independent of the testing policy, choosing $\pi$ to minimize the expected number of infecting days is equivalent to minimize the conditional expectation $\EE[N_{inf}(\pi)\mid \text{Contact \textbf{is} infected}]$. Hence, the optimal testing policy is independent on the prior probability that the contact is infected. We focus the optimization to minimize the number of infected days given that the individual is infected; scaling this objective by the prior probability of infection yields the (unconditional) expected number of infecting days.
	
	\item \textbf{Number of infecting days of the infected contact:} The generative model for the viral load was described in Appendix \ref{evol} and modeled as a function of time, generating multiple curves randomly based on the controls points described in Figure 1
	. The expected number of infecting days ($N_{inf}$) is calculated for each simulated viral load path, considering the period before isolation (either through a positive test or self-isolation of symptomatic cases at the onset of symptoms). Hence, this methodology is flexible to accommodate alternative approaches to generate the viral load curve.
	
	\item \textbf{Day of infection of the contact:} Our methodology incorporates uncertainty on the day in which the contact has been infected, assuming a set of days where the index case and the contact had significant interaction and the day the index case was confirmed as infected. This assumption is more realistic in settings with structured contact networks that interact frequently (e.g. school and workplace).
	
	To incorporate this uncertainty in the model, we build a probabilistic distribution of the days in which the infection may have happened, and use this probability distribution when simulating the viral loads of the contacts. Transmission from the index case to the contact occurs on exposure day $t$ when: (i) the index case is infectious on day $t$ (defined as the event $I_t$); (ii) the contact has not yet been infected by the index, that is, is susceptible at the beginning of day $t$ (defined as the event $S_t$). Conditional on the events $I_t$ and $S_t$, infection occurs with probability $\beta$, referred to as the infectivity parameter. Using these definitions, the probability that the index case transmits the disease to the contact on day $t$ is given by:
	
	\begin{equation}
	r_t =\beta \Pr(I_t| S_{t}) \cdot \Pr(S_t), \label{eq:rt_full}
	\end{equation}
	
	Note that the events $I_t$ and $S_t$ are not independent, because observing no infection prior to $t$ provides some evidence that the index case may have not yet been infectious. Hence, we use simulation methods to compute equation \eqref{eq:rt_full}.
	
	Define the event $U_d=$ index case was infected on day $d$. For each day $d$ prior to the index case confirmation, we simulate many viral load paths representing the evolution of the disease for the index case when he/she was infected on day $d$. Each viral load path $V^{k,d}$ is a vector specifying the viral load of the index case on each day, denoted $V_t^{k,d}$ (set to zero for $t<d$ because the index was not yet infected); with some abuse of notation, we also use $V^{k,d}$ to denote the event that the index case follows this viral load path. A priori, all the paths $V^{k,d}$ have the same probability, but conditioning on index case confirmation at $t=0$ generates a filter that removes paths that are not consistent with the confirmation event. For example, when the index case is detected via a random LFA test, the filter drops all the paths with $V^{k,d}_0<10^5$; for the weekly LFA test detection, an additional filter is used to drop all the paths with $V^{k,d}_{-6}>10^5$. Denote $\tilde V^{k,d}$ all the paths that \textit{remain} after the filters, $\tilde N$ the total number of remaining paths and $u_d$ the number of these paths that start on day $d$. Note that confirmation at $t=0$ implies that in all the surviving paths the contact has not yet self-isolated during $t<0$. The conditional probability that the index case was infected on day $d$, is the proportion of paths $\tilde V^{k,d}$ that start on day $d$, defined as $q_d = u_d/\tilde N$.
	
	Conditioning on the filtered paths $\tilde V^{k,d}$ and using the indicator function $\mathbbm{1}(\tilde V_t^{k,d}>10^6)$ to represent a viral path that is infectious on day $t$, equation \eqref{eq:rt_full} can be expressed as:
	\begin{align}
	r_t &=\frac{1}{\tilde N}\sum_{k,d} \beta \Pr(I_t| S_{t},\tilde V^{k,d}) \cdot \Pr(S_t|\tilde V^{k,d}) \nonumber \\
	&=\frac{1}{\tilde N}\sum_{k,d} \beta \mathbbm{1}(\tilde V_t^{k,d}>10^6) \cdot \prod_{j\leq t-1} (1 - \beta \mathbbm{1}(\tilde V_j^{k,d}>10^6)) \nonumber \\
	&= \sum_{d\leq t} q_d \cdot \frac{1}{u_d} \sum_k \beta \mathbbm{1}(\tilde V_t^{k,d}>10^6) \cdot \prod_{j\leq t-1} (1 - \beta \mathbbm{1}(\tilde V_j^{k,d}>10^6)) \label{eq:rt_v}
	\end{align}
	
	To facilitate computations, we used the following approximation for equation \eqref{eq:rt_v}:
	\begin{align*}
	r_t &\approx \sum_{d\leq t} q_d \cdot  \beta \frac{1}{u_d} \sum_k\mathbbm{1}(\tilde V_t^{k,d}>10^6) \cdot \prod_{j\leq t-1} (1 - \beta \frac{1}{u_d} \sum_k\mathbbm{1}(\tilde V_j^{k,d}>10^6)) \\
	&= \sum_d q_d \cdot \beta \Pr(I_t|U_d) \cdot \prod_{j\leq t-1}(1-\beta \Pr(I_j|U_d) ),
	\end{align*}
	
	where the values $\Pr(I_j|U_d)=\frac{1}{u_d} \sum_k\mathbbm{1}(\tilde V_t^{k,d}>10^6)$ can be computed once and used for all the simulations including different values of the infectivity parameter $\beta$.
	Finally, we compute the normalized probabilities by conditioning that the contact was infected. For the results shown in this paper, we have computed the exact and approximate values of the normalized $r_t$  for all $t$ and high and low values of $\beta$, and obtained good approximations, within 1\% of the probability values.

\end{itemize}

\section{Simulation based optimization}
In what follows, we present a detailed mathematical formulation for the optimization problem. We consider a standard probability space in which we measure the viral load of an individual who has been infected. This randomness could be attributed to the random variations in viral load evolution for different individuals. All random variables and filters are defined with respect to this probability space. We use the following notation:

\begin{itemize}
	\item $V_t$, $t = 1, \ldots, T$ = Viral load on day $t$ after the index case is discovered. If we know the exact day the individual was infected, then $V_t$ would be completely described by the process explained in Section \ref{evol}. However, since we do not necessarily know the exact day but only a probabilistic distribution over the days of infection, we take $V_t$ to be the random process conditioned on that infection day distribution.
	\item $x^A_t, x^P_t$ = Variables indicating that a test result was \textbf{observed} at day $t$ ($x^A$ for LFA test, and $x^P$ for PCR), and they have a value of 1 if a result is observed (independent of its value) and 0 otherwise.
	\item $L^A=10^5$, $L^P=10^3$ Levels of detection for each test type. We use and $L^I=10^6$ to represent the viral load threshold above which an individual is infectious.
	\item $\mathcal{D}^A$ = $\{d \mid x_d^A= 1\}$, $\mathcal{D}^P=\{d \mid x_d^P= 1\}$. Sets of days where a LFA and PCR test result were observed. Note that for the LFA test, this value coincides with the day of the test, whereas for PCR, it corresponds to one day later (we assume that PCR test results are obtained 24 hours after they are taken, while for LFA tests, these are obtained immediately). Thus,

	\item $\mathcal{D}^A_t \{d \in \mathcal{D}^A \mid d\leq t\}$:  Days of LFA test results up to day $t$. Similar for $\mathcal{D}^P_t$ with PCR test days.
	
	
	\item $R_t^A, R_t^P$ = Random variables indicating the result of an LFA or PCR test observed on day $t$. The distributions of $R_t^A$ and $R_t^P$ depend on $x_t^A$ and $x_t^P$, respectively. If a test result was not observed that day, then we assume that $R_t$ takes the value of -1.
	\[
	R_t^A = \begin{cases} 
	- 1 & x_t^A=0 \\
	\mathds{1}\{V_t \geq L^A\} & x_t^A=1 
	\end{cases},\ \ \
	R_t^P = \begin{cases} 
	- 1 & x_t^P=0 \\
	\mathds{1}\{V_{t-1} \geq L^P\} & x_t^P=1 
	\end{cases},
	\]
	Denoting $R_t=\max(R_t^A,R_t^P)$ as a random variable that indicates the presence of any test, then:
	\[
	R_t = \begin{cases} 
	- 1 & x_t^A=x_t^P=0 \\
	\mathds{1}\{V_t \geq L^A\} & x_t^A=1 \text{ and } x_t^P=0 \\
	\mathds{1}\{V_{t-1} \geq L^P\} & x_t^A=0 \text{ and } x_t^P=1 \\
	\mathds{1}\{V_t \geq L^A \text{ or } V_{t-1} \geq L^P\} & x_t^A=x_t^P=1.
	\end{cases}
	\]
	

	\item $\mathcal{H}=(\mathcal{H}_t)_t$ = Filtration with respect to the process $((R^A_t, R^P_t))_{t}$, i.e., $\mathcal{H}_t= \sigma(R^A_k, R_k^P\mid k\leq t)$.
	
	\item $\mathcal{S}_t$ = Observable state at the beginning of time $t$. Note that $\mathcal{H}_t$ corresponds to all the information the decision maker has at time $t$ about the state of the infection in the target individual. $\mathcal{S}_t = (\mathcal{H}_{t-1}, (x_\tau)_{\tau \leq t-1})$.
\end{itemize}

An individual who is infected will be contagious only when the viral load surpasses $L^I=10^6$ cp/ml and remains active (not isolated) until positive test results emerge or at symptoms onset. This means that an infectious day will occur if and only if the following three events happen at day $t$:
\begin{itemize}
	\item $I_t = \{V_t \geq L^I\}$: The individual is infecting.
	\item $N_t = \{(R_k)_{k\leq t} \in \{-1, 0\}^t\}$: All test results up to day $t$ have been negative.
	\item $Z_t$: No symptoms at day $t$.
\end{itemize}
Thus, the total number of days where the agent is infecting is equal to:
\[
\sum_{t=0}^T \mathds{1}\{I_t \cap N_t \cap Z_t\}.
\]

The decision maker designing the test schedule does not know the value of $\mathds{1}\{I_t \cap N_t \cap Z_t\}$ and can only infer the distribution of the event $I_t \cap N_t \cap Z_t$ based on prior knowledge of the distribution of the viral load for an infected individual as well as the information obtained through the testing policy, which allows to update the belief on the viral load distribution each time a test result is observed.

Define $\mathcal{J}_T = \mathds{1}\{I_T \cap N_T \cap Z_T\}$ and $ \mathcal{J}_t = \mathds{1}\{I_t \cap N_t \cap Z_t\} + \mathcal{J}_{t+1}$ recursively. Therefore, the decision maker will try at the beginning of each day $t$ to minimize the following quantity:
\begin{equation}
\EE[\mathcal{J}_t \mid \mathcal{S}_t] = \EE[\mathds{1}\{I_t \cap N_t \cap Z_t\} \mid \mathcal{S}_{t}] + \EE[\mathcal{J}_{t+1} \mid  \mathcal{S}_{t}]. 
\label{eq:2} 
\end{equation}


Recall that $\mathcal{H}_t=\sigma(R^A_k, R^P_k\mid k\leq t)$, which means that $N_{t-1}$ is $\mathcal{H}_{t-1}$-measurable, and since $N_t = N_{t-1}\cap \{R_t \in \{-1,0\}\}$, we can rewrite the first term of Equation~\eqref{eq:2} as
\begin{align*}
\EE[\mathds{1}\{I_t \cap N_t \cap Z_t\} \mid \mathcal{S}_{t}] = \mathds{1}\{N_{t-1},Z_t\} \EE[\mathds{1} \{I_t \cap \{R_t \in \{-1,0\}\} \cap Z_t\} \mid \mathcal{S}_{t}] = \PP(I_t, R_t \in \{-1,0\}\mid \mathcal{S}_t),
\end{align*}
because if a positive test is observed at some point in the past, the individual is taken to quarantine and the risk is over, then $\mathds{1}\{N_{t-1}\}$ must be equal to one if the decision maker is making a decision at time $t$. The same happens if the individual presents symptoms on day $t$. Given that the distribution of $R_t$ is determined by $x_t^A$ and $x_{t}^P$, we have
\begin{equation}
\PP(I_t, R_t \in \{-1,0\}, Z_t \mid \mathcal{S}_t) = 
\begin{cases}
\PP(V_t \geq L^I \mid \mathcal{S}_t) & x_{t}^P = x_t^A =0 \\
\PP(V_t \geq L^I, V_t < L^A\mid \mathcal{S}_t) = 0 & x_t^A=1 \text{ and } x_t^P=0 \\
\PP(V_t \geq L^I, V_{t-1} < L^P\mid \mathcal{S}_t) & x_t^A=0 \text{ and } x_t^P=1 \\
\PP(V_t \geq L^I, V_{t-1} < L^P, V_t < L^A \mid \mathcal{S}_t\}=0 & x_t^A=x_t^P=1
\end{cases}
\label{eq:1}
\end{equation}

The second and fourth cases are equal to zero since a negative LFA test immediately discards the event that the agent may be infecting. Let us look at the first case in more detail. We have
\begin{equation}
\PP(V_t \geq L^I \mid \mathcal{H}_{t-1}, x_1,\ldots, x_{t-1}) = \PP(V_t \geq L^I \mid V_{d^A} < L^A\ \forall\in\mathcal{D}^A,\ V_{d^P-1} < L^P\ \forall d^P\in \mathcal{D}^P)
\label{eq:3}
\end{equation}
The third term in Equation~\eqref{eq:1} can be written similarly.


Using the probability distributions for the times of LOD and the times for peak viral load and end of contagious period described in Section \ref{evol}, we determine the probability distribution for $V_t$ for each $t$. Thus, we can use Monte Carlo simulations to compute the value of \eqref{eq:3} using the identity:
\begin{align}
&\PP(V_t \geq L^I \mid V_{d^A} < L^A\ \forall\in\mathcal{D}^A,\ V_{d^P-1} < L^P\ \forall d^P\in \mathcal{D}^P) \nonumber
\\ & = \frac{\PP(V_t \geq L^I, V_{d^A} < L^A\ \forall\in\mathcal{D}^A,\ V_{d^P-1} < L^P\ \forall d^P\in \mathcal{D}^P)}{\PP(V_{d^A} < L^A\ \forall\in\mathcal{D}^A,\ V_{d^P-1} < L^P\ \forall d^P\in \mathcal{D}^P)}.
\label{eq:5}
\end{align}

To conclude, we recall that the expected number of infected days is given by
\[
\EE\left[\sum_{t=0}^T \mathds{1}\{I_t \cap N_t \cap Z_t\}\right] =\sum_{t=0}^T \PP(I_t, N_t, Z_t).
\]
By total probability, we can condition each of the probabilities $\PP(I_t, N_t, Z_t)$ by the state up to time $t$, which indicates the probability of being in such a state if we follow a certain policy. Each of these terms is of the form $\PP(I_t, N_t, Z_t\mid \mathcal{S}_t)\PP(\mathcal{S}_t)$; thus, in Equation \eqref{eq:5}, the expectation can be written as
\[\EE\left[\sum_{t=0}^T \mathds{1}\{I_t \cap N_t \cap Z_t\}\right]=\sum_{t=0}^T \PP(V_t \geq L^I, V_{d^A} < L^A\ \forall\in\mathcal{D}^A,\ V_{d^P-1} < L^P\ \forall d^P\in \mathcal{D}^P)\]
where each of the terms in the summation can be computed using Monte Carlo simulations.

\section{Additional results} \label{sec:add_results}
\subsection{Analysis with other types of index case detection} \label{app:sec:agcad}

\begin{figure}[H]
	\centering
	\includegraphics[scale=0.25]{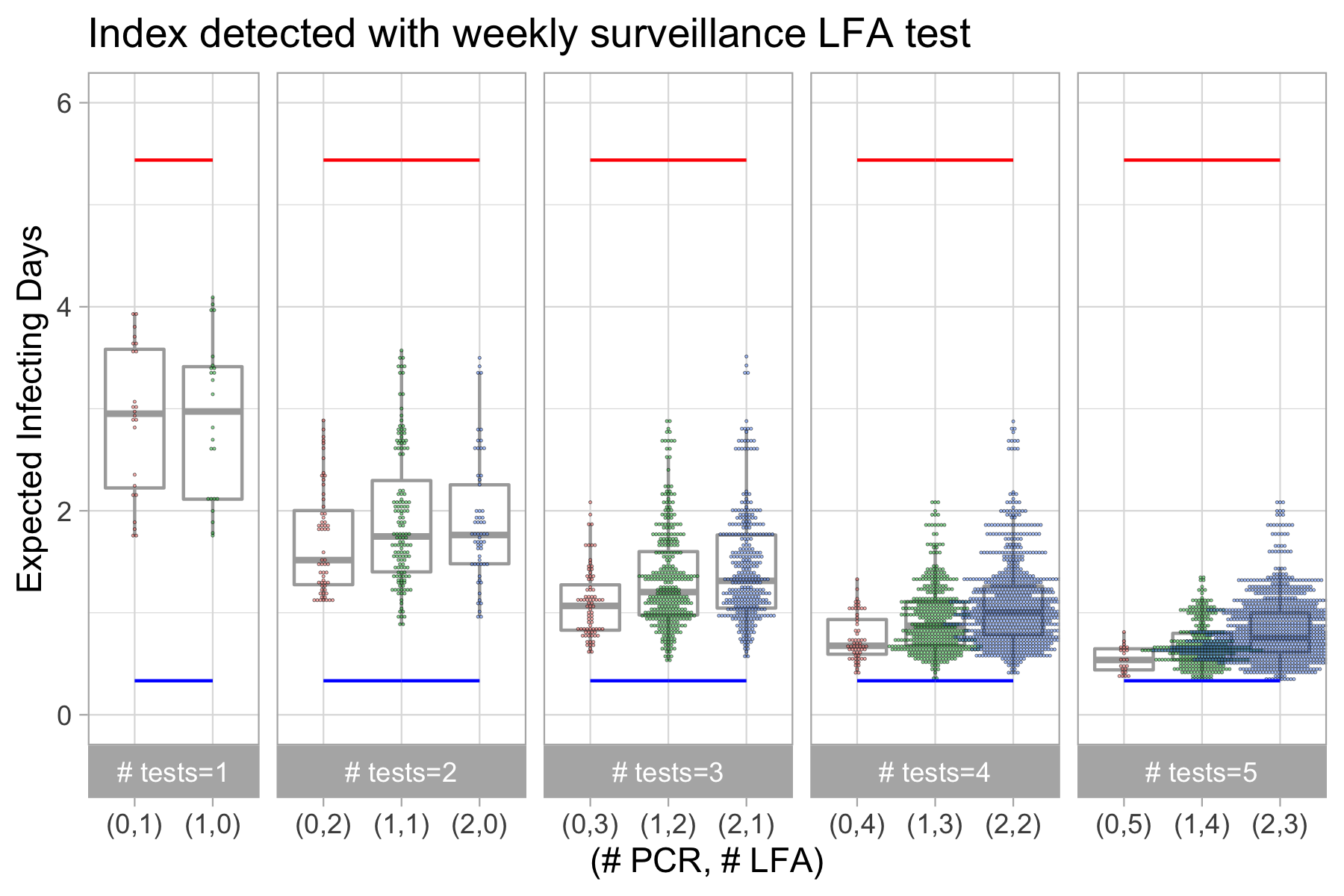}
	\begin{overpic}[scale=0.25,trim=0 0 0 75,clip]{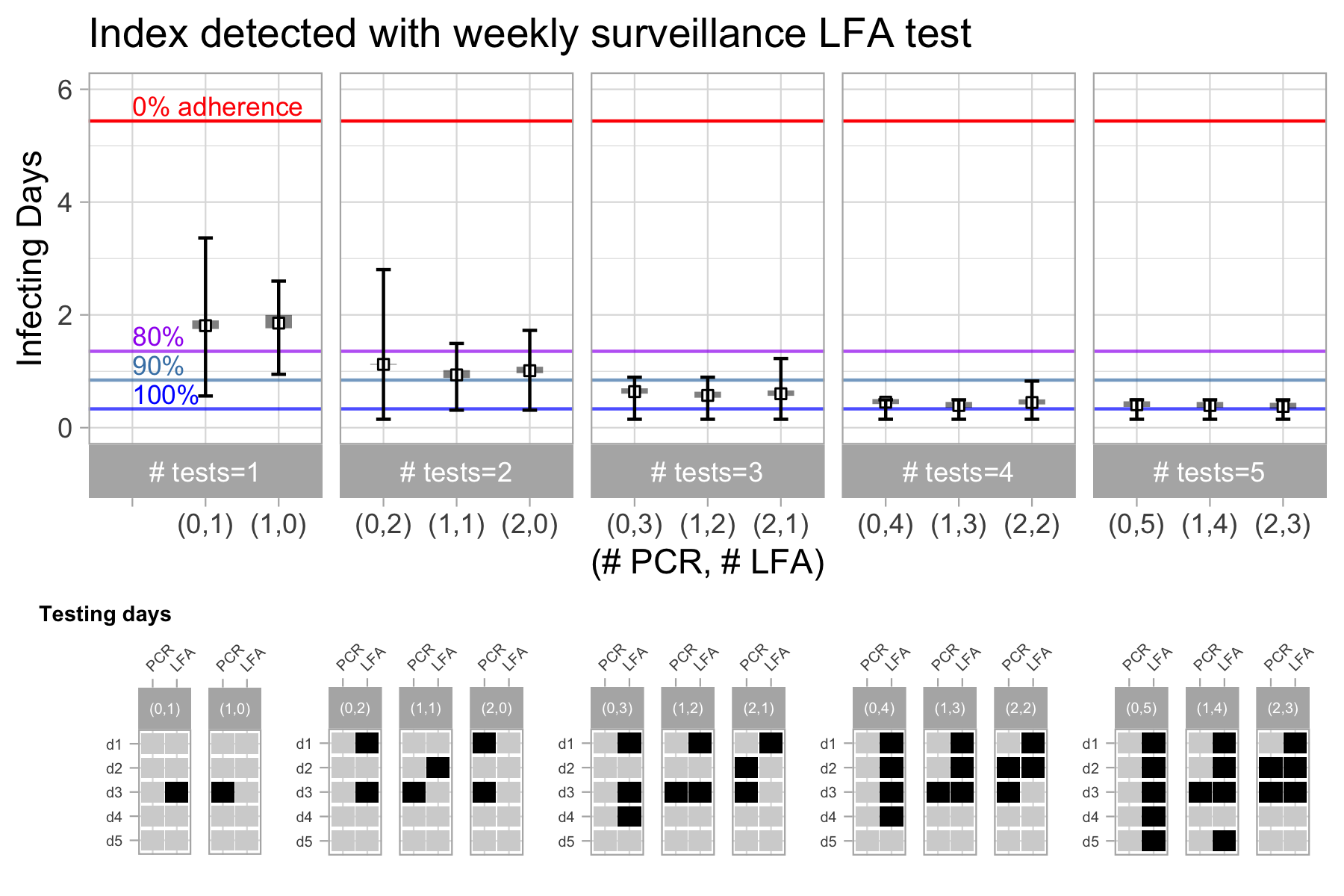}
		\put(85,46){\includegraphics[scale=0.32]{figures/legend_results.png}} 
	\end{overpic}
	\caption{\textbf{Evaluation of testing strategies for a traced contact exposed to an index case detected with weekly surveillance LFA tests.}
	}
	\centering
	\label{fig:LFAweekly}
\end{figure}

\subsection{Results with different scenarios of LFA test sensitivity} \label{app:sec:test_sensitivity}

\begin{figure}[H]
	\centering
	\includegraphics[width=0.8\columnwidth]{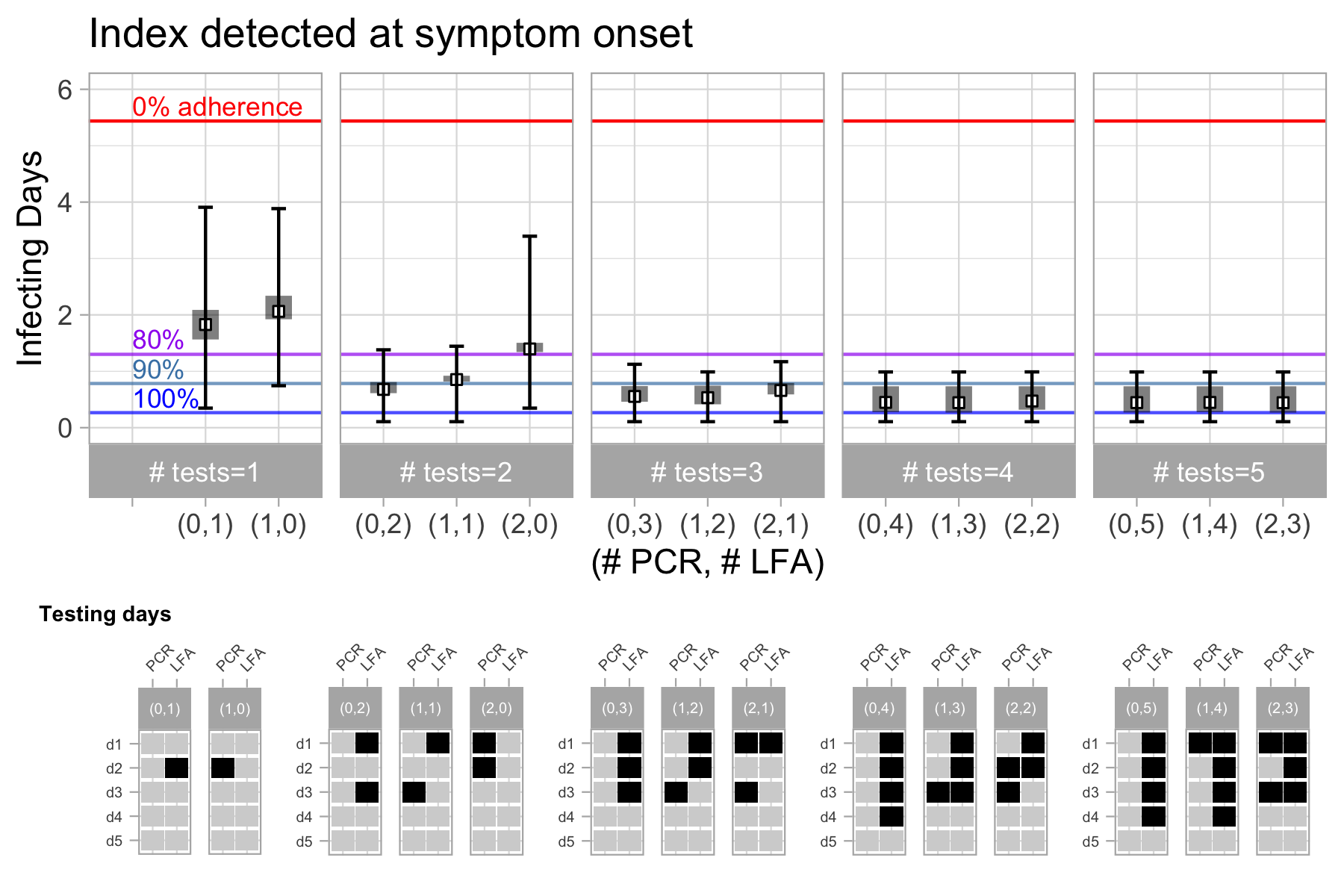}
	\includegraphics[width=0.8\columnwidth]{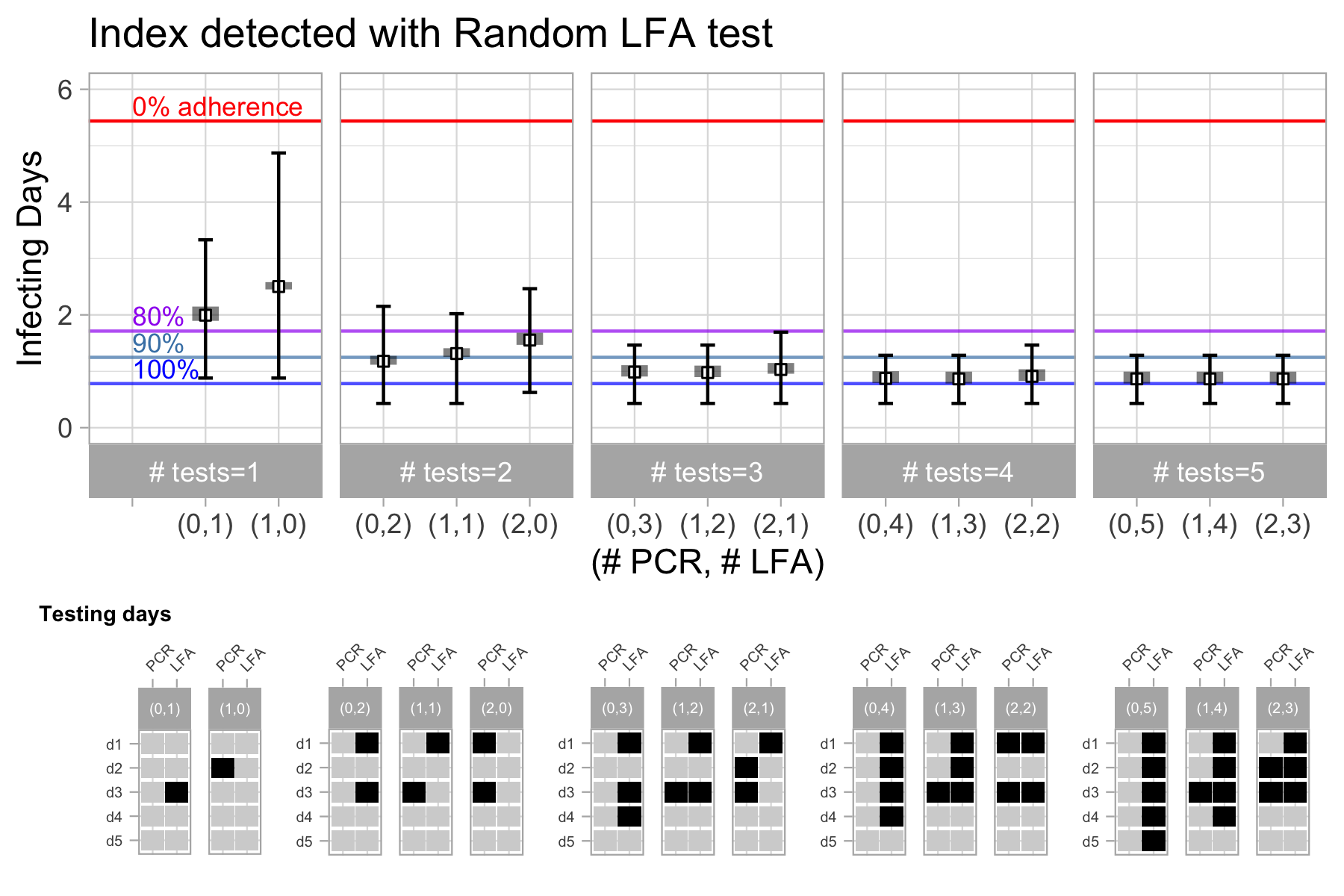}
	\caption{Robust testing policies for the \textit{LFA High} test sensitivity scenario. For the test schedule with (PCR=0,LFA=5) in the top panel, the last LFA test is on day 6 (not shown in the figure).}
	\label{app:fig:results_test_High}
\end{figure}

\begin{figure}[H]
	\centering
	\includegraphics[width=0.8\columnwidth]{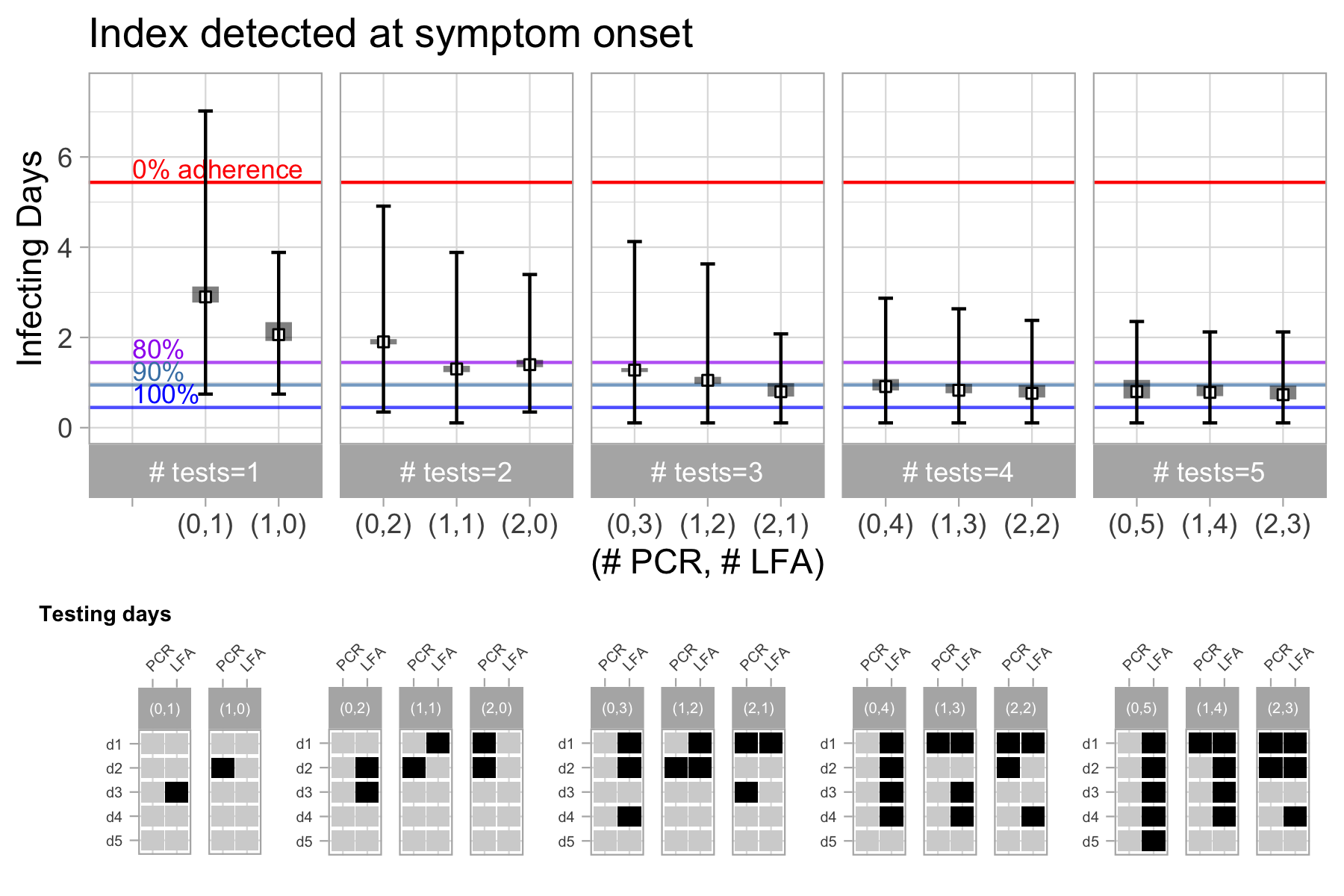}
	\includegraphics[width=0.8\columnwidth]{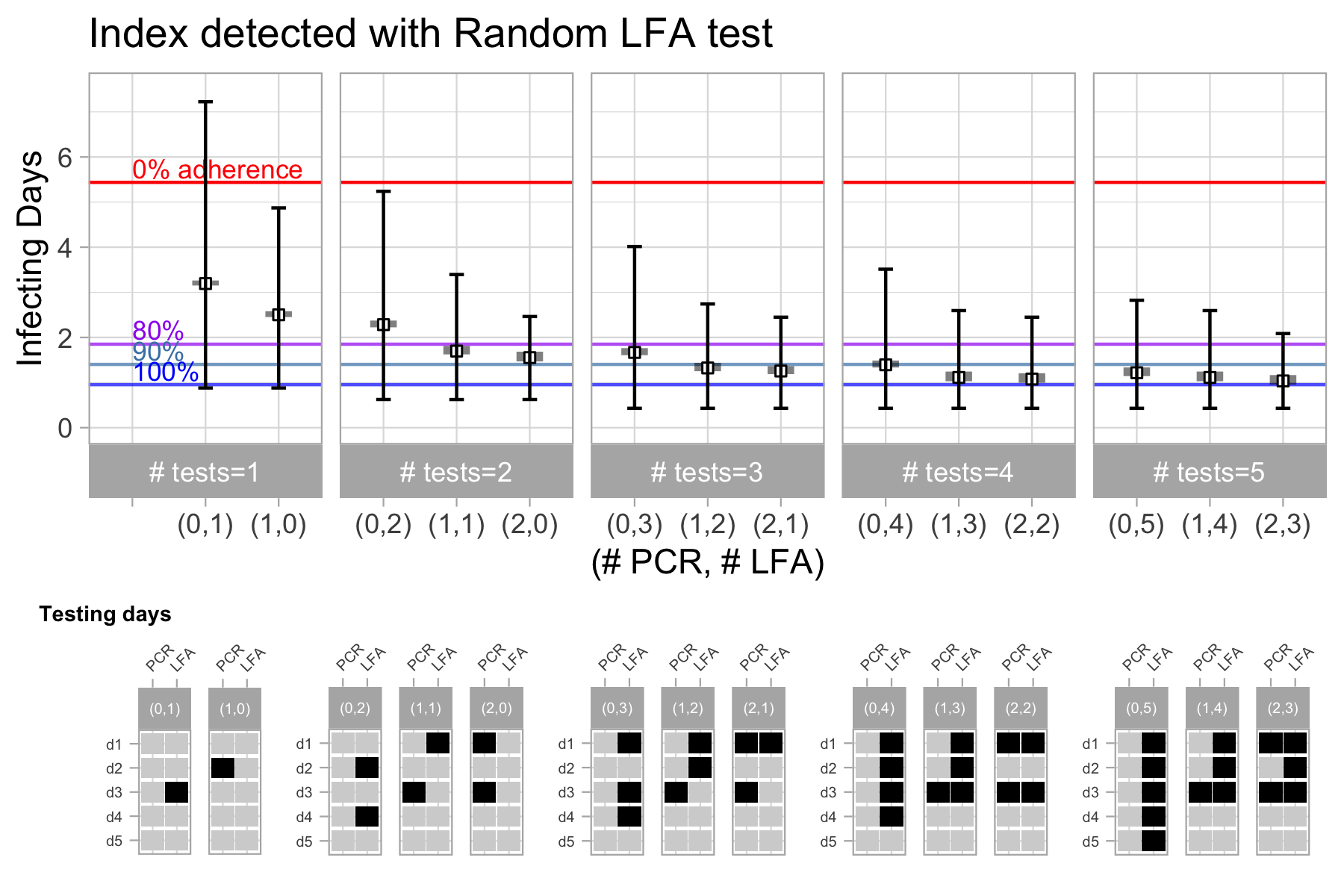}
	\caption{Robust testing policies for the \textit{LFA Low} test sensitivity scenario. For the test schedule (PCR=1,LFA=4) in the bottom panel, the last LFA is taken on day 6 (not shown in the figure).}
	\label{app:fig:results_test_Low}
\end{figure}

\subsection{Results with alternative models of viral load trajectories}\label{app:sec:jones}

\begin{figure}[H]
	\centering
	\includegraphics[width=0.8\columnwidth]{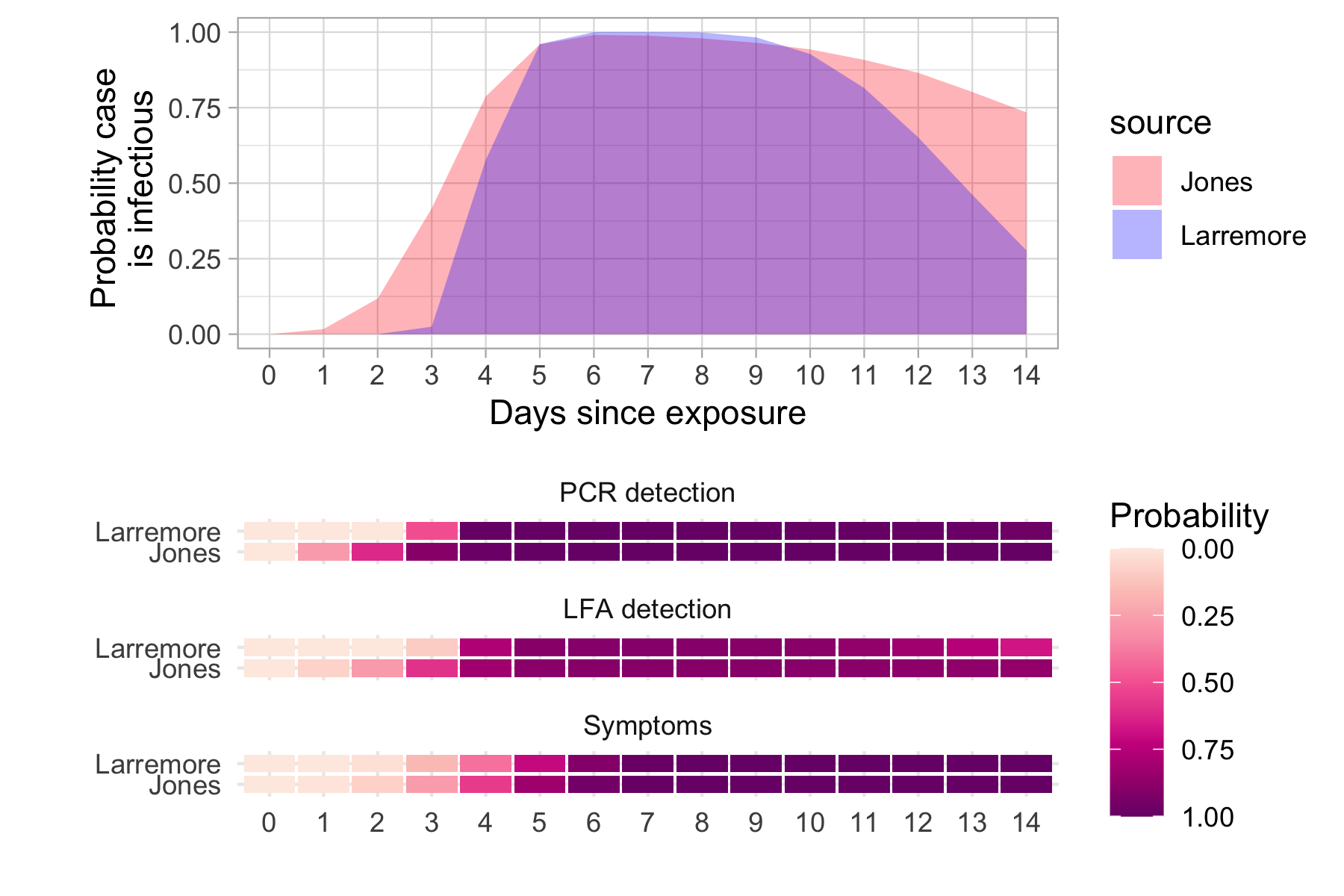}
	\caption{Comparison of infectiousness and test sensitivity dynamics between the viral load models of \cite{larremore2021test} and \cite{jones2021estimating}}.
	\label{app:fig:comp_jones}
\end{figure}

Figure \ref{app:fig:comp_jones} compares the infectiousness and test sensitivity of the grouped estimates of viral load trajectories reported in \cite{jones2021estimating}, with those used in our main analysis (based on \cite{larremore2021test}). Specifically, we simulated viral load paths using the group estimates reported in Figure S5 in the Supplementary material of \cite{jones2021estimating}, summarized in Table \ref{app:tab:jones} (each viral load path was generated by simulating each parameter from a Normal distribution with the indicated mean and standard deviation). These viral load trajectories suggest a more extended period of infectiousness, but also earlier detection with PCR and LFA tests and symptoms onset. To study whether this affected the main conclusions of our analysis, we repeated all the simulations using the viral model of \cite{jones2021estimating}. 

\begin{table}[]
	\centering
	\begin{tabular}{lcc}
		\hline
		Parameter         & Mean  & Std. Dev. \\ \hline
		Increasing slope  & 2     & 0.39      \\
		Days to peak load & 4.3   & 0.92      \\
		Peak viral load   & 8.1   & 0.7       \\
		Decreasing slope  & -0.17 & 0.02      \\ \hline
	\end{tabular}
	
	\caption{Parameters to simulate viral load trajectories based on the results reported in \cite{jones2021estimating}}
	\label{app:tab:jones}
\end{table}

\begin{figure}[H]
	\centering
	\includegraphics[width=0.8\columnwidth]{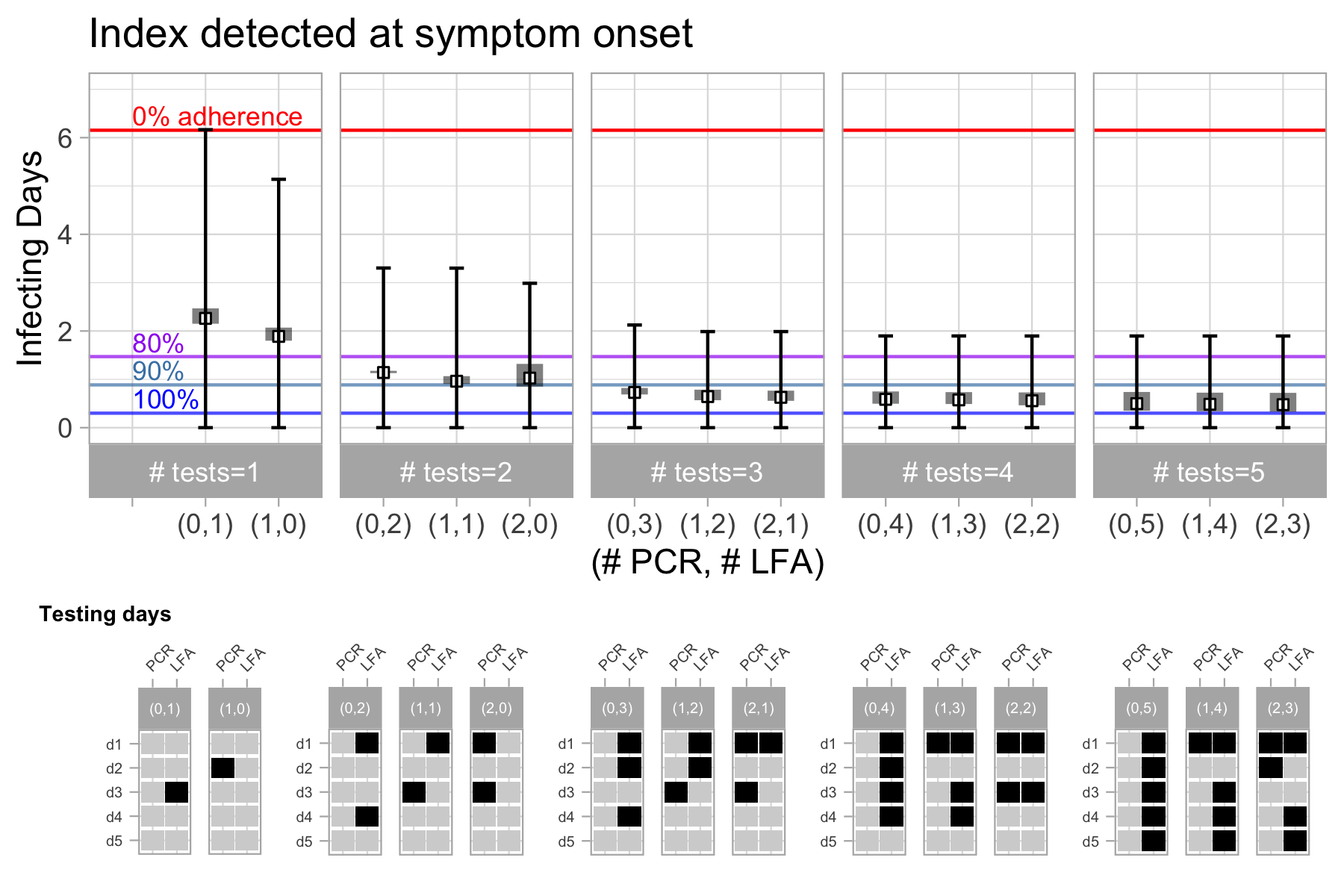}
	\includegraphics[width=0.8\columnwidth]{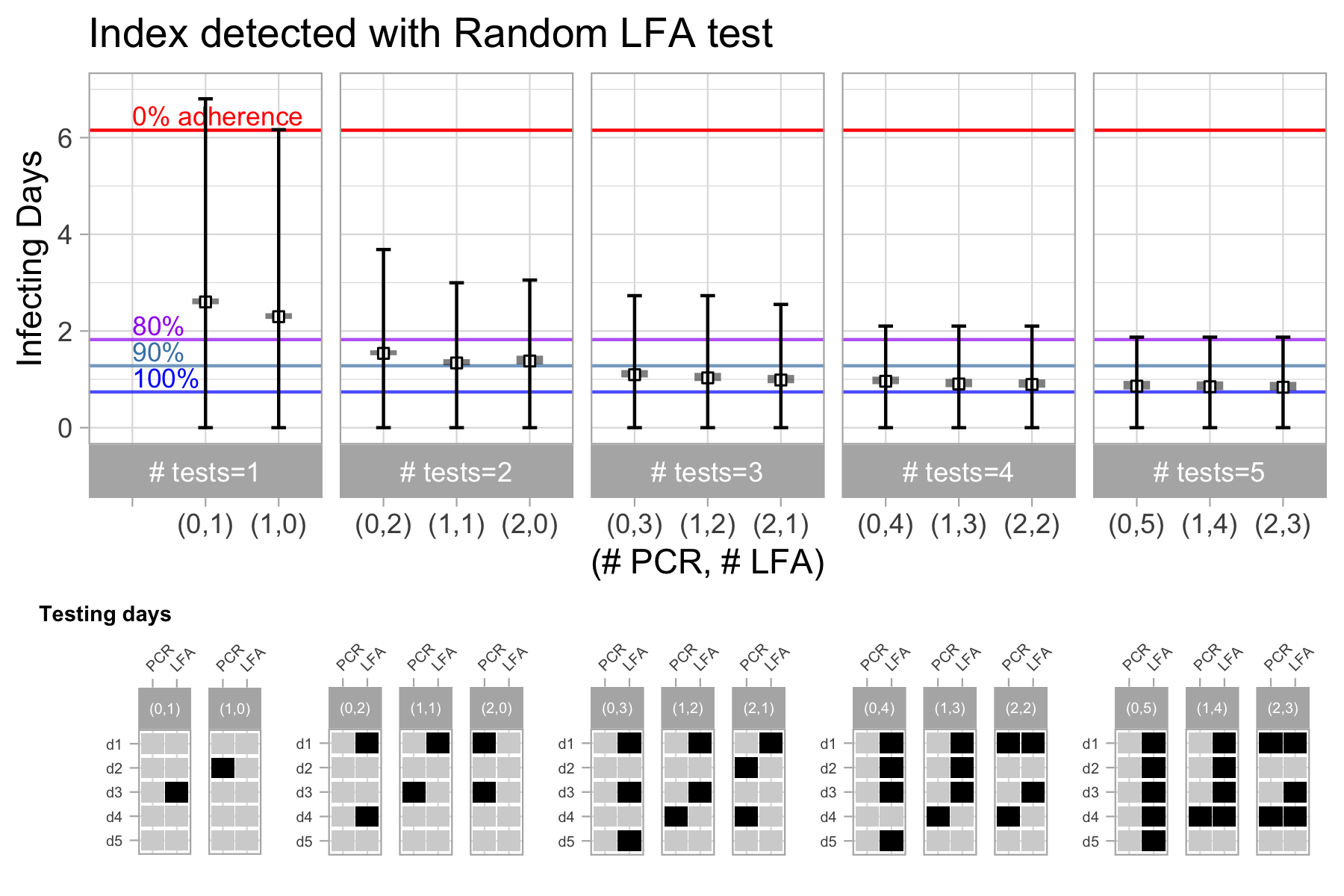}
	\caption{Robust testing policies for the \textit{LFA Med} test sensitivity scenario using viral load trajectories from \cite{jones2021estimating}.}
	\label{app:fig:results_jones_Med}
\end{figure}

\begin{figure}[H]
	\centering
	\includegraphics[width=0.8\columnwidth]{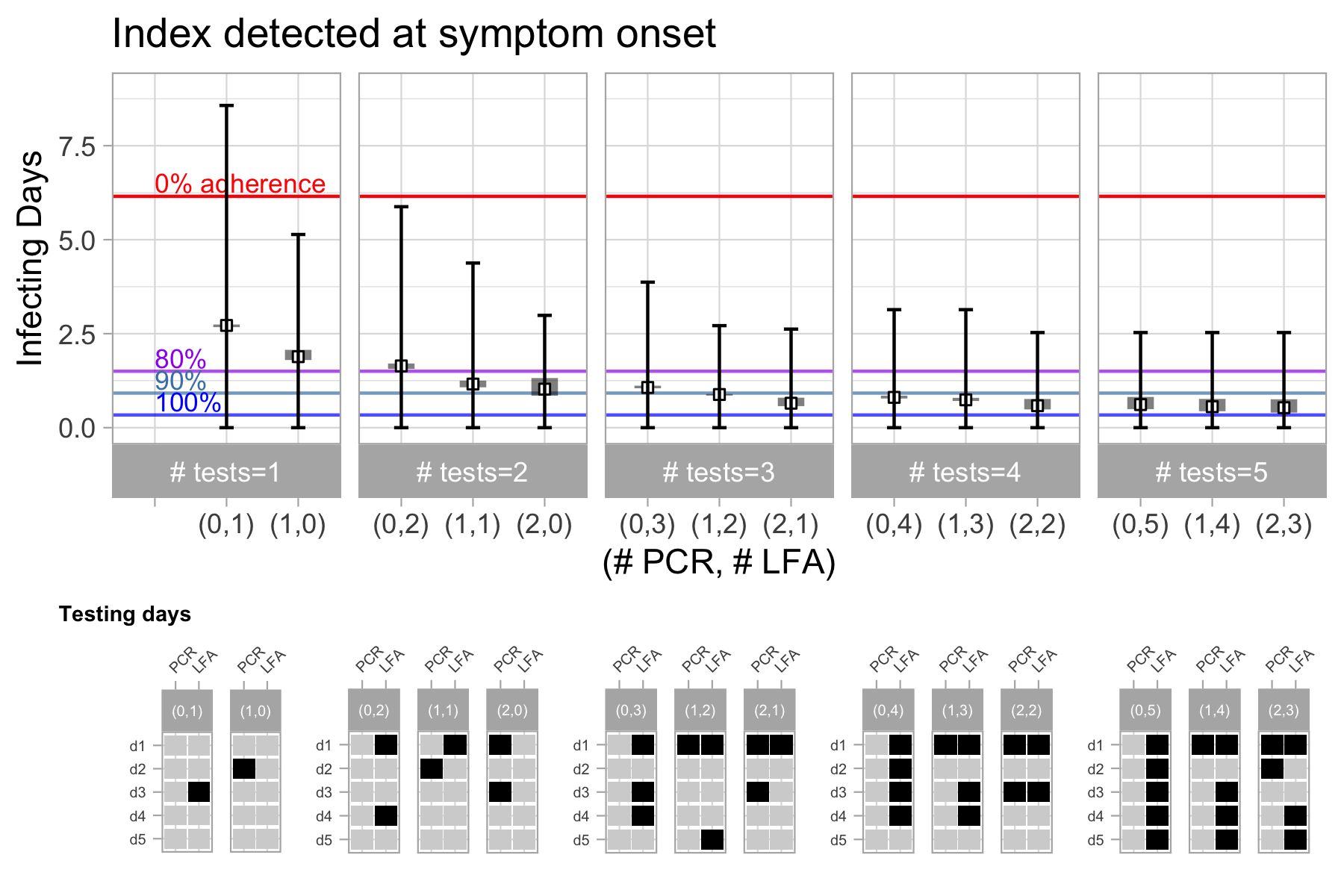}
	\includegraphics[width=0.8\columnwidth]{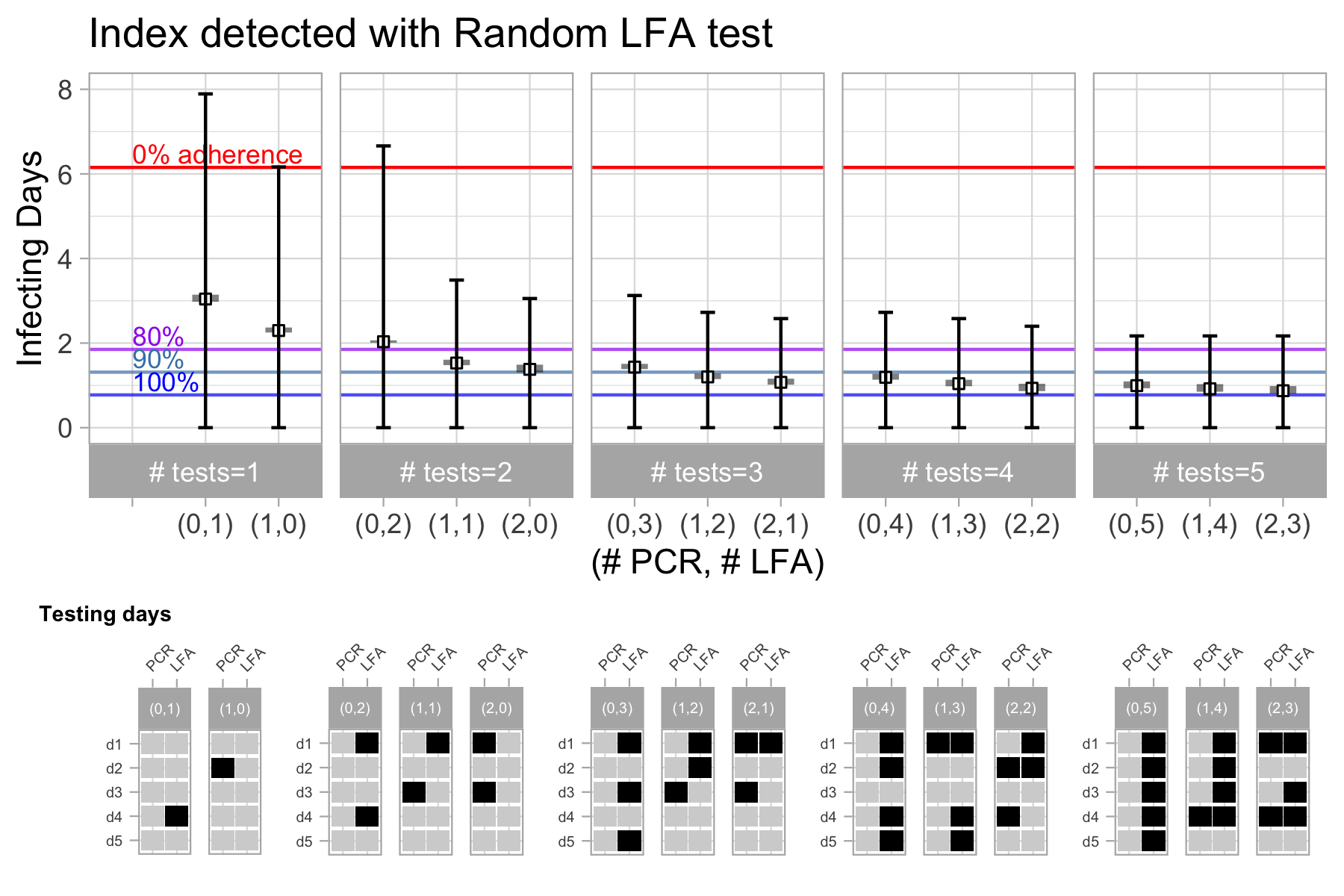}
	\caption{Robust testing policies for the \textit{LFA Med-Low} test sensitivity scenario using viral load trajectories from \cite{jones2021estimating}.}
	\label{app:fig:results_jones_MedLow}
\end{figure}

Figures \ref{app:fig:results_jones_Med} and \ref{app:fig:results_jones_MedLow} show the robust testing policies obtained through the simulations using these alternative viral load paths, considering two scenarios of LFA test sensitivity (\textit{LFA Med} and \textit{LFA Med-Low}).

\subsection{Detailed results of robust  policies}

In  what follows, we present the expected infecting days and the false negative rates at the times when  tests are performed, for all values of the infectivity parameter $\beta$ and  all combinations of tests considered. For Tables \ref{TT1}, \ref{TT2}, and \ref{TT3}, we considered the  LFA \textit{Med} test sensitivity scenario.

\begin{table}
	\tiny{
		\centering
		\begin{tabular}{l|l|l|l|l|l|l|l}
			\# LFA & \# PCR & LFA schedule & PCR schedule & Exp.infecting days & $\beta$ & FNR (LFA)  & FNR (PCR)\\
			
			\hline\hline
			1 & 0 & [2] & ~ & 2.07 & 0.01 & [0.14] & ~ \\
			1 & 0 & [2] & ~ & 2.13 & 0.1 & [0.14] & ~ \\ 
			1 & 0 & [2] & ~ & 2.42 & 0.5 & [0.12]    & ~ \\ 
			1 & 0 & [2] & ~ & 2.82 & 1 & [0.10]  & ~ \\ 
			2 & 0 & [1, 3] & ~ & 1.08 & 0.01 & [0.31, 0.11]  & ~ \\ 
			2 & 0 & [1, 3] & ~ & 1.11 & 0.1 & [0.29, 0.11]   & ~ \\ 
			2 & 0 & [1, 3] & ~ & 1.34 & 0.5 & [0.20, 0.10]   & ~ \\ 
			2 & 0 & [1, 3] & ~ & 1.68 & 1 & [0.12, 0.10]  & ~ \\ 
			3 & 0 & [1, 2, 3] & ~ & 0.77 & 0.01 & [0.31, 0.16, 0.12] & ~ \\ 
			3 & 0 & [1, 2, 3] & ~ & 0.83 & 0.1 & [0.29, 0.16, 0.12] & ~ \\ 
			3 & 0 & [1, 2, 3] & ~ & 1.13 & 0.5 & [0.20, 0.13, 0.11] & ~ \\ 
			3 & 0 & [1, 2, 3] & ~ & 1.55 & 1 & [0.12, 0.11, 0.11] & ~ \\
			4 & 0 & [1, 2, 3, 4] & ~ & 0.71 & 0.01 & [0.31, 0.16, 0.12, 0.11] & ~ \\ 
			4 & 0 & [1, 2, 3, 4] & ~ & 0.78 & 0.1 & [0.29, 0.16, 0.12, 0.11]  & ~ \\ 
			4 & 0 & [1, 2, 3, 4] & ~ & 1.11 & 0.5 & [0.20, 0.13, 0.11, 0.10]  & ~ \\ 
			4 & 0 & [1, 2, 3, 4] & ~ & 1.55 & 1 & [0.12, 0.11, 0.10, 0.11]   & ~ \\ 
			5 & 0 & [1, 2, 3, 4, 5] & ~ & 0.71 & 0.01 & [0.31, 0.16, 0.12, 0.10, 0.06]  & ~ \\ 
			5 & 0 & [1, 2, 3, 4, 5] & ~ & 0.77 & 0.1 & [0.29, 0.16, 0.12, 0.10 , 0.06]   & ~ \\ 
			5 & 0 & [1, 2, 3, 4, 5] & ~ & 1.11 & 0.5 & [0.20, 0.13, 0.11, 0.09, 0.04] & ~ \\ 
			5 & 0 & [1, 2, 3, 4, 5] & ~ & 1.54 & 1 & [0.12, 0.11, 0.11, 0.09, 0.02] & ~ \\ \hline \hline
			0 & 1 & ~ & [2] & 1.37 & 0.01 & ~ & [0.02] \\ 
			0 & 1 & ~ & [2] & 1.44 & 0.1 & ~ & [0.02] \\ 
			0 & 1 & ~ & [2] & 1.83 & 0.5 & ~ & [0.01] \\
			0 & 1 & ~ & [2] & 2.33 & 1 & ~ & [0.00] \\
			1 & 1 & [1] & [3] & 0.87 & 0.01 & [0.31] & [0.00] \\ 
			1 & 1 & [1] & [3] & 0.92 & 0.1 & [0.29] & [0.00] \\
			1 & 1 & [1] & [3] & 1.22 & 0.5 & [0.20]   & [0.00] \\
			1 & 1 & [1] & [3] & 1.63 & 1 & [0.12]  & [0.00]  \\
			2 & 1 & [1, 2] & [3] & 0.71 & 0.01 & [0.31, 0.16]   & [0.01]   \\ 
			2 & 1 & [1, 2] & [3] & 0.77 & 0.1 & [0.29, 0.16] & [0.01]  \\ 
			2 & 1 & [1, 2] & [3] & 1.11 & 0.5 & [0.20, 0.13] & [0.00] \\ 
			2 & 1 & [1, 2] & [3] & 1.54 & 1 & [0.12, 0.11] & [0.00] \\ 
			3 & 1 & [1, 2, 3] & [1] & 0.7 & 0.01 & [0.31, 0.19, 0.10] & [0.35] \\ 
			3 & 1 & [1, 2, 3] & [1] & 0.75 & 0.1 & [0.29, 0.18, 0.09] & [0.32]  \\ 
			3 & 1 & [1, 2, 3] & [1] & 1.04 & 0.5 & [0.20, 0.13, 0.07] & [0.17] \\ 
			3 & 1 & [1, 2, 3] & [1] & 1.45 & 1 & [0.12, 0.08, 0.04] & [0.04] \\ 
			4 & 1 & [1, 2, 3, 4] & [1] & 0.64 & 0.01 & [0.31, 0.19 , 0.10, 0.05] & [0.35]  \\ 
			4 & 1 & [1, 2, 3, 4] & [1] & 0.7 & 0.1 & [0.29, 0.18, 0.09, 0.05] & [0.32]  \\ 
			4 & 1 & [1, 2, 3, 4] & [1] & 1.02 & 0.5 & [0.20, 0.14, 0.07, 0.03] & [0.17] \\ 
			4 & 1 & [1, 2, 3, 4] & [1] & 1.45 & 1 & [0.12, 0.08, 0.07, 0.01] & [0.04]  \\ 
			5 & 1 & [1, 2, 3, 4, 5] & [1] & 0.64 & 0.01 & [0.31, 0.18, 0.10, 0.05, 0.02] & [0.35] \\ 
			5 & 1 & [1, 2, 3, 4, 5] & [1] & 0.7 & 0.1 & [0.29, 0.18, 0.09, 0.05, 0.02] & [0.32]  \\ 
			5 & 1 & [1, 2, 3, 4, 5] & [1] & 1.02 & 0.5 & [0.20, 0.14, 0.07, 0.03, 0.04] & [0.17] \\ 
			5 & 1 & [1, 2, 3, 4, 5] & [1] & 1.45 & 1 & [0.12, 0.08, 0.04, 0.01, 0.00] & [0.04]   \\ \hline \hline
			0 & 2 & ~ & [1, 2] & 0.77 & 0.01 & ~ & [0.28, 0.05] \\ 
			0 & 2 & ~ & [1, 2] & 0.82 & 0.1 & ~ & [0.25, 0.04] \\ 
			0 & 2 & ~ & [1, 2] & 1.07 & 0.5 & ~ & [0.11, 0.03] \\ 
			0 & 2 & ~ & [1, 2] & 1.45 & 1 & ~ & [0.01, 0.02]   \\ 
			1 & 2 & [1] & [1, 3] & 0.72 & 0.01 & [0.31] & [0.35, 0.00] \\ 
			1 & 2 & [1] & [1, 3] & 0.77 & 0.1 & [0.29] & [0.32, 0.00] \\ 
			1 & 2 & [1] & [1, 3] & 1.05 & 0.5 & [0.20]  & [0.17, 0.00] \\ 
			1 & 2 & [1] & [1, 3] & 1.45 & 1 & [0.12] & [0.04, 0.00] \\
			2 & 2 & [1, 2] & [1, 3] & 0.64 & 0.01 & [0.31, 0.19] & [0.35 0.01] \\ 
			2 & 2 & [1, 2] & [1, 3] & 0.7 & 0.1 & [0.29, 0.18] & [0.31, 0.01]  \\ 
			2 & 2 & [1, 2] & [1, 3] & 1.02 & 0.5 & [0.20, 0.143] & [0.17, 0.00] \\ 
			2 & 2 & [1, 2] & [1, 3] & 1.45 & 1 & [0.12, 0.08] & [0.04, 0.00] \\ 
			3 & 2 & [1, 2, 4] & [1, 2] & 0.64 & 0.01 & [0.31, 0.18, 0.06] & [0.35, 0.18] \\ 
			3 & 2 & [1, 2, 4] & [1, 2] & 0.7 & 0.1 & [0.29, 0.18, 0.05] & [0.32, 0.16] \\ 
			3 & 2 & [1, 2, 4] & [1, 2] & 1.02 & 0.5 & [0.20, 0.14, 0.03] & [0.17, 0.11] \\ 
			3 & 2 & [1, 2, 4] & [1, 2] & 1.45 & 1 & [0.12, 0.08, 0.01] & [0.04, 0.06] \\ 
			4 & 2 & [1, 2, 3, 4] & [1, 2] & 0.63 & 0.01 & [0.31, 0.18, 0.12, 0.08] & [0.35, 0.17] \\
			4 & 2 & [1, 2, 3, 4] & [1, 2] & 0.69 & 0.1 & [0.29, 0.18, 0.11, 0.05] & [0.32, 0.16] \\ 
			4 & 2 & [1, 2, 3, 4] & [1, 2] & 1.01 & 0.5 & [0.20, 0.14, 0.07, 0.03] & [0.17, 0.11] \\
			4 & 2 & [1, 2, 3, 4] & [1, 2] & 1.45 & 1 & [0.12, 0.08, 0.02, 0.01] & [0.04, 0.06] \\ 
			5 & 2 & [1, 2, 3, 4, 6] & [1, 2] & 0.62 & 0.01 & [0.31 0.19, 0.12, 0.06, 0.02] & [0.35, 0.17] \\ 
			5 & 2 & [1, 2, 3, 4, 6] & [1, 2] & 0.69 & 0.1 & [0.29, 0.18, 0.11, 0.06, 0.02] & [0.32, 0.16] \\ 
			5 & 2 & [1, 2, 3, 4, 6] & [1, 2] & 1.01 & 0.5 & [0.20, 0.13, 0.07, 0.03, 0.01] & [0.17, 0.14] \\ 
			5 & 2 & [1, 2, 3, 4, 6] & [1, 2] & 1.45 & 1 & [0.12, 0.08, 0.01, 0.01, 0.00]  & [0.04, 0.06] \\ \hline
		\end{tabular}
	}
	\caption{ Index detected at symptom onset}
	\label{TT1}
\end{table}

\begin{table}[H]
	\centering
	\tiny{
		\begin{tabular}{l|l|l|l|l|l|l|l}
			\# LFA & \# PCR & LFA schedule & PCR schedule & Exp.infecting days & $\beta$ & FNR (LFA)  & FNR (PCR)\\
			\hline\hline
			1 & 0 & [3] & ~ & 2.93 & 0.01 & [0.11] & ~ \\ 
			1 & 0 & [3] & ~ & 2.96 & 0.1 & [0.11] & ~ \\ 
			1 & 0 & [3] & ~ & 3.07 & 0.5 & [0.11] & ~ \\ 
			1 & 0 & [3] & ~ & 3.21 & 1 & [0.11] & ~ \\ 
			2 & 0 & [1, 3] & ~ & 1.62 & 0.01 & [0.33, 0.11] & ~ \\ 
			2 & 0 & [1, 3] & ~ & 1.64 & 0.1 & [0.33, 0.11] & ~ \\ 
			2 & 0 & [1, 3] & ~ & 1.73 & 0.5 & [0.31, 0.11] & ~ \\ 
			2 & 0 & [1, 3] & ~ & 1.84 & 1 & [0.29, 0.11] & ~ \\ 
			3 & 0 & [1, 3, 4] & ~ & 1.45 & 0.01 & [0.33, 0.11, 0.11] & ~ \\ 
			3 & 0 & [1, 3, 4] & ~ & 1.47 & 0.1 & [0.33, 0.11, 0.11] & ~ \\ 
			3 & 0 & [1, 3, 4] & ~ & 1.57 & 0.5 & [0.31, 0.11, 0.12] & ~ \\ 
			3 & 0 & [1, 3, 4] & ~ & 1.7 & 1 & [0.29, 0.11, 0.12] & ~ \\ 
			4 & 0 & [1, 2, 3, 4] & ~ & 1.25 & 0.01 & [0.33, 0.17, 0.13, 0.11] & ~ \\ 
			4 & 0 & [1, 2, 3, 4] & ~ & 1.28 & 0.1 & [0.33, 0.17, 0.13, 0.12] & ~ \\ 
			4 & 0 & [1, 2, 3, 4] & ~ & 1.38 & 0.5 & [0.31, 0.16, 0.13, 0.12] & ~ \\ 
			4 & 0 & [1, 2, 3, 4] & ~ & 1.53 & 1 & [0.29, 0.16, 0.13, 0.12] & ~ \\ 
			5 & 0 & [1, 2, 3, 4, 5] & ~ & 1.25 & 0.01 & [0.33, 0.17, 0.13, 0.12, 0.06] & ~ \\ 
			5 & 0 & [1, 2, 3, 4, 5] & ~ & 1.27 & 0.1 & [0.33, 0.17, 0.13, 0.12, 0.06] & ~ \\ 
			5 & 0 & [1, 2, 3, 4, 5] & ~ & 1.38 & 0.5 & [0.31, 0.16, 0.13, 0.12, 0.07] & ~ \\ 
			5 & 0 & [1, 2, 3, 4, 5] & ~ & 1.52 & 1 & [0.29, 0.16, 0.13, 0.12, 0.09] & ~ \\ \hline\hline
			0 & 1 &  & [2] & 1.86 & 0.01 & ~ & [0.03] \\ 
			0 & 1 &  & [2] & 1.89 & 0.1 & ~ & [0.03] \\ 
			0 & 1 &  & [2] & 2 & 0.5 & ~ & [0.02] \\ 
			0 & 1 &  & [2] & 2.14 & 1 & ~ & [0.02] \\ 
			1 & 1 & [1] & [3] & 1.4 & 0.01 & [0.33] & [0.00] \\ 
			1 & 1 & [1] & [3] & 1.43 & 0.1 & [0.33] & [0.00] \\ 
			1 & 1 & [1] & [3] & 1.53 & 0.5 & [0.31] & [0.00] \\ 
			1 & 1 & [1] & [3] & 1.66 & 1 & [0.29] & [0.00] \\ 
			2 & 1 & [1, 2] & [3] & 1.25 & 0.01 & [0.33, 0.17] & [0.01] \\ 
			2 & 1 & [1, 2] & [3] & 1.27 & 0.1 & [0.33, 0.17] & [0.01] \\ 
			2 & 1 & [1, 2] & [3] & 1.38 & 0.5 & [0.31, 0.16] & [0.01] \\ 
			2 & 1 & [1, 2] & [3] & 1.52 & 1 & [0.29, 0.16] & [0.01] \\ 
			3 & 1 & [1, 2, 3] & [3] & 1.25 & 0.01 & [0.33, 0.17, 0.13] & [0.03] \\ 
			3 & 1 & [1, 2, 3] & [3] & 1.27 & 0.1 & [0.33, 0.17, 0.13] & [0.03] \\ 
			3 & 1 & [1, 2, 3] & [3] & 1.38 & 0.5 & [0.31, 0.16, 0.13] & [0.02] \\ 
			3 & 1 & [1, 2, 3] & [3] & 1.52 & 1 & [0.29, 0.16, 0.13] & [0.02] \\ 
			4 & 1 & [1, 2, 3, 5] & [3] & 1.24 & 0.01 & [0.33, 0.17, 0.13, 0.03] & [0.03] \\ 
			4 & 1 & [1, 2, 3, 5] & [3] & 1.27 & 0.1 & [0.33, 0.17, 0.13, 0.03] & [0.03] \\ 
			4 & 1 & [1, 2, 3, 5] & [3] & 1.37 & 0.5 & [0.31, 0.16, 0.13, 0.03] & [0.02] \\ 
			4 & 1 & [1, 2, 3, 5] & [3] & 1.52 & 1 & [0.29, 0.16, 0.13, 0.02] & [0.02] \\ 
			5 & 1 & [1, 2, 3, 4, 5] & [2] & 1.22 & 0.01 & [0.33, 0.17, 0.12, 0.06, 0.03] & [0.13] \\ 
			5 & 1 & [1, 2, 3, 4, 5] & [2] & 1.25 & 0.1 & [0.33, 0.17, 0.11, 0.05, 0.03] & [0.13] \\
			5 & 1 & [1, 2, 3, 4, 5] & [2] & 1.36 & 0.5 & [0.31, 0.16, 0.11, 0.05, 0.03] & [0.12] \\ 
			5 & 1 & [1, 2, 3, 4, 5] & [2] & 1.5 & 1 & [0.29, 0.16, 0.10, 0.05, 0.02] & [0.10] \\ \hline\hline
			0 & 2 &  & [1, 3] & 1.35 & 0.01 & ~ & [0.31, 0.00] \\ 
			0 & 2 &  & [1, 3] & 1.37 & 0.1 & ~ & [0.30, 0.00] \\ 
			0 & 2 &  & [1, 3] & 1.46 & 0.5 & ~ & [0.28, 0.00] \\ 
			0 & 2 &  & [1, 3] & 1.58 & 1 & ~ & [0.24, 0.00] \\ 
			1 & 2 & [1] & [2, 3] & 1.23 & 0.01 & [0.33] & [0.04, 0.02] \\ 
			1 & 2 & [1] & [2, 3] & 1.25 & 0.1 & [0.33] & [0.04, 0.02] \\ 
			1 & 2 & [1] & [2, 3] & 1.36 & 0.5 & [0.31] & [0.03, 0.02] \\ 
			1 & 2 & [1] & [2, 3] & 1.51 & 1 & [0.29] & [0.03, 0.02] \\ 
			2 & 2 & [1, 2] & [2, 3] & 1.23 & 0.01 & [0.33, 0.17] & [0.13, 0.03] \\ 
			2 & 2 & [1, 2] & [2, 3] & 1.25 & 0.1 & [0.33, 0.17] & [0.13, 0.03] \\ 
			2 & 2 & [1, 2] & [2, 3] & 1.36 & 0.5 & [0.31, 0.16] & [0.12, 0.03] \\ 
			2 & 2 & [1, 2] & [2, 3] & 1.5 & 1 & [0.29, 0.16] & [0.10, 0.03] \\ 
			3 & 2 & [1, 2, 3] & [1, 3] & 1.18 & 0.01 & [0.33, 0.16, 0.09] & [0.37, 0.03] \\ 
			3 & 2 & [1, 2, 3] & [1, 3] & 1.21 & 0.1 & [0.33, 0.16, 0.09] & [0.36, 0.03] \\ 
			3 & 2 & [1, 2, 3] & [1, 3] & 1.31 & 0.5 & [0.31, 0.15, 0.08] & [0.33 0.03] \\ 
			3 & 2 & [1, 2, 3] & [1, 3] & 1.45 & 1 & [0.29, 0.14, 0.08] & [0.30, 0.02] \\ 
			4 & 2 & [1, 2, 3, 5] & [1, 3] & 1.18 & 0.01 & [0.33, 0.16, 0.09, 0.03] & [0.37, 0.03] \\ 
			4 & 2 & [1, 2, 3, 5] & [1, 3] & 1.2 & 0.1 & [0.33, 0.16, 0.09, 0.03] & [0.36, 0.03] \\ 
			4 & 2 & [1, 2, 3, 5] & [1, 3] & 1.31 & 0.5 & [0.31, 0.15, 0.08, 0.03] & [0.33, 0.03] \\ 
			4 & 2 & [1, 2, 3, 5] & [1, 3] & 1.45 & 1 & [0.29, 0.14, 0.07, 0.02] & [0.30, 0.02] \\ 
			5 & 2 & [1, 2, 3, 4, 5] & [1, 2] & 1.17 & 0.01 & [0.33, 0.16, 0.12, 0.06, 0.03] & [0.37, 0.16] \\ 
			5 & 2 & [1, 2, 3, 4, 5] & [1, 2] & 1.19 & 0.1 & [0.33, 0.16, 0.12, 0.07, 0.03] & [0.36, 0.17] \\ 
			5 & 2 & [1, 2, 3, 4, 5] & [1, 2] & 1.3 & 0.5 & [0.31, 0.15, 0.11, 0.05, 0.02] & [0.33, 0.15] \\ 
			5 & 2 & [1, 2, 3, 4, 5] & [1, 2] & 1.44 & 1 & [0.29, 0.14, 0.10, 0.04, 0.03] & [0.30, 0.16] \\ \hline
		\end{tabular}
	}
	\caption{Index detected with random LFA test}
	\label{TT2}
\end{table}

\newpage

\begin{table}[H]
	\centering
	\tiny{
		\begin{tabular}{l|l|l|l|l|l|l|l}
			\# LFA & \# PCR & LFA schedule & PCR schedule & Exp.infecting days & $\beta$ & FNR (LFA)  & FNR (PCR)\\
			\hline\hline
			
			1 & 0 & [3] & ~ & 2.34 & 0.01 & [0.11] & ~ \\ 
			1 & 0 & [3] & ~ & 2.35 & 0.1 & [0.11] & ~ \\ 
			1 & 0 & [3] & ~ & 2.43 & 0.5 & [0.11] & ~ \\ 
			1 & 0 & [3] & ~ & 2.56 & 1 & [0.11] & ~ \\ 
			2 & 0 & [1, 3] & ~ & 1.11 & 0.01 & [0.42, 0.11] & ~ \\ 
			2 & 0 & [1, 3] & ~ & 1.11 & 0.1 & [0.41, 0.11] & ~ \\
			2 & 0 & [1, 3] & ~ & 1.15 & 0.5 & [0.39, 0.11] & ~ \\ 
			2 & 0 & [1, 3] & ~ & 1.22 & 1 & [0.36, 0.11] & ~ \\ 
			3 & 0 & [1, 3, 4] & ~ & 0.87 & 0.01 & [0.42, 0.11, 0.11] & ~ \\ 
			3 & 0 & [1, 3, 4] & ~ & 0.88 & 0.1 & [0.41, 0.11, 0.11] & ~ \\
			3 & 0 & [1, 3, 4] & ~ & 0.94 & 0.5 & [0.39, 0.11, 0.11] & ~ \\ 
			3 & 0 & [1, 3, 4] & ~ & 1.01 & 1 & [0.36, 0.11, 0.11] & ~ \\ 
			4 & 0 & [1, 2, 3, 4] & ~ & 0.63 & 0.01 & [0.42, 0.19, 0.13, 0.11] & ~ \\
			4 & 0 & [1, 2, 3, 4] & ~ & 0.65 & 0.1 & [0.41, 0.19, 0.13, 0.11] & ~ \\ 
			4 & 0 & [1, 2, 3, 4] & ~ & 0.71 & 0.5 & [0.39, 0.18, 0.13, 0.11] & ~ \\ 
			4 & 0 & [1, 2, 3, 4] & ~ & 0.80 & 1 & [0.36, 0.17, 0.13, 0.11] & ~ \\ 
			5 & 0 & [1, 2, 3, 4, 5] & ~ & 0.63 & 0.01 & [0.42, 0.19, 0.13, 0.11, 0.04] & ~ \\ 
			5 & 0 & [1, 2, 3, 4, 5] & ~ & 0.64 & 0.1 & [0.41, 0.19, 0.13, 0.11, 0.07] & ~ \\ 
			5 & 0 & [1, 2, 3, 4, 5] & ~ & 0.70 & 0.5 & [0.39, 0.18, 0.13, 0.11, 0.06] & ~ \\ 
			5 & 0 & [1, 2, 3, 4, 5] & ~ & 0.79 & 1 & [0.36, 0.17, 0.13, 0.11, 0.06] & ~ \\ \hline\hline
			0 & 1 & ~ & [1] & 1.76 & 0.01 & ~ & [0.0] \\ 
			0 & 1 & ~ & [1] & 1.78 & 0.1 & ~ & [0.0] \\ 
			0 & 1 & ~ & [1] & 1.87 & 0.5 & ~ & [0.0] \\ 
			0 & 1 & ~ & [1] & 2.00 & 1 & ~ & [0.0] \\ 
			1 & 1 & [2] & [3] & 1.09 & 0.01 & [0.16] & [0.01] \\ 
			1 & 1 & [2] & [3] & 1.11 & 0.1 & [0.16] & [0.01] \\ 
			1 & 1 & [2] & [3] & 1.20 & 0.5 & [0.16] & [0.01] \\ 
			1 & 1 & [2] & [3] & 1.32 & 1 & [0.15] & [0.01] \\ 
			2 & 1 & [1, 3] & [3] & 0.81 & 0.01 & [0.42, 0.11] & [0.01] \\ 
			2 & 1 & [1, 3] & [3] & 0.82 & 0.1 & [0.41, 0.11] & [0.01] \\ 
			2 & 1 & [1, 3] & [3] & 0.88 & 0.5 & [0.39, 0.11] & [0.01] \\ 
			2 & 1 & [1, 3] & [3] & 0.96 & 1 & [0.36, 0.11] & [0.01] \\ 
			3 & 1 & [1, 2, 3] & [3] & 0.62 & 0.01 & [0.42, 0.19, 0.13] & [0.04] \\
			3 & 1 & [1, 2, 3] & [3] & 0.64 & 0.1 & [0.41, 0.19, 0.13] & [0.04] \\
			3 & 1 & [1, 2, 3] & [3] & 0.70 & 0.5 & [0.39, 0.18, 0.13] & [0.03] \\ 
			3 & 1 & [1, 2, 3] & [3] & 0.79 & 1 & [0.36, 0.17, 0.13] & [0.03] \\
			4 & 1 & [1, 2, 3, 5] & [3] & 0.62 & 0.01 & [0.42, 0.19, 0.13, 0.04] & [0.04] \\ 
			4 & 1 & [1, 2, 3, 5] & [3] & 0.63 & 0.1 & [0.41, 0.19, 0.13, 0.04] & [0.04] \\ 
			4 & 1 & [1, 2, 3, 5] & [3] & 0.70 & 0.5 & [0.39, 0.18, 0.13, 0.04] & [0.03] \\ 
			4 & 1 & [1, 2, 3, 5] & [3] & 0.78 & 1 & [0.36, 0.17, 0.13, 0.04] & [0.03] \\ 
			5 & 1 & [1, 2, 3, 4, 5] & [2] & 0.60 & 0.01 & [0.42, 0.19, 0.16, 0.08, 0.04] & [0.18] \\
			5 & 1 & [1, 2, 3, 4, 5] & [2] & 0.61 & 0.1 & [0.41, 0.19, 0.16, 0.09, 0.05] & [0.17] \\ 
			5 & 1 & [1, 2, 3, 4, 5] & [2] & 0.68 & 0.5 & [0.39, 0.18, 0.14, 0.07, 0.04] & [0.16] \\ 
			5 & 1 & [1, 2, 3, 4, 5] & [2] & 0.76 & 1 & [0.36, 0.17, 0.13, 0.06, 0.03] & [0.15] \\ \hline\hline
			0 & 2 & ~ & [1, 3] & 0.79 & 0.01 & ~ & [0.43, 0.00] \\ 
			0 & 2 & ~ & [1, 3] & 0.80 & 0.1 & ~ & [0.42, 0.00] \\ 
			0 & 2 & ~ & [1, 3] & 0.85 & 0.5 & ~ & [0.39, 0.00] \\ 
			0 & 2 & ~ & [1, 3] & 0.91 & 1 & ~ & [0.35, 0.00] \\ 
			1 & 2 & [1] & [2, 3] & 0.61 & 0.01 & [0.42] & [0.05, 0.03] \\ 
			1 & 2 & [1] & [2, 3] & 0.62 & 0.1 & [0.41] & [0.05, 0.03] \\
			1 & 2 & [1] & [2, 3] & 0.69 & 0.5 & [0.39] & [0.05, 0.03] \\ 
			1 & 2 & [1] & [2, 3] & 0.77 & 1 & [0.36] & [0.04, 0.03] \\ 
			2 & 2 & [1, 2] & [2, 3] & 0.60 & 0.01 & [0.42, 0.19] & [0.18, 0.04] \\ 
			2 & 2 & [1, 2] & [2, 3] & 0.62 & 0.1 & [0.41, 0.19] & [0.17, 0.04] \\
			2 & 2 & [1, 2] & [2, 3] & 0.68 & 0.5 & [0.39, 0.18] & [0.16, 0.04] \\ 
			2 & 2 & [1, 2] & [2, 3] & 0.77 & 1 & [0.36, 0.17] & [0.15, 0.03] \\ 
			3 & 2 & [1, 2, 3] & [2, 3] & 0.60 & 0.01 & [0.42, 0.19, 0.16] & [0.18, 0.09] \\ 
			3 & 2 & [1, 2, 3] & [2, 3] & 0.61 & 0.1 & [0.41, 0.19, 0.16] & [0.17, 0.09] \\ 
			3 & 2 & [1, 2, 3] & [2, 3] & 0.68 & 0.5 & [0.39, 0.18, 0.15] & [0.16, 0.11] \\ 
			3 & 2 & [1, 2, 3] & [2, 3] & 0.77 & 1 & [0.36, 0.17, 0.13] & [0.15, 0.08] \\ 
			4 & 2 & [1, 2, 3, 6] & [2, 3] & 0.60 & 0.01 & [0.42, 0.19, 0.15, 0.04] & [0.18, 0.13] \\ 
			4 & 2 & [1, 2, 3, 6] & [2, 3] & 0.61 & 0.1 & [0.41, 0.19, 0.15, 0.04] & [0.18, 0.09] \\
			4 & 2 & [1, 2, 3, 6] & [2, 3] & 0.67 & 0.5 & [0.39, 0.18, 0.15, 0.03] & [0.16, 0.09] \\
			4 & 2 & [1, 2, 3, 6] & [2, 3] & 0.76 & 1 & [0.36, 0.17, 0.14, 0.03] & [0.15, 0.08] \\
			5 & 2 & [1, 2, 3, 4, 5] & [1, 2] & 0.55 & 0.01 & [0.42, 0.21, 0.16, 0.07, 0.04] & [0.50, 0.22] \\ 
			5 & 2 & [1, 2, 3, 4, 5] & [1, 2] & 0.56 & 0.1 & [0.41, 0.2, 0.15, 0.08, 0.05] & [0.49, 0.21] \\ 
			5 & 2 & [1, 2, 3, 4, 5] & [1, 2] & 0.63 & 0.5 & [0.39, 0.2, 0.15, 0.09, 0.04] & [0.46, 0.20] \\ 
			5 & 2 & [1, 2, 3, 4, 5] & [1, 2] & 0.71 & 1 & [0.36, 0.19, 0.14, 0.06, 0.03] & [0.42, 0.19] \\ \hline
		\end{tabular}
	}
	\caption{Index case detected with weekly surveillance LFA test}
	\label{TT3}
\end{table}

\end{document}


\appendix

\noindent \textbf{\Large {Appendix: Mathematical Formulation}}

\section{Disease Evolution and Infectiousness}
\label{evol}

Since the moment a susceptible person becomes infected with COVID-19, the viral load steadily increases until it reaches the limit of detection (LOD), which is the minimum viral load that can be detected with a PCR test. Then, the viral load keeps growing until reaching its peak value. After that, it continuously decreases until it becomes undetectable.
Following \citealt{larremore2021test}, we use a LOD of $10^3$ cp/ml for PCR tests and $10^5$ cp/ml for LFA tests. This methodology adjusts the parameters to simulate viral load for symptomatic and asymptomatic patients.

For the viral load evolution, we use the model described in \citealt{larremore2021test}, and use  $V_t$ to denote the viral load of an infected individual at time $t$ since exposure to the virus. The following parameters describe the control points used to generate sample paths of $V_t$ (see Figure 1): 

\begin{itemize}
\item $t_0$= Time when an infected person reaches the LOD of $10^3$ cp/ml of viral load. $t_0\sim U[2.5, 3.5]$.
\item $t_{peak}$ = Time when an infected individual reaches peak viral load. $t_{peak} = t_0 + \min(3, (0.5 + \gamma))$, where $\gamma$ is a random variable that follows a gamma distribution with a shape of 1.5 and scale of 1.
\item $V_{peak}$ = Value of the logarithm of viral load at its peak. $V_{peak} = V_{t_{peak}} \sim U[7,11]$.
\item $t_{sympt}$= Time when symptoms appear for a symptomatic individual. An individual is symptomatic with probability $1/2$. $t_{sympt} = t_{peak} + s$, where $s\sim U[0, 3]$.
\item $t_f$: Ending time of infectious period. The probability distribution of $t_f$ depends on whether the individual is symptomatic or not. Thus, if it is symptomatic, then $t_f=t_{sympt} + f$; if it is asymptomatic, then $t_f= t_{peak} + f$, where $f\sim U[4,9]$. Note that this is the time when the viral load drops below $10^6$ and the individual is no longer infectious.
\end{itemize}

We assume that the logarithm of the viral load follows a linear function (constant growth) between the time it reaches the LOD for PCR and the time of peak load (from $3$ to $V_{peak})$) and another linear function (constant decrease) from the peak time to the end time of infection, from ${V_{peak}}$ to $6$.
These model parameters fully characterize the viral load evolution for both a symptomatic and an asymptomatic patient in the course of the disease.

\section{Mathematical Formulation}\label{sec:app_math}
Our goal is to determine the most effective testing policies for traced contact to minimize the secondary infection risk in the community,
without the need of an immediate isolation. Thus, we minimized the expected number of contagious days (i.e. days in which the contact is infecting, with viral load greater than or equal to $10^6$ cp/ml), and the traced contact was active in the community before detection and isolation.

Let $T$ be the time horizon in days in which we monitor the traced contact (set to $T=14$). Thus, a test can be scheduled on any day $t=0,1,\ldots, T$, where $t=0$ corresponds to the day on which the contact is identified. A testing schedule is defined by the specific days in which the LFA and/or PCR tests are taken: we denote a schedule as a tuple $(P, A)$, where $A$ is the set of days an LFA test is performed, and $P$ is the set of days a PCR test is performed. If we fix the number of LFA and PCR tests that can be used, then the set $\Pi(i,j)$ consists of all feasible schedules to perform exactly $j$ LFA tests and $i$ PCR tests. We notice that all policies within this set use exactly the same number of tests of each type. For each schedule, we compute the expected number of infectious days a traced contact is active in the community before being isolated. Additionally, we denote $\Pi=\cup_{i,j\in \mathbb{N}} \Pi(i, j)$ as the set of all possible policies.

The dynamics of the testing and isolation process are as follows. At the beginning of $t$, an LFA or PCR test is taken if they are scheduled at $t$. In the case of a LFA test, we assume that its result is observed immediately. In practice, this would take at most 30 minutes, but we assume that the individual is isolated until the LFA test result is back, which makes this is a realistic assumption; in the case of a PCR test, we observe the test result at the beginning of the next day ($t+1$). When positive test results are observed, the individual is immediately isolated and starts quarantine. If the test result is negative, then the individual remains active in the community until the next scheduled test or if symptoms develop.

We evaluated the performance of a test schedule based on the number of days a suspected infected individual was contagious before being identified as such and therefore imposed a risk to the community. For this, we take the perspective of a decision maker who has a budget that specifies the number of LFA and PCR tests that can be performed and needs to decide on which days to take these tests to minimize the number of days an infected individual was contagious in the workplace before being isolated. We remark on the difference between being infected and being contagious, and only the latter imposes an exposure risk to others.


There are several sources of randomness when measuring the number of infectious days of a traced contact before isolation. In what follows, we explain how we consider this randomness in our model and its effect on the computation of the expected number of infectious days.

\begin{itemize}

\item \textbf{Uncertainty on whether or not a traced contact has been infected:} in our analysis, we assume the individual \emph{is} actually infected (i.e., we condition on the event that the contact was infected by the index case at some date of exposure). Although this may seem paradoxical at first since an infected individual should always be isolated, this is methodologically correct in our setting. Our optimization minimizes the number of infecting days subject to the individual not being isolated until confirmed by a positive test or self-isolated at symptom onset. In this optimization, a non infected individual will always contribute zero to the infecting days regardless of the selected policy; therefore, the expected value is conditional on the event of infection. Another reason is that given the applications we consider, we believe decision makers are more concerned with measuring the performance of a testing policy with respect to how good it is at isolating infecting individuals rather than focusing on cases in which the individual is actually not infected. Additionally, taking this approach makes us indifferent to the underlying probability of being infected, which is difficult to estimate and context dependent.

Formally, the objective is to minimize is the expected number of infecting days, and our control is the schedule of the tests (given a number of tests). Denote $\pi$ the testing policy (days in which the tests are performed), $N_{inf}(\pi)$ a random variable representing the number of infecting days under that policy and $\EE[N_{inf}(\pi)]$ its expectation. Because $N_{inf}(\pi)$ equals to zero when the contact was not infected, conditioning on the event that the contact is infected yields:  
\begin{align*}
    \EE[N_{inf}(\pi)] 
    &= \EE[N_{inf}(\pi)\mid \text{Contact \textbf{is} infected}]\times\Pr(\text{Contact \textbf{is} infected})
\end{align*}
Because the probability that the contact is infected  ($\Pr(\text{Contact \textbf{is} infected})$) is independent of the testing policy, choosing $\pi$ to minimize the expected number of infecting days is equivalent to minimize the conditional expectation $\EE[N_{inf}(\pi)\mid \text{Contact \textbf{is} infected}]$. Hence, the optimal testing policy is independent on the prior probability that the contact is infected. We focus the optimization to minimize the number of infected days given that the individual is infected; scaling this objective by the prior probability of infection yields the (unconditional) expected number of infecting days.

\item \textbf{Number of infecting days of the infected contact:} The generative model for the viral load was described in Appendix \ref{evol} and modeled as a function of time, generating multiple curves randomly based on the controls points described in Figure 1
. The expected number of infecting days ($N_{inf}$) is calculated for each simulated viral load path, considering the period before isolation (either through a positive test or self-isolation of symptomatic cases at the onset of symptoms). Hence, this methodology is flexible to accommodate alternative approaches to generate the viral load curve.

\item \textbf{Day of infection of the contact:} Our methodology incorporates uncertainty on the day in which the contact has been infected, assuming a set of days where the index case and the contact had significant interaction and the day the index case was confirmed as infected. This assumption is more realistic in settings with structured contact networks that interact frequently (e.g. school and workplace).

To incorporate this uncertainty in the model, we build a probabilistic distribution of the days in which the infection may have happened, and use this probability distribution when simulating the viral loads of the contacts. Transmission from the index case to the contact occurs on exposure day $t$ when: (i) the index case is infectious on day $t$ (defined as the event $I_t$); (ii) the contact has not yet been infected by the index, that is, is susceptible at the beginning of day $t$ (defined as the event $S_t$). Conditional on the events $I_t$ and $S_t$, infection occurs with probability $\beta$, referred to as the infectivity parameter. Using these definitions, the probability that the index case transmits the disease to the contact on day $t$ is given by:

\begin{equation}
    r_t =\beta \Pr(I_t| S_{t}) \cdot \Pr(S_t), \label{eq:rt_full}
\end{equation}

Note that the events $I_t$ and $S_t$ are not independent, because observing no infection prior to $t$ provides some evidence that the index case may have not yet been infectious. Hence, we use simulation methods to compute equation \eqref{eq:rt_full}.

Define the event $U_d=$ index case was infected on day $d$. For each day $d$ prior to the index case confirmation, we simulate many viral load paths representing the evolution of the disease for the index case when he/she was infected on day $d$. Each viral load path $V^{k,d}$ is a vector specifying the viral load of the index case on each day, denoted $V_t^{k,d}$ (set to zero for $t<d$ because the index was not yet infected); with some abuse of notation, we also use $V^{k,d}$ to denote the event that the index case follows this viral load path. A priori, all the paths $V^{k,d}$ have the same probability, but conditioning on index case confirmation at $t=0$ generates a filter that removes paths that are not consistent with the confirmation event. For example, when the index case is detected via a random LFA test, the filter drops all the paths with $V^{k,d}_0<10^5$; for the weekly LFA test detection, an additional filter is used to drop all the paths with $V^{k,d}_{-6}>10^5$. Denote $\tilde V^{k,d}$ all the paths that \textit{remain} after the filters, $\tilde N$ the total number of remaining paths and $u_d$ the number of these paths that start on day $d$. Note that confirmation at $t=0$ implies that in all the surviving paths the contact has not yet self-isolated during $t<0$. The conditional probability that the index case was infected on day $d$, is the proportion of paths $\tilde V^{k,d}$ that start on day $d$, defined as $q_d = u_d/\tilde N$.

Conditioning on the filtered paths $\tilde V^{k,d}$ and using the indicator function $\mathbbm{1}(\tilde V_t^{k,d}>10^6)$ to represent a viral path that is infectious on day $t$, equation \eqref{eq:rt_full} can be expressed as:
\begin{align}
    r_t &=\frac{1}{\tilde N}\sum_{k,d} \beta \Pr(I_t| S_{t},\tilde V^{k,d}) \cdot \Pr(S_t|\tilde V^{k,d}) \nonumber \\
        &=\frac{1}{\tilde N}\sum_{k,d} \beta \mathbbm{1}(\tilde V_t^{k,d}>10^6) \cdot \prod_{j\leq t-1} (1 - \beta \mathbbm{1}(\tilde V_j^{k,d}>10^6)) \nonumber \\
        &= \sum_{d\leq t} q_d \cdot \frac{1}{u_d} \sum_k \beta \mathbbm{1}(\tilde V_t^{k,d}>10^6) \cdot \prod_{j\leq t-1} (1 - \beta \mathbbm{1}(\tilde V_j^{k,d}>10^6)) \label{eq:rt_v}
\end{align}

To facilitate computations, we used the following approximation for equation \eqref{eq:rt_v}:
 \begin{align*}
    r_t &\approx \sum_{d\leq t} q_d \cdot  \beta \frac{1}{u_d} \sum_k\mathbbm{1}(\tilde V_t^{k,d}>10^6) \cdot \prod_{j\leq t-1} (1 - \beta \frac{1}{u_d} \sum_k\mathbbm{1}(\tilde V_j^{k,d}>10^6)) \\
        &= \sum_d q_d \cdot \beta \Pr(I_t|U_d) \cdot \prod_{j\leq t-1}(1-\beta \Pr(I_j|U_d) ),
\end{align*}

where the values $\Pr(I_j|U_d)=\frac{1}{u_d} \sum_k\mathbbm{1}(\tilde V_t^{k,d}>10^6)$ can be computed once and used for all the simulations including different values of the infectivity parameter $\beta$.
Finally, we compute the normalized probabilities by conditioning that the contact was infected. For the results shown in this paper, we have computed the exact and approximate values of the normalized $r_t$  for all $t$ and high and low values of $\beta$, and obtained good approximations, within 1\% of the probability values.

\end{itemize}

\section{Simulation based optimization}
In what follows, we present a detailed mathematical formulation for the optimization problem. We consider a standard probability space in which we measure the viral load of an individual who has been infected. This randomness could be attributed to the random variations in viral load evolution for different individuals. All random variables and filters are defined with respect to this probability space. We use the following notation:

\begin{itemize}
\item $V_t$, $t = 1, \ldots, T$ = Viral load on day $t$ after the index case is discovered. If we know the exact day the individual was infected, then $V_t$ would be completely described by the process explained in Section \ref{evol}. However, since we do not necessarily know the exact day but only a probabilistic distribution over the days of infection, we take $V_t$ to be the random process conditioned on that infection day distribution.
\item $x^A_t, x^P_t$ = Variables indicating that a test result was \textbf{observed} at day $t$ ($x^A$ for LFA test, and $x^P$ for PCR), and they have a value of 1 if a result is observed (independent of its value) and 0 otherwise.
\item $L^A=10^5$, $L^P=10^3$ Levels of detection for each test type. We use and $L^I=10^6$ to represent the viral load threshold above which an individual is infectious.
\item $\mathcal{D}^A$ = $\{d \mid x_d^A= 1\}$, $\mathcal{D}^P=\{d \mid x_d^P= 1\}$. Sets of days where a LFA and PCR test result were observed. Note that for the LFA test, this value coincides with the day of the test, whereas for PCR, it corresponds to one day later (we assume that PCR test results are obtained 24 hours after they are taken, while for LFA tests, these are obtained immediately). Thus,

\item $\mathcal{D}^A_t \{d \in \mathcal{D}^A \mid d\leq t\}$:  Days of LFA test results up to day $t$. Similar for $\mathcal{D}^P_t$ with PCR test days.


\item $R_t^A, R_t^P$ = Random variables indicating the result of an LFA or PCR test observed on day $t$. The distributions of $R_t^A$ and $R_t^P$ depend on $x_t^A$ and $x_t^P$, respectively. If a test result was not observed that day, then we assume that $R_t$ takes the value of -1.
\[
      R_t^A = \begin{cases} 
      - 1 & x_t^A=0 \\
      \mathds{1}\{V_t \geq L^A\} & x_t^A=1 
      \end{cases},\ \ \
      R_t^P = \begin{cases} 
      - 1 & x_t^P=0 \\
      \mathds{1}\{V_{t-1} \geq L^P\} & x_t^P=1 
      \end{cases},
    \]
Denoting $R_t=\max(R_t^A,R_t^P)$ as a random variable that indicates the presence of any test, then:
\[
      R_t = \begin{cases} 
      - 1 & x_t^A=x_t^P=0 \\
      \mathds{1}\{V_t \geq L^A\} & x_t^A=1 \text{ and } x_t^P=0 \\
      \mathds{1}\{V_{t-1} \geq L^P\} & x_t^A=0 \text{ and } x_t^P=1 \\
      \mathds{1}\{V_t \geq L^A \text{ or } V_{t-1} \geq L^P\} & x_t^A=x_t^P=1.
      \end{cases}
    \]


\item $\mathcal{H}=(\mathcal{H}_t)_t$ = Filtration with respect to the process $((R^A_t, R^P_t))_{t}$, i.e., $\mathcal{H}_t= \sigma(R^A_k, R_k^P\mid k\leq t)$.

\item $\mathcal{S}_t$ = Observable state at the beginning of time $t$. Note that $\mathcal{H}_t$ corresponds to all the information the decision maker has at time $t$ about the state of the infection in the target individual. $\mathcal{S}_t = (\mathcal{H}_{t-1}, (x_\tau)_{\tau \leq t-1})$.
\end{itemize}

An individual who is infected will be contagious only when the viral load surpasses $L^I=10^6$ cp/ml and remains active (not isolated) until positive test results emerge or at symptoms onset. This means that an infectious day will occur if and only if the following three events happen at day $t$:
\begin{itemize}
\item $I_t = \{V_t \geq L^I\}$: The individual is infecting.
\item $N_t = \{(R_k)_{k\leq t} \in \{-1, 0\}^t\}$: All test results up to day $t$ have been negative.
\item $Z_t$: No symptoms at day $t$.
\end{itemize}
Thus, the total number of days where the agent is infecting is equal to:
\[
\sum_{t=0}^T \mathds{1}\{I_t \cap N_t \cap Z_t\}.
\]

The decision maker designing the test schedule does not know the value of $\mathds{1}\{I_t \cap N_t \cap Z_t\}$ and can only infer the distribution of the event $I_t \cap N_t \cap Z_t$ based on prior knowledge of the distribution of the viral load for an infected individual as well as the information obtained through the testing policy, which allows to update the belief on the viral load distribution each time a test result is observed.

Define $\mathcal{J}_T = \mathds{1}\{I_T \cap N_T \cap Z_T\}$ and $ \mathcal{J}_t = \mathds{1}\{I_t \cap N_t \cap Z_t\} + \mathcal{J}_{t+1}$ recursively. Therefore, the decision maker will try at the beginning of each day $t$ to minimize the following quantity:
\begin{equation}
	\EE[\mathcal{J}_t \mid \mathcal{S}_t] = \EE[\mathds{1}\{I_t \cap N_t \cap Z_t\} \mid \mathcal{S}_{t}] + \EE[\mathcal{J}_{t+1} \mid  \mathcal{S}_{t}]. 
\label{eq:2} 
\end{equation}


Recall that $\mathcal{H}_t=\sigma(R^A_k, R^P_k\mid k\leq t)$, which means that $N_{t-1}$ is $\mathcal{H}_{t-1}$-measurable, and since $N_t = N_{t-1}\cap \{R_t \in \{-1,0\}\}$, we can rewrite the first term of Equation~\eqref{eq:2} as
\begin{align*}
\EE[\mathds{1}\{I_t \cap N_t \cap Z_t\} \mid \mathcal{S}_{t}] = \mathds{1}\{N_{t-1},Z_t\} \EE[\mathds{1} \{I_t \cap \{R_t \in \{-1,0\}\} \cap Z_t\} \mid \mathcal{S}_{t}] = \PP(I_t, R_t \in \{-1,0\}\mid \mathcal{S}_t),
\end{align*}
because if a positive test is observed at some point in the past, the individual is taken to quarantine and the risk is over, then $\mathds{1}\{N_{t-1}\}$ must be equal to one if the decision maker is making a decision at time $t$. The same happens if the individual presents symptoms on day $t$. Given that the distribution of $R_t$ is determined by $x_t^A$ and $x_{t}^P$, we have
\begin{equation}
\PP(I_t, R_t \in \{-1,0\}, Z_t \mid \mathcal{S}_t) = 
\begin{cases}
\PP(V_t \geq L^I \mid \mathcal{S}_t) & x_{t}^P = x_t^A =0 \\
\PP(V_t \geq L^I, V_t < L^A\mid \mathcal{S}_t) = 0 & x_t^A=1 \text{ and } x_t^P=0 \\
\PP(V_t \geq L^I, V_{t-1} < L^P\mid \mathcal{S}_t) & x_t^A=0 \text{ and } x_t^P=1 \\
\PP(V_t \geq L^I, V_{t-1} < L^P, V_t < L^A \mid \mathcal{S}_t\}=0 & x_t^A=x_t^P=1
\end{cases}
\label{eq:1}
\end{equation}

The second and fourth cases are equal to zero since a negative LFA test immediately discards the event that the agent may be infecting. Let us look at the first case in more detail. We have
\begin{equation}
	\PP(V_t \geq L^I \mid \mathcal{H}_{t-1}, x_1,\ldots, x_{t-1}) = \PP(V_t \geq L^I \mid V_{d^A} < L^A\ \forall\in\mathcal{D}^A,\ V_{d^P-1} < L^P\ \forall d^P\in \mathcal{D}^P)
\label{eq:3}
\end{equation}
The third term in Equation~\eqref{eq:1} can be written similarly.


Using the probability distributions for the times of LOD and the times for peak viral load and end of contagious period described in Section \ref{evol}, we determine the probability distribution for $V_t$ for each $t$. Thus, we can use Monte Carlo simulations to compute the value of \eqref{eq:3} using the identity:
\begin{align}
&\PP(V_t \geq L^I \mid V_{d^A} < L^A\ \forall\in\mathcal{D}^A,\ V_{d^P-1} < L^P\ \forall d^P\in \mathcal{D}^P) \nonumber
\\ & = \frac{\PP(V_t \geq L^I, V_{d^A} < L^A\ \forall\in\mathcal{D}^A,\ V_{d^P-1} < L^P\ \forall d^P\in \mathcal{D}^P)}{\PP(V_{d^A} < L^A\ \forall\in\mathcal{D}^A,\ V_{d^P-1} < L^P\ \forall d^P\in \mathcal{D}^P)}.
\label{eq:5}
\end{align}

To conclude, we recall that the expected number of infected days is given by
\[
\EE\left[\sum_{t=0}^T \mathds{1}\{I_t \cap N_t \cap Z_t\}\right] =\sum_{t=0}^T \PP(I_t, N_t, Z_t).
\]
By total probability, we can condition each of the probabilities $\PP(I_t, N_t, Z_t)$ by the state up to time $t$, which indicates the probability of being in such a state if we follow a certain policy. Each of these terms is of the form $\PP(I_t, N_t, Z_t\mid \mathcal{S}_t)\PP(\mathcal{S}_t)$; thus, in Equation \eqref{eq:5}, the expectation can be written as
\[\EE\left[\sum_{t=0}^T \mathds{1}\{I_t \cap N_t \cap Z_t\}\right]=\sum_{t=0}^T \PP(V_t \geq L^I, V_{d^A} < L^A\ \forall\in\mathcal{D}^A,\ V_{d^P-1} < L^P\ \forall d^P\in \mathcal{D}^P)\]
where each of the terms in the summation can be computed using Monte Carlo simulations.

\section{Additional results} \label{sec:add_results}
\subsection{Analysis with other types of index case detection} \label{app:sec:agcad}

\begin{figure}[H]
\centering
        \includegraphics[scale=0.25]{figures/high_home_plot_allalpha_agcad.png}
    \begin{overpic}[scale=0.25,trim=0 0 0 75,clip]{figures/high_home_plot_combined_robust_agcad.png}
        \put(85,46){\includegraphics[scale=0.32]{figures/legend_results.png}} 
    \end{overpic}
\caption{\textbf{Evaluation of testing strategies for a traced contact exposed to an index case detected with weekly surveillance LFA tests.}
}
\centering
\label{fig:LFAweekly}
\end{figure}

\subsection{Results with different scenarios of LFA test sensitivity} \label{app:sec:test_sensitivity}

\begin{figure}[H]
    \centering
    \includegraphics[width=0.8\columnwidth]{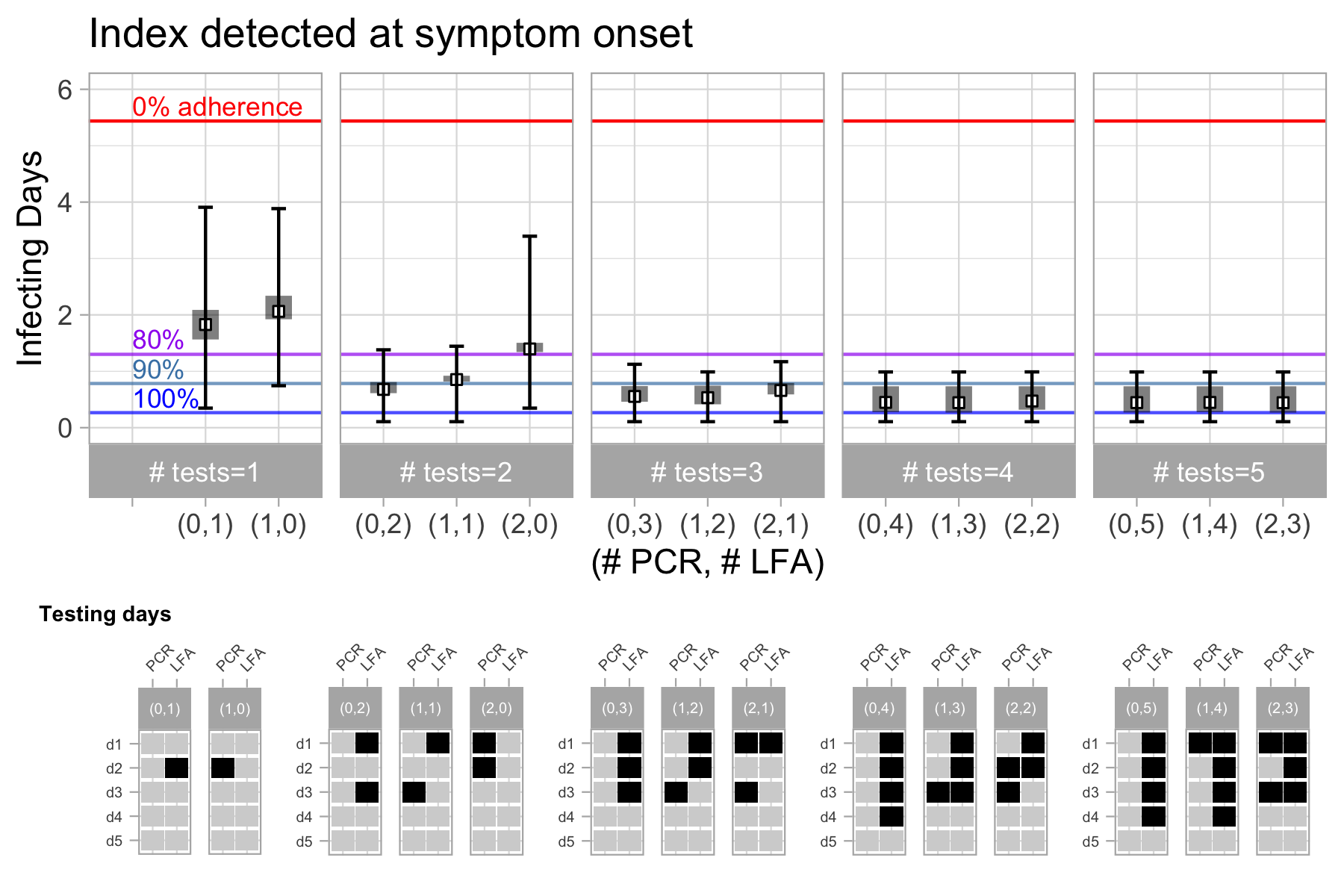}
    \includegraphics[width=0.8\columnwidth]{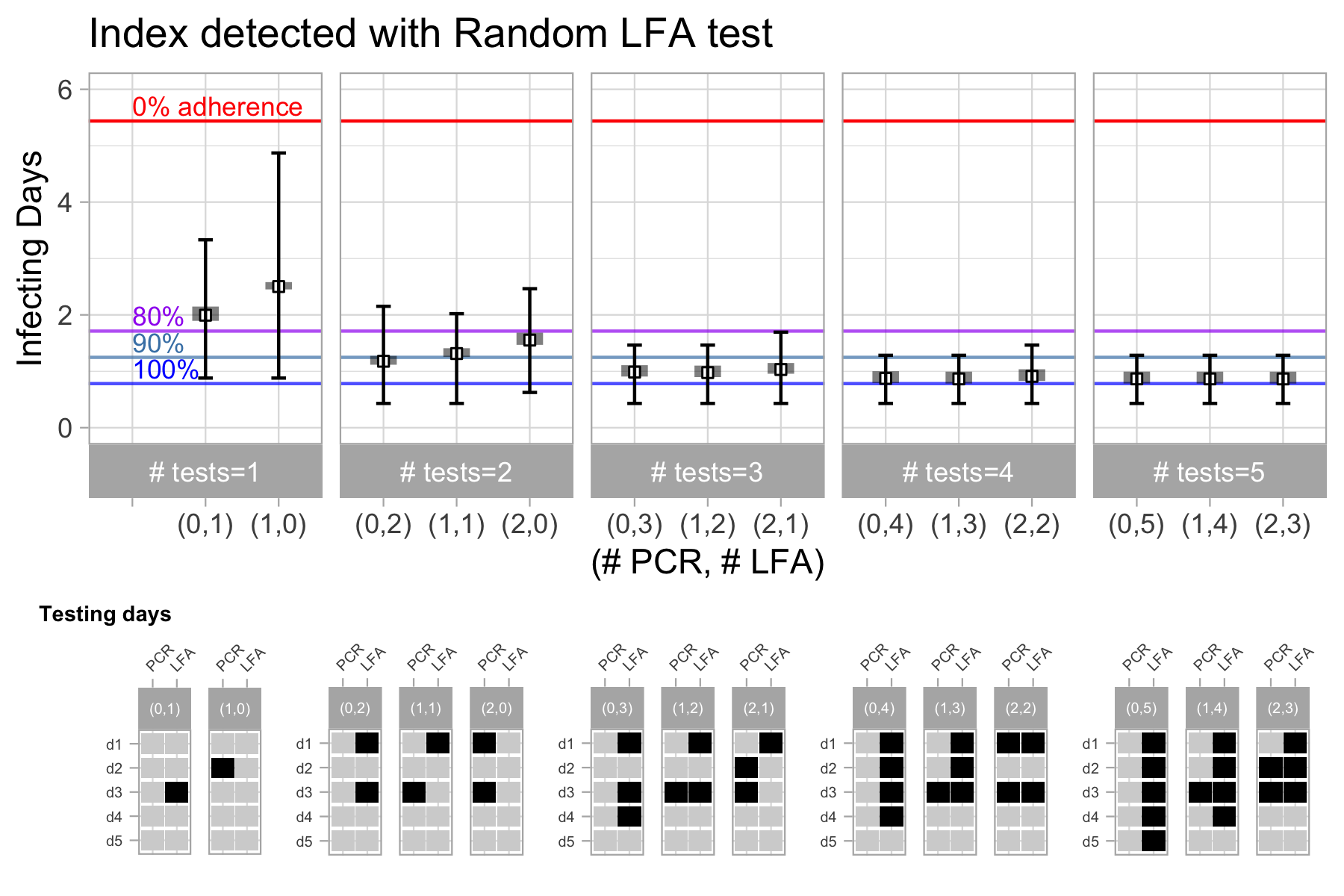}
    \caption{Robust testing policies for the \textit{LFA High} test sensitivity scenario. For the test schedule with (PCR=0,LFA=5) in the top panel, the last LFA test is on day 6 (not shown in the figure).}
    \label{app:fig:results_test_High}
\end{figure}

\begin{figure}[H]
    \centering
    \includegraphics[width=0.8\columnwidth]{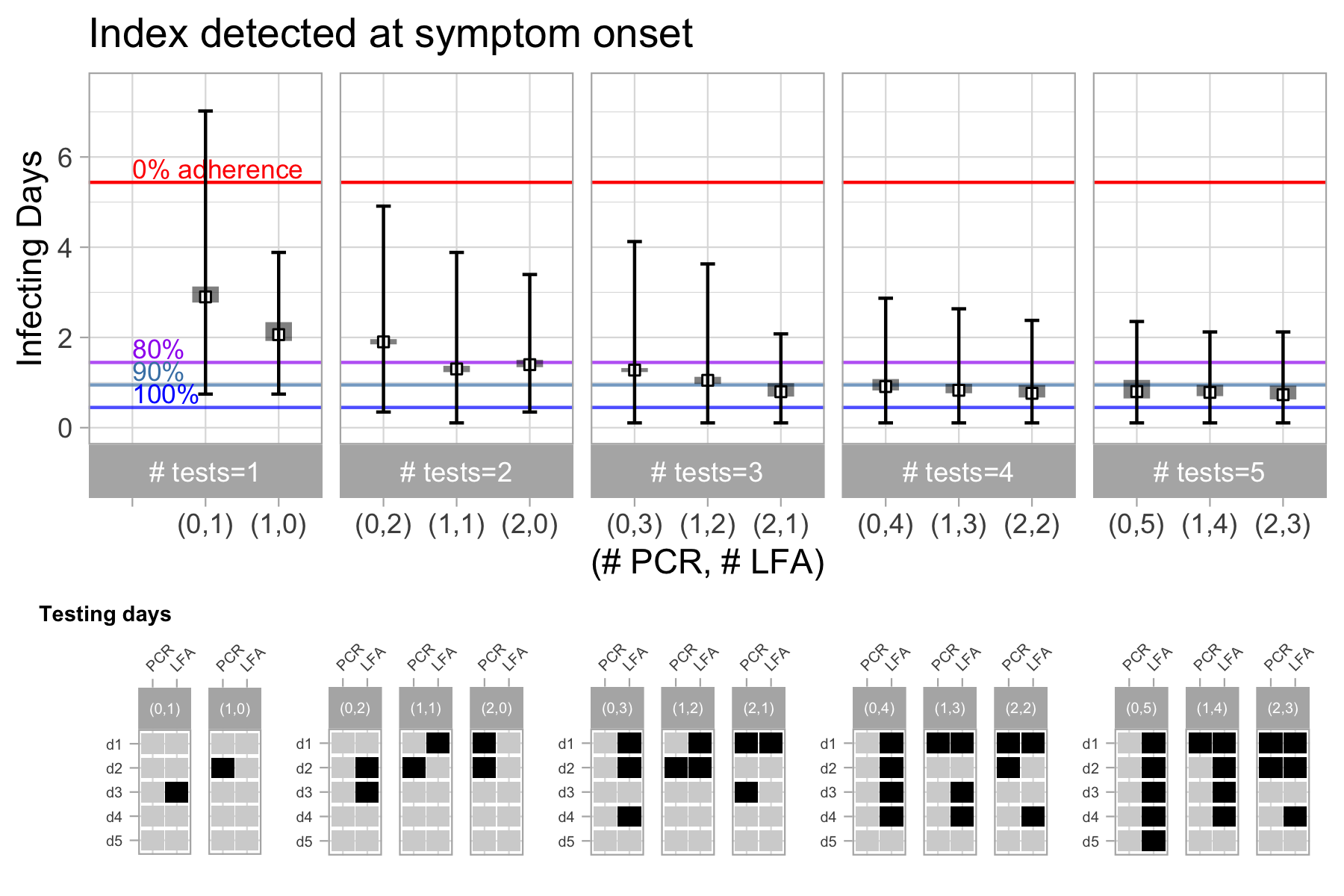}
    \includegraphics[width=0.8\columnwidth]{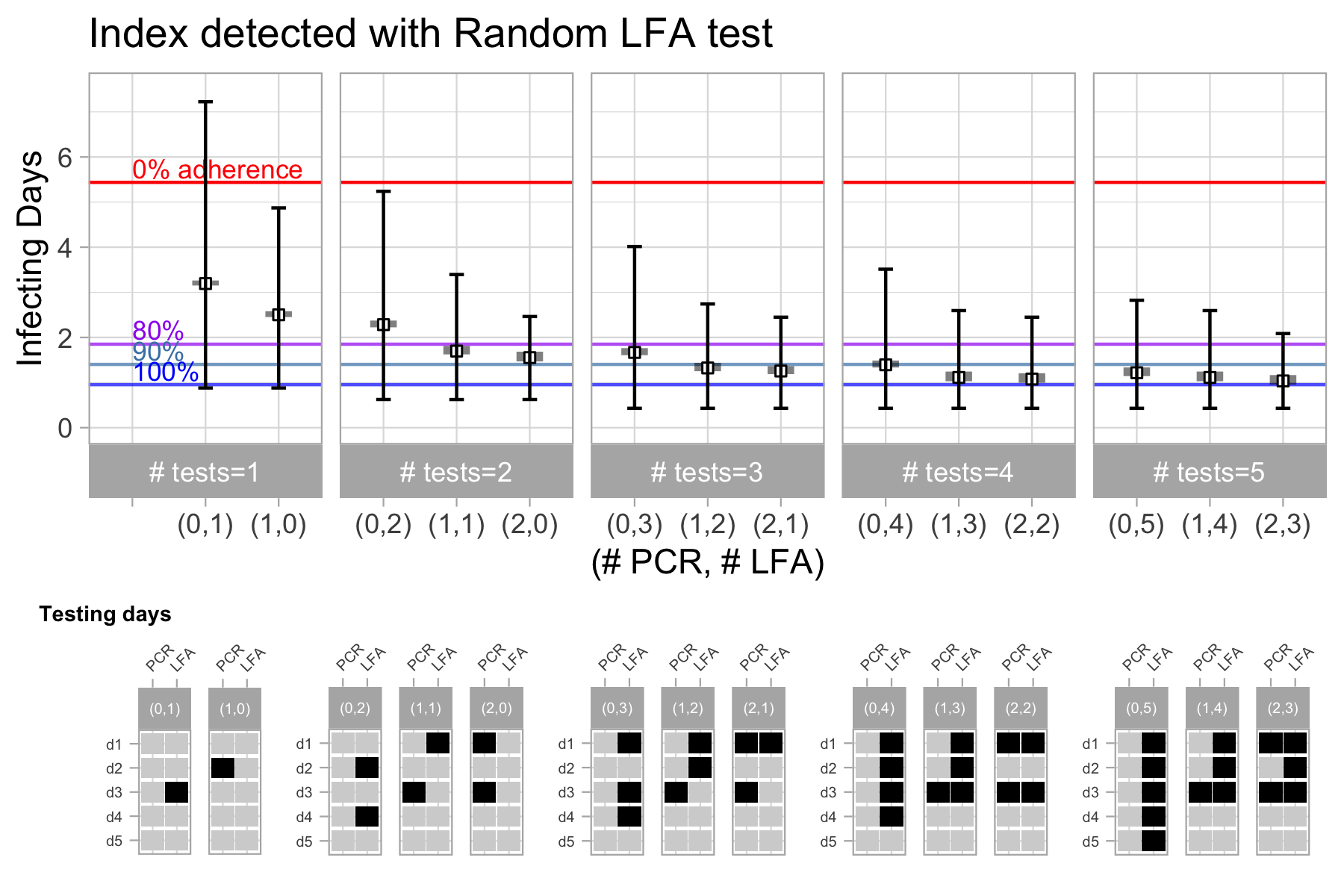}
    \caption{Robust testing policies for the \textit{LFA Low} test sensitivity scenario. For the test schedule (PCR=1,LFA=4) in the bottom panel, the last LFA is taken on day 6 (not shown in the figure).}
    \label{app:fig:results_test_Low}
\end{figure}

\subsection{Results with alternative models of viral load trajectories}\label{app:sec:jones}

\begin{figure}[H]
    \centering
    \includegraphics[width=0.8\columnwidth]{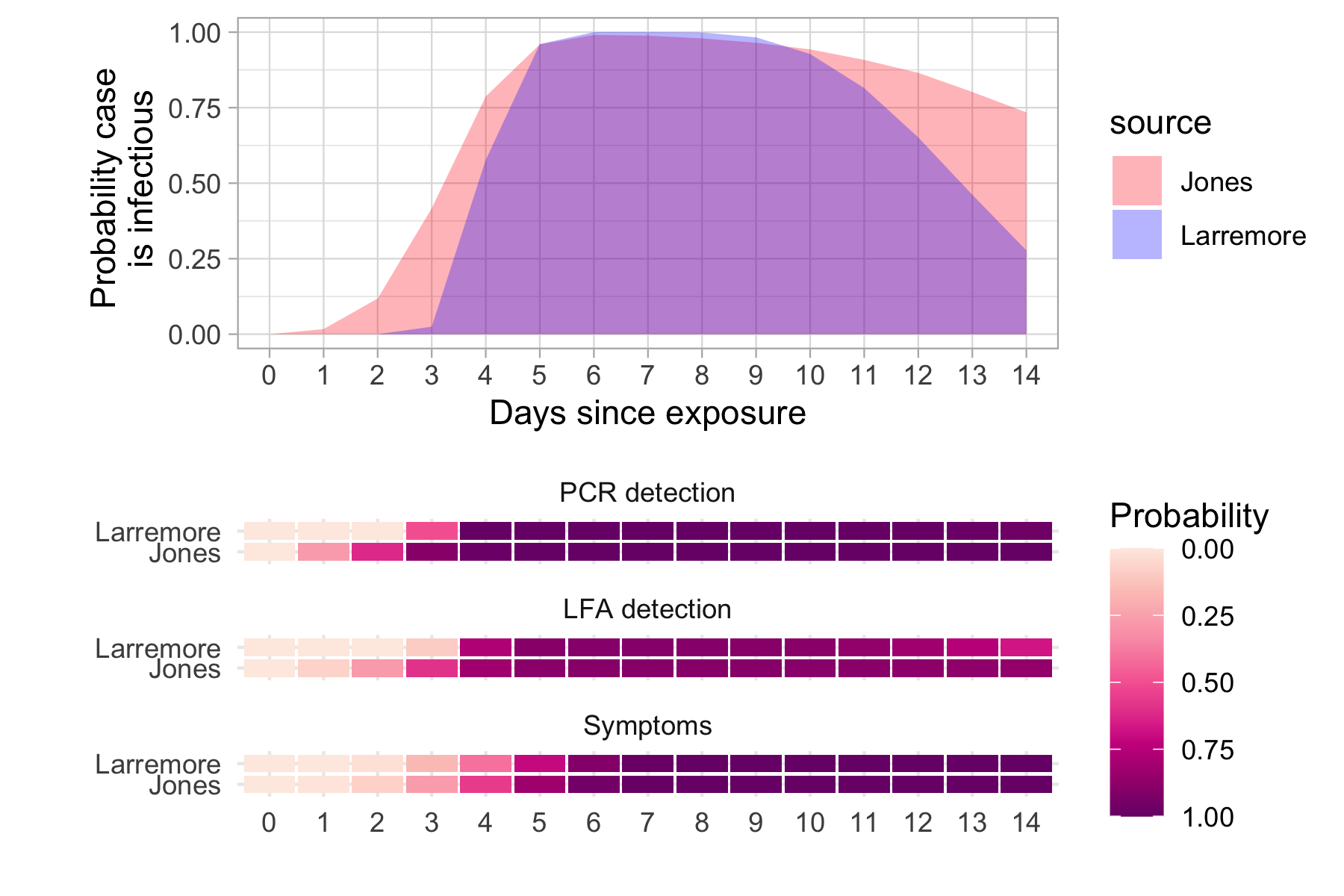}
    \caption{Comparison of infectiousness and test sensitivity dynamics between the viral load models of \citealt{larremore2021test} and \citealt{jones2021estimating}}.
    \label{app:fig:comp_jones}
\end{figure}

Figure \ref{app:fig:comp_jones} compares the infectiousness and test sensitivity of the grouped estimates of viral load trajectories reported in \citealt{jones2021estimating}, with those used in our main analysis (based on \citealt{larremore2021test}. Specifically, we simulated viral load paths using the group estimates reported in Figure S5 in the Supplementary material of \citealt{jones2021estimating}, summarized in Table \ref{app:tab:jones} (each viral load path was generated by simulating each parameter from a Normal distribution with the indicated mean and standard deviation). These viral load trajectories suggest a more extended period of infectiousness, but also earlier detection with PCR and LFA tests and symptoms onset. To study whether this affected the main conclusions of our analysis, we repeated all the simulations using the viral model of \citealt{jones2021estimating}. 

\begin{table}[]
\centering
\begin{tabular}{lcc}
\hline
Parameter         & Mean  & Std. Dev. \\ \hline
Increasing slope  & 2     & 0.39      \\
Days to peak load & 4.3   & 0.92      \\
Peak viral load   & 8.1   & 0.7       \\
Decreasing slope  & -0.17 & 0.02      \\ \hline
\end{tabular}

\caption{Parameters to simulate viral load trajectories based on the results reported in \citealt{jones2021estimating}}
\label{app:tab:jones}
\end{table}

\begin{figure}[H]
    \centering
    \includegraphics[width=0.8\columnwidth]{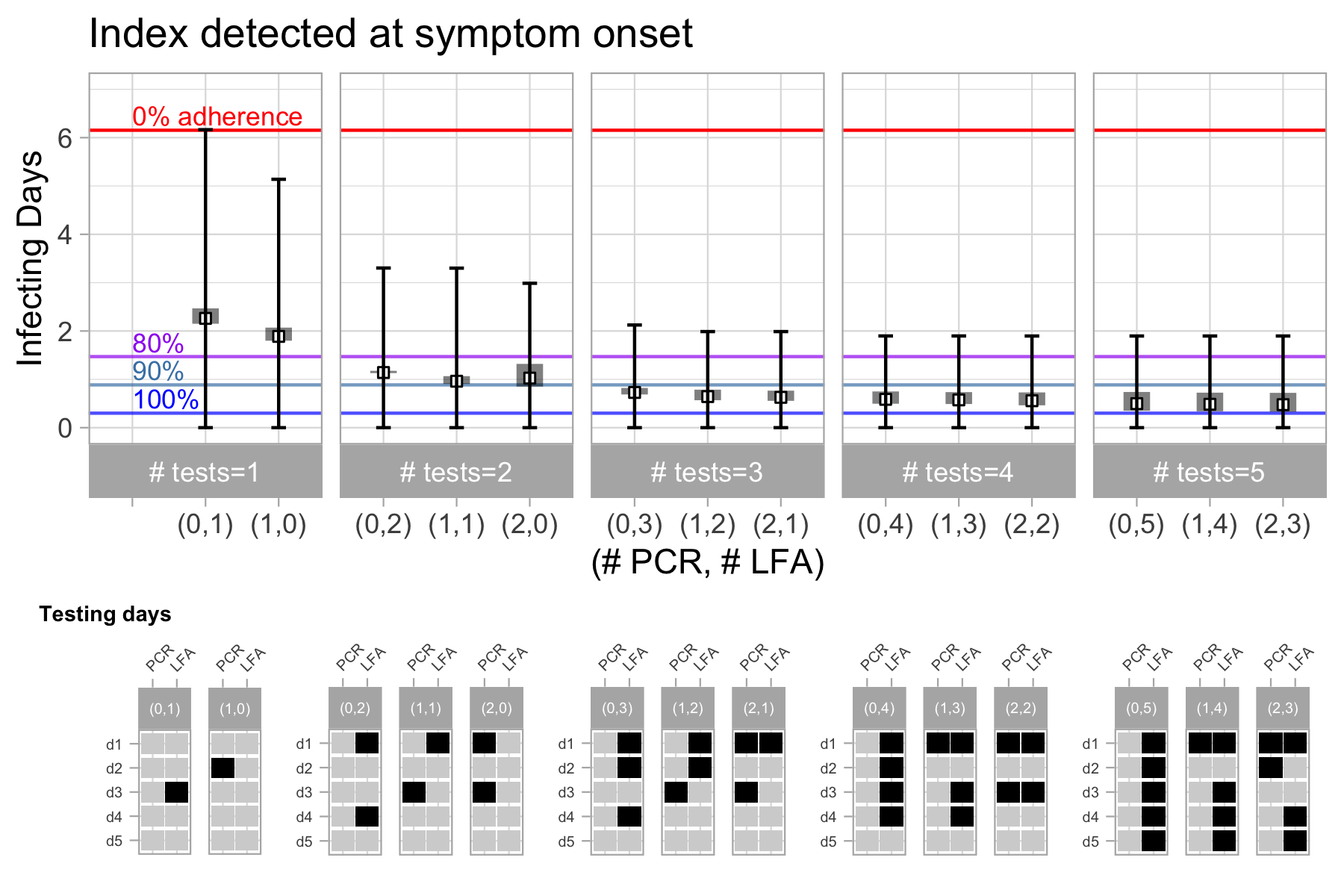}
    \includegraphics[width=0.8\columnwidth]{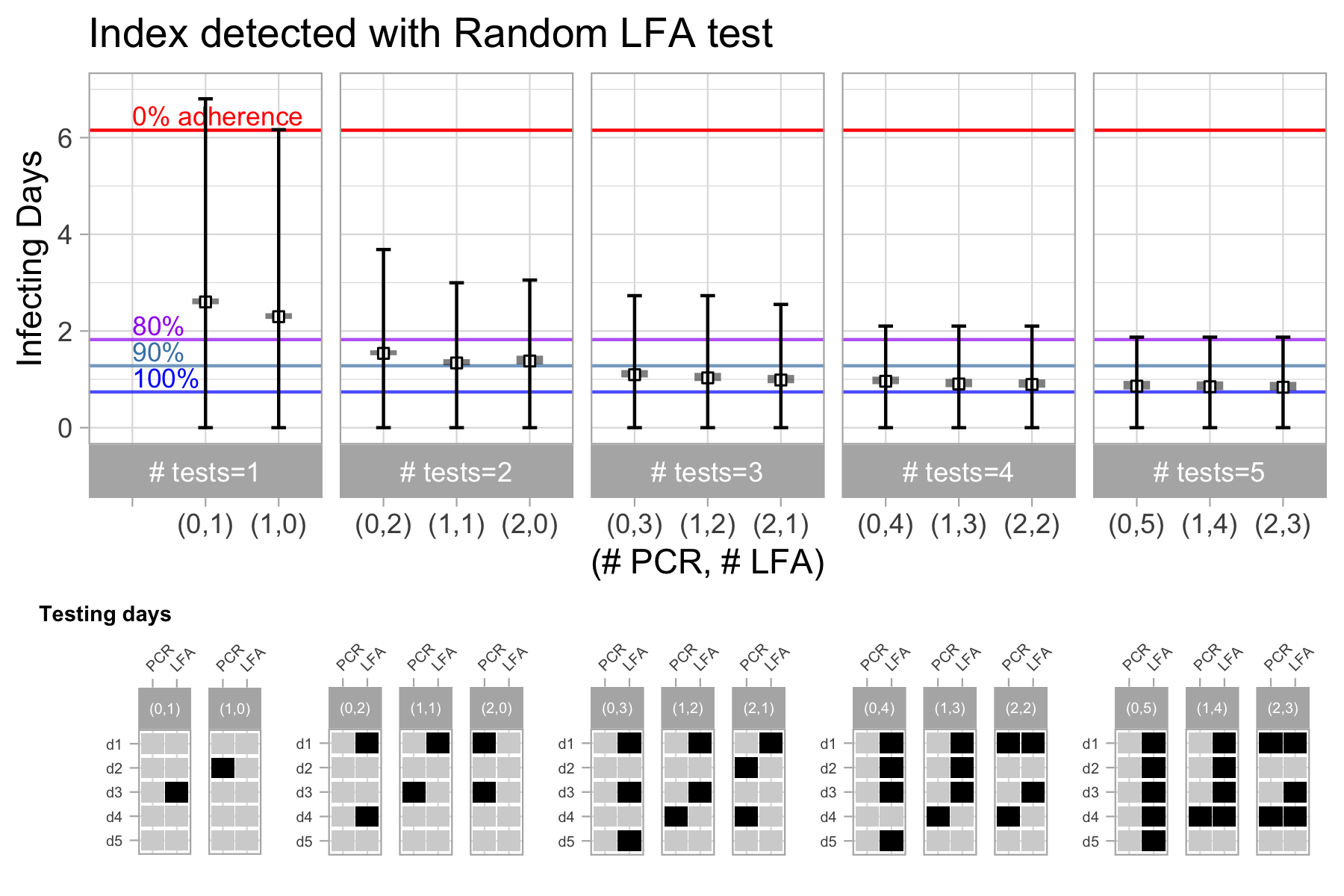}
    \caption{Robust testing policies for the \textit{LFA Med} test sensitivity scenario using viral load trajectories from \citealt{jones2021estimating}.}
    \label{app:fig:results_jones_Med}
\end{figure}

\begin{figure}[H]
    \centering
    \includegraphics[width=0.8\columnwidth]{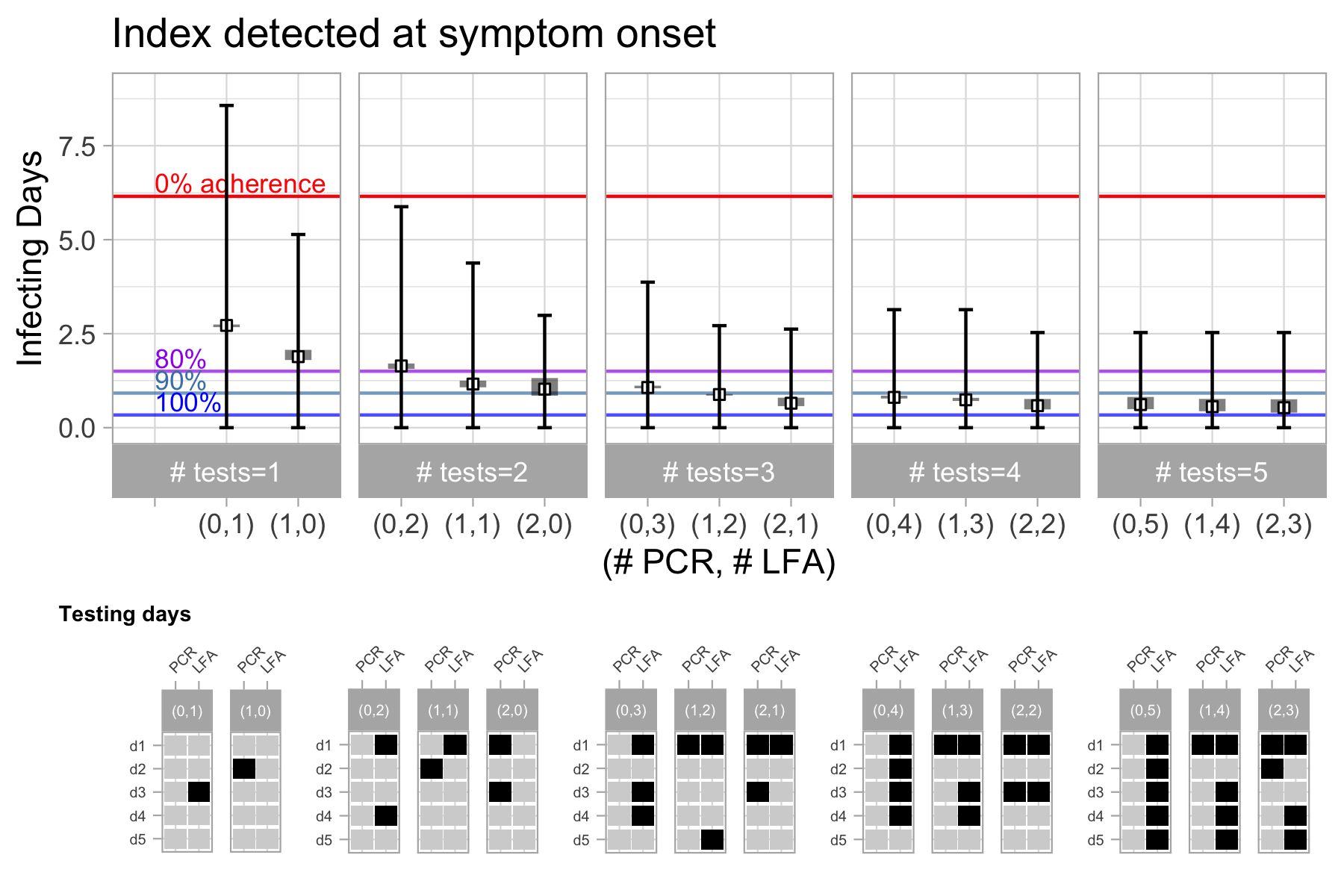}
    \includegraphics[width=0.8\columnwidth]{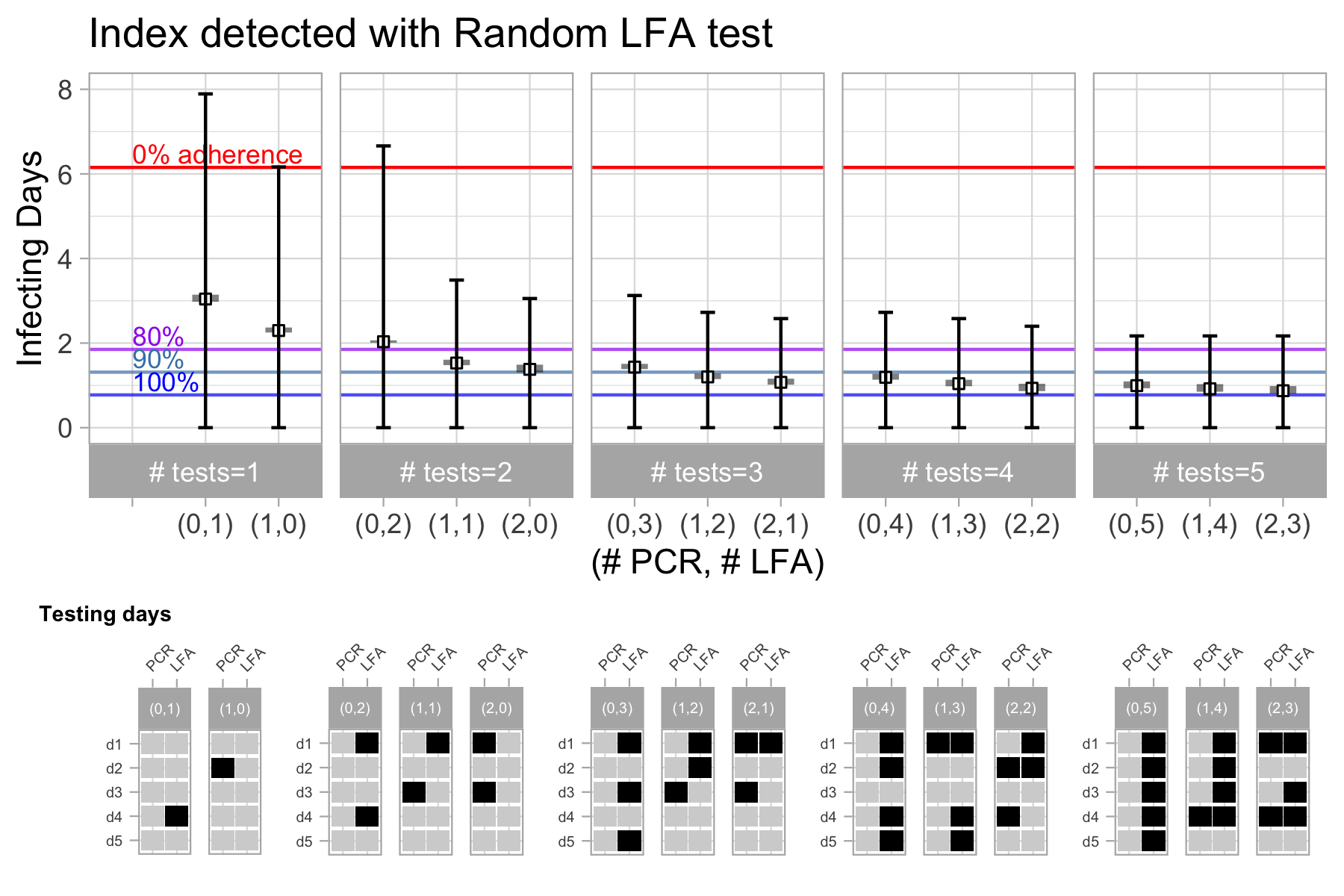}
    \caption{Robust testing policies for the \textit{LFA Med-Low} test sensitivity scenario using viral load trajectories from \citealt{jones2021estimating}.}
    \label{app:fig:results_jones_MedLow}
\end{figure}

Figures \ref{app:fig:results_jones_Med} and \ref{app:fig:results_jones_MedLow} show the robust testing policies obtained through the simulations using these alternative viral load paths, considering two scenarios of LFA test sensitivity (\textit{LFA Med} and \textit{LFA Med-Low}).

\subsection{Detailed results of robust  policies}

In  what follows, we present the expected infecting days and the false negative rates at the times when  tests are performed, for all values of the infectivity parameter $\beta$ and  all combinations of tests considered. For Tables \ref{TT1}, \ref{TT2}, and \ref{TT3}, we considered the  LFA \textit{Med} test sensitivity scenario.

\begin{table}
\tiny{
    \centering
    \begin{tabular}{l|l|l|l|l|l|l|l}
    \# LFA & \# PCR & LFA schedule & PCR schedule & Exp.infecting days & $\beta$ & FNR (LFA)  & FNR (PCR)\\
 
    \hline\hline
        1 & 0 & [2] & ~ & 2.07 & 0.01 & [0.14] & ~ \\
        1 & 0 & [2] & ~ & 2.13 & 0.1 & [0.14] & ~ \\ 
        1 & 0 & [2] & ~ & 2.42 & 0.5 & [0.12]    & ~ \\ 
        1 & 0 & [2] & ~ & 2.82 & 1 & [0.10]  & ~ \\ 
        2 & 0 & [1, 3] & ~ & 1.08 & 0.01 & [0.31, 0.11]  & ~ \\ 
        2 & 0 & [1, 3] & ~ & 1.11 & 0.1 & [0.29, 0.11]   & ~ \\ 
        2 & 0 & [1, 3] & ~ & 1.34 & 0.5 & [0.20, 0.10]   & ~ \\ 
        2 & 0 & [1, 3] & ~ & 1.68 & 1 & [0.12, 0.10]  & ~ \\ 
        3 & 0 & [1, 2, 3] & ~ & 0.77 & 0.01 & [0.31, 0.16, 0.12] & ~ \\ 
        3 & 0 & [1, 2, 3] & ~ & 0.83 & 0.1 & [0.29, 0.16, 0.12] & ~ \\ 
        3 & 0 & [1, 2, 3] & ~ & 1.13 & 0.5 & [0.20, 0.13, 0.11] & ~ \\ 
        3 & 0 & [1, 2, 3] & ~ & 1.55 & 1 & [0.12, 0.11, 0.11] & ~ \\
        4 & 0 & [1, 2, 3, 4] & ~ & 0.71 & 0.01 & [0.31, 0.16, 0.12, 0.11] & ~ \\ 
        4 & 0 & [1, 2, 3, 4] & ~ & 0.78 & 0.1 & [0.29, 0.16, 0.12, 0.11]  & ~ \\ 
        4 & 0 & [1, 2, 3, 4] & ~ & 1.11 & 0.5 & [0.20, 0.13, 0.11, 0.10]  & ~ \\ 
        4 & 0 & [1, 2, 3, 4] & ~ & 1.55 & 1 & [0.12, 0.11, 0.10, 0.11]   & ~ \\ 
        5 & 0 & [1, 2, 3, 4, 5] & ~ & 0.71 & 0.01 & [0.31, 0.16, 0.12, 0.10, 0.06]  & ~ \\ 
        5 & 0 & [1, 2, 3, 4, 5] & ~ & 0.77 & 0.1 & [0.29, 0.16, 0.12, 0.10 , 0.06]   & ~ \\ 
        5 & 0 & [1, 2, 3, 4, 5] & ~ & 1.11 & 0.5 & [0.20, 0.13, 0.11, 0.09, 0.04] & ~ \\ 
        5 & 0 & [1, 2, 3, 4, 5] & ~ & 1.54 & 1 & [0.12, 0.11, 0.11, 0.09, 0.02] & ~ \\ \hline \hline
        0 & 1 & ~ & [2] & 1.37 & 0.01 & ~ & [0.02] \\ 
        0 & 1 & ~ & [2] & 1.44 & 0.1 & ~ & [0.02] \\ 
        0 & 1 & ~ & [2] & 1.83 & 0.5 & ~ & [0.01] \\
        0 & 1 & ~ & [2] & 2.33 & 1 & ~ & [0.00] \\
        1 & 1 & [1] & [3] & 0.87 & 0.01 & [0.31] & [0.00] \\ 
        1 & 1 & [1] & [3] & 0.92 & 0.1 & [0.29] & [0.00] \\
        1 & 1 & [1] & [3] & 1.22 & 0.5 & [0.20]   & [0.00] \\
        1 & 1 & [1] & [3] & 1.63 & 1 & [0.12]  & [0.00]  \\
        2 & 1 & [1, 2] & [3] & 0.71 & 0.01 & [0.31, 0.16]   & [0.01]   \\ 
        2 & 1 & [1, 2] & [3] & 0.77 & 0.1 & [0.29, 0.16] & [0.01]  \\ 
        2 & 1 & [1, 2] & [3] & 1.11 & 0.5 & [0.20, 0.13] & [0.00] \\ 
        2 & 1 & [1, 2] & [3] & 1.54 & 1 & [0.12, 0.11] & [0.00] \\ 
        3 & 1 & [1, 2, 3] & [1] & 0.7 & 0.01 & [0.31, 0.19, 0.10] & [0.35] \\ 
        3 & 1 & [1, 2, 3] & [1] & 0.75 & 0.1 & [0.29, 0.18, 0.09] & [0.32]  \\ 
        3 & 1 & [1, 2, 3] & [1] & 1.04 & 0.5 & [0.20, 0.13, 0.07] & [0.17] \\ 
        3 & 1 & [1, 2, 3] & [1] & 1.45 & 1 & [0.12, 0.08, 0.04] & [0.04] \\ 
        4 & 1 & [1, 2, 3, 4] & [1] & 0.64 & 0.01 & [0.31, 0.19 , 0.10, 0.05] & [0.35]  \\ 
        4 & 1 & [1, 2, 3, 4] & [1] & 0.7 & 0.1 & [0.29, 0.18, 0.09, 0.05] & [0.32]  \\ 
        4 & 1 & [1, 2, 3, 4] & [1] & 1.02 & 0.5 & [0.20, 0.14, 0.07, 0.03] & [0.17] \\ 
        4 & 1 & [1, 2, 3, 4] & [1] & 1.45 & 1 & [0.12, 0.08, 0.07, 0.01] & [0.04]  \\ 
        5 & 1 & [1, 2, 3, 4, 5] & [1] & 0.64 & 0.01 & [0.31, 0.18, 0.10, 0.05, 0.02] & [0.35] \\ 
        5 & 1 & [1, 2, 3, 4, 5] & [1] & 0.7 & 0.1 & [0.29, 0.18, 0.09, 0.05, 0.02] & [0.32]  \\ 
        5 & 1 & [1, 2, 3, 4, 5] & [1] & 1.02 & 0.5 & [0.20, 0.14, 0.07, 0.03, 0.04] & [0.17] \\ 
        5 & 1 & [1, 2, 3, 4, 5] & [1] & 1.45 & 1 & [0.12, 0.08, 0.04, 0.01, 0.00] & [0.04]   \\ \hline \hline
        0 & 2 & ~ & [1, 2] & 0.77 & 0.01 & ~ & [0.28, 0.05] \\ 
        0 & 2 & ~ & [1, 2] & 0.82 & 0.1 & ~ & [0.25, 0.04] \\ 
        0 & 2 & ~ & [1, 2] & 1.07 & 0.5 & ~ & [0.11, 0.03] \\ 
        0 & 2 & ~ & [1, 2] & 1.45 & 1 & ~ & [0.01, 0.02]   \\ 
        1 & 2 & [1] & [1, 3] & 0.72 & 0.01 & [0.31] & [0.35, 0.00] \\ 
        1 & 2 & [1] & [1, 3] & 0.77 & 0.1 & [0.29] & [0.32, 0.00] \\ 
        1 & 2 & [1] & [1, 3] & 1.05 & 0.5 & [0.20]  & [0.17, 0.00] \\ 
        1 & 2 & [1] & [1, 3] & 1.45 & 1 & [0.12] & [0.04, 0.00] \\
        2 & 2 & [1, 2] & [1, 3] & 0.64 & 0.01 & [0.31, 0.19] & [0.35 0.01] \\ 
        2 & 2 & [1, 2] & [1, 3] & 0.7 & 0.1 & [0.29, 0.18] & [0.31, 0.01]  \\ 
        2 & 2 & [1, 2] & [1, 3] & 1.02 & 0.5 & [0.20, 0.143] & [0.17, 0.00] \\ 
        2 & 2 & [1, 2] & [1, 3] & 1.45 & 1 & [0.12, 0.08] & [0.04, 0.00] \\ 
        3 & 2 & [1, 2, 4] & [1, 2] & 0.64 & 0.01 & [0.31, 0.18, 0.06] & [0.35, 0.18] \\ 
        3 & 2 & [1, 2, 4] & [1, 2] & 0.7 & 0.1 & [0.29, 0.18, 0.05] & [0.32, 0.16] \\ 
        3 & 2 & [1, 2, 4] & [1, 2] & 1.02 & 0.5 & [0.20, 0.14, 0.03] & [0.17, 0.11] \\ 
        3 & 2 & [1, 2, 4] & [1, 2] & 1.45 & 1 & [0.12, 0.08, 0.01] & [0.04, 0.06] \\ 
        4 & 2 & [1, 2, 3, 4] & [1, 2] & 0.63 & 0.01 & [0.31, 0.18, 0.12, 0.08] & [0.35, 0.17] \\
        4 & 2 & [1, 2, 3, 4] & [1, 2] & 0.69 & 0.1 & [0.29, 0.18, 0.11, 0.05] & [0.32, 0.16] \\ 
        4 & 2 & [1, 2, 3, 4] & [1, 2] & 1.01 & 0.5 & [0.20, 0.14, 0.07, 0.03] & [0.17, 0.11] \\
        4 & 2 & [1, 2, 3, 4] & [1, 2] & 1.45 & 1 & [0.12, 0.08, 0.02, 0.01] & [0.04, 0.06] \\ 
        5 & 2 & [1, 2, 3, 4, 6] & [1, 2] & 0.62 & 0.01 & [0.31 0.19, 0.12, 0.06, 0.02] & [0.35, 0.17] \\ 
        5 & 2 & [1, 2, 3, 4, 6] & [1, 2] & 0.69 & 0.1 & [0.29, 0.18, 0.11, 0.06, 0.02] & [0.32, 0.16] \\ 
        5 & 2 & [1, 2, 3, 4, 6] & [1, 2] & 1.01 & 0.5 & [0.20, 0.13, 0.07, 0.03, 0.01] & [0.17, 0.14] \\ 
        5 & 2 & [1, 2, 3, 4, 6] & [1, 2] & 1.45 & 1 & [0.12, 0.08, 0.01, 0.01, 0.00]  & [0.04, 0.06] \\ \hline
    \end{tabular}
    }
    \caption{ Index detected at symptom onset}
    \label{TT1}
\end{table}

\begin{table}[!ht]
    \centering
    \tiny{
    \begin{tabular}{l|l|l|l|l|l|l|l}
       \# LFA & \# PCR & LFA schedule & PCR schedule & Exp.infecting days & $\beta$ & FNR (LFA)  & FNR (PCR)\\
    \hline\hline
        1 & 0 & [3] & ~ & 2.93 & 0.01 & [0.11] & ~ \\ 
        1 & 0 & [3] & ~ & 2.96 & 0.1 & [0.11] & ~ \\ 
        1 & 0 & [3] & ~ & 3.07 & 0.5 & [0.11] & ~ \\ 
        1 & 0 & [3] & ~ & 3.21 & 1 & [0.11] & ~ \\ 
        2 & 0 & [1, 3] & ~ & 1.62 & 0.01 & [0.33, 0.11] & ~ \\ 
        2 & 0 & [1, 3] & ~ & 1.64 & 0.1 & [0.33, 0.11] & ~ \\ 
        2 & 0 & [1, 3] & ~ & 1.73 & 0.5 & [0.31, 0.11] & ~ \\ 
        2 & 0 & [1, 3] & ~ & 1.84 & 1 & [0.29, 0.11] & ~ \\ 
        3 & 0 & [1, 3, 4] & ~ & 1.45 & 0.01 & [0.33, 0.11, 0.11] & ~ \\ 
        3 & 0 & [1, 3, 4] & ~ & 1.47 & 0.1 & [0.33, 0.11, 0.11] & ~ \\ 
        3 & 0 & [1, 3, 4] & ~ & 1.57 & 0.5 & [0.31, 0.11, 0.12] & ~ \\ 
        3 & 0 & [1, 3, 4] & ~ & 1.7 & 1 & [0.29, 0.11, 0.12] & ~ \\ 
        4 & 0 & [1, 2, 3, 4] & ~ & 1.25 & 0.01 & [0.33, 0.17, 0.13, 0.11] & ~ \\ 
        4 & 0 & [1, 2, 3, 4] & ~ & 1.28 & 0.1 & [0.33, 0.17, 0.13, 0.12] & ~ \\ 
        4 & 0 & [1, 2, 3, 4] & ~ & 1.38 & 0.5 & [0.31, 0.16, 0.13, 0.12] & ~ \\ 
        4 & 0 & [1, 2, 3, 4] & ~ & 1.53 & 1 & [0.29, 0.16, 0.13, 0.12] & ~ \\ 
        5 & 0 & [1, 2, 3, 4, 5] & ~ & 1.25 & 0.01 & [0.33, 0.17, 0.13, 0.12, 0.06] & ~ \\ 
        5 & 0 & [1, 2, 3, 4, 5] & ~ & 1.27 & 0.1 & [0.33, 0.17, 0.13, 0.12, 0.06] & ~ \\ 
        5 & 0 & [1, 2, 3, 4, 5] & ~ & 1.38 & 0.5 & [0.31, 0.16, 0.13, 0.12, 0.07] & ~ \\ 
        5 & 0 & [1, 2, 3, 4, 5] & ~ & 1.52 & 1 & [0.29, 0.16, 0.13, 0.12, 0.09] & ~ \\ \hline\hline
        0 & 1 &  & [2] & 1.86 & 0.01 & ~ & [0.03] \\ 
        0 & 1 &  & [2] & 1.89 & 0.1 & ~ & [0.03] \\ 
        0 & 1 &  & [2] & 2 & 0.5 & ~ & [0.02] \\ 
        0 & 1 &  & [2] & 2.14 & 1 & ~ & [0.02] \\ 
        1 & 1 & [1] & [3] & 1.4 & 0.01 & [0.33] & [0.00] \\ 
        1 & 1 & [1] & [3] & 1.43 & 0.1 & [0.33] & [0.00] \\ 
        1 & 1 & [1] & [3] & 1.53 & 0.5 & [0.31] & [0.00] \\ 
        1 & 1 & [1] & [3] & 1.66 & 1 & [0.29] & [0.00] \\ 
        2 & 1 & [1, 2] & [3] & 1.25 & 0.01 & [0.33, 0.17] & [0.01] \\ 
        2 & 1 & [1, 2] & [3] & 1.27 & 0.1 & [0.33, 0.17] & [0.01] \\ 
        2 & 1 & [1, 2] & [3] & 1.38 & 0.5 & [0.31, 0.16] & [0.01] \\ 
        2 & 1 & [1, 2] & [3] & 1.52 & 1 & [0.29, 0.16] & [0.01] \\ 
        3 & 1 & [1, 2, 3] & [3] & 1.25 & 0.01 & [0.33, 0.17, 0.13] & [0.03] \\ 
        3 & 1 & [1, 2, 3] & [3] & 1.27 & 0.1 & [0.33, 0.17, 0.13] & [0.03] \\ 
        3 & 1 & [1, 2, 3] & [3] & 1.38 & 0.5 & [0.31, 0.16, 0.13] & [0.02] \\ 
        3 & 1 & [1, 2, 3] & [3] & 1.52 & 1 & [0.29, 0.16, 0.13] & [0.02] \\ 
        4 & 1 & [1, 2, 3, 5] & [3] & 1.24 & 0.01 & [0.33, 0.17, 0.13, 0.03] & [0.03] \\ 
        4 & 1 & [1, 2, 3, 5] & [3] & 1.27 & 0.1 & [0.33, 0.17, 0.13, 0.03] & [0.03] \\ 
        4 & 1 & [1, 2, 3, 5] & [3] & 1.37 & 0.5 & [0.31, 0.16, 0.13, 0.03] & [0.02] \\ 
        4 & 1 & [1, 2, 3, 5] & [3] & 1.52 & 1 & [0.29, 0.16, 0.13, 0.02] & [0.02] \\ 
        5 & 1 & [1, 2, 3, 4, 5] & [2] & 1.22 & 0.01 & [0.33, 0.17, 0.12, 0.06, 0.03] & [0.13] \\ 
        5 & 1 & [1, 2, 3, 4, 5] & [2] & 1.25 & 0.1 & [0.33, 0.17, 0.11, 0.05, 0.03] & [0.13] \\
        5 & 1 & [1, 2, 3, 4, 5] & [2] & 1.36 & 0.5 & [0.31, 0.16, 0.11, 0.05, 0.03] & [0.12] \\ 
        5 & 1 & [1, 2, 3, 4, 5] & [2] & 1.5 & 1 & [0.29, 0.16, 0.10, 0.05, 0.02] & [0.10] \\ \hline\hline
        0 & 2 &  & [1, 3] & 1.35 & 0.01 & ~ & [0.31, 0.00] \\ 
        0 & 2 &  & [1, 3] & 1.37 & 0.1 & ~ & [0.30, 0.00] \\ 
        0 & 2 &  & [1, 3] & 1.46 & 0.5 & ~ & [0.28, 0.00] \\ 
        0 & 2 &  & [1, 3] & 1.58 & 1 & ~ & [0.24, 0.00] \\ 
        1 & 2 & [1] & [2, 3] & 1.23 & 0.01 & [0.33] & [0.04, 0.02] \\ 
        1 & 2 & [1] & [2, 3] & 1.25 & 0.1 & [0.33] & [0.04, 0.02] \\ 
        1 & 2 & [1] & [2, 3] & 1.36 & 0.5 & [0.31] & [0.03, 0.02] \\ 
        1 & 2 & [1] & [2, 3] & 1.51 & 1 & [0.29] & [0.03, 0.02] \\ 
        2 & 2 & [1, 2] & [2, 3] & 1.23 & 0.01 & [0.33, 0.17] & [0.13, 0.03] \\ 
        2 & 2 & [1, 2] & [2, 3] & 1.25 & 0.1 & [0.33, 0.17] & [0.13, 0.03] \\ 
        2 & 2 & [1, 2] & [2, 3] & 1.36 & 0.5 & [0.31, 0.16] & [0.12, 0.03] \\ 
        2 & 2 & [1, 2] & [2, 3] & 1.5 & 1 & [0.29, 0.16] & [0.10, 0.03] \\ 
        3 & 2 & [1, 2, 3] & [1, 3] & 1.18 & 0.01 & [0.33, 0.16, 0.09] & [0.37, 0.03] \\ 
        3 & 2 & [1, 2, 3] & [1, 3] & 1.21 & 0.1 & [0.33, 0.16, 0.09] & [0.36, 0.03] \\ 
        3 & 2 & [1, 2, 3] & [1, 3] & 1.31 & 0.5 & [0.31, 0.15, 0.08] & [0.33 0.03] \\ 
        3 & 2 & [1, 2, 3] & [1, 3] & 1.45 & 1 & [0.29, 0.14, 0.08] & [0.30, 0.02] \\ 
        4 & 2 & [1, 2, 3, 5] & [1, 3] & 1.18 & 0.01 & [0.33, 0.16, 0.09, 0.03] & [0.37, 0.03] \\ 
        4 & 2 & [1, 2, 3, 5] & [1, 3] & 1.2 & 0.1 & [0.33, 0.16, 0.09, 0.03] & [0.36, 0.03] \\ 
        4 & 2 & [1, 2, 3, 5] & [1, 3] & 1.31 & 0.5 & [0.31, 0.15, 0.08, 0.03] & [0.33, 0.03] \\ 
        4 & 2 & [1, 2, 3, 5] & [1, 3] & 1.45 & 1 & [0.29, 0.14, 0.07, 0.02] & [0.30, 0.02] \\ 
        5 & 2 & [1, 2, 3, 4, 5] & [1, 2] & 1.17 & 0.01 & [0.33, 0.16, 0.12, 0.06, 0.03] & [0.37, 0.16] \\ 
        5 & 2 & [1, 2, 3, 4, 5] & [1, 2] & 1.19 & 0.1 & [0.33, 0.16, 0.12, 0.07, 0.03] & [0.36, 0.17] \\ 
        5 & 2 & [1, 2, 3, 4, 5] & [1, 2] & 1.3 & 0.5 & [0.31, 0.15, 0.11, 0.05, 0.02] & [0.33, 0.15] \\ 
        5 & 2 & [1, 2, 3, 4, 5] & [1, 2] & 1.44 & 1 & [0.29, 0.14, 0.10, 0.04, 0.03] & [0.30, 0.16] \\ \hline
    \end{tabular}
    }
\caption{Index detected with random LFA test}
\label{TT2}
\end{table}

\begin{table}[!ht]
    \centering
     \tiny{
    \begin{tabular}{l|l|l|l|l|l|l|l}
       \# LFA & \# PCR & LFA schedule & PCR schedule & Exp.infecting days & $\beta$ & FNR (LFA)  & FNR (PCR)\\
    \hline\hline
   
        1 & 0 & [3] & ~ & 2.34 & 0.01 & [0.11] & ~ \\ 
        1 & 0 & [3] & ~ & 2.35 & 0.1 & [0.11] & ~ \\ 
        1 & 0 & [3] & ~ & 2.43 & 0.5 & [0.11] & ~ \\ 
        1 & 0 & [3] & ~ & 2.56 & 1 & [0.11] & ~ \\ 
        2 & 0 & [1, 3] & ~ & 1.11 & 0.01 & [0.42, 0.11] & ~ \\ 
        2 & 0 & [1, 3] & ~ & 1.11 & 0.1 & [0.41, 0.11] & ~ \\
        2 & 0 & [1, 3] & ~ & 1.15 & 0.5 & [0.39, 0.11] & ~ \\ 
        2 & 0 & [1, 3] & ~ & 1.22 & 1 & [0.36, 0.11] & ~ \\ 
        3 & 0 & [1, 3, 4] & ~ & 0.87 & 0.01 & [0.42, 0.11, 0.11] & ~ \\ 
        3 & 0 & [1, 3, 4] & ~ & 0.88 & 0.1 & [0.41, 0.11, 0.11] & ~ \\
        3 & 0 & [1, 3, 4] & ~ & 0.94 & 0.5 & [0.39, 0.11, 0.11] & ~ \\ 
        3 & 0 & [1, 3, 4] & ~ & 1.01 & 1 & [0.36, 0.11, 0.11] & ~ \\ 
        4 & 0 & [1, 2, 3, 4] & ~ & 0.63 & 0.01 & [0.42, 0.19, 0.13, 0.11] & ~ \\
        4 & 0 & [1, 2, 3, 4] & ~ & 0.65 & 0.1 & [0.41, 0.19, 0.13, 0.11] & ~ \\ 
        4 & 0 & [1, 2, 3, 4] & ~ & 0.71 & 0.5 & [0.39, 0.18, 0.13, 0.11] & ~ \\ 
        4 & 0 & [1, 2, 3, 4] & ~ & 0.80 & 1 & [0.36, 0.17, 0.13, 0.11] & ~ \\ 
        5 & 0 & [1, 2, 3, 4, 5] & ~ & 0.63 & 0.01 & [0.42, 0.19, 0.13, 0.11, 0.04] & ~ \\ 
        5 & 0 & [1, 2, 3, 4, 5] & ~ & 0.64 & 0.1 & [0.41, 0.19, 0.13, 0.11, 0.07] & ~ \\ 
        5 & 0 & [1, 2, 3, 4, 5] & ~ & 0.70 & 0.5 & [0.39, 0.18, 0.13, 0.11, 0.06] & ~ \\ 
        5 & 0 & [1, 2, 3, 4, 5] & ~ & 0.79 & 1 & [0.36, 0.17, 0.13, 0.11, 0.06] & ~ \\ \hline\hline
        0 & 1 & ~ & [1] & 1.76 & 0.01 & ~ & [0.0] \\ 
        0 & 1 & ~ & [1] & 1.78 & 0.1 & ~ & [0.0] \\ 
        0 & 1 & ~ & [1] & 1.87 & 0.5 & ~ & [0.0] \\ 
        0 & 1 & ~ & [1] & 2.00 & 1 & ~ & [0.0] \\ 
        1 & 1 & [2] & [3] & 1.09 & 0.01 & [0.16] & [0.01] \\ 
        1 & 1 & [2] & [3] & 1.11 & 0.1 & [0.16] & [0.01] \\ 
        1 & 1 & [2] & [3] & 1.20 & 0.5 & [0.16] & [0.01] \\ 
        1 & 1 & [2] & [3] & 1.32 & 1 & [0.15] & [0.01] \\ 
        2 & 1 & [1, 3] & [3] & 0.81 & 0.01 & [0.42, 0.11] & [0.01] \\ 
        2 & 1 & [1, 3] & [3] & 0.82 & 0.1 & [0.41, 0.11] & [0.01] \\ 
        2 & 1 & [1, 3] & [3] & 0.88 & 0.5 & [0.39, 0.11] & [0.01] \\ 
        2 & 1 & [1, 3] & [3] & 0.96 & 1 & [0.36, 0.11] & [0.01] \\ 
        3 & 1 & [1, 2, 3] & [3] & 0.62 & 0.01 & [0.42, 0.19, 0.13] & [0.04] \\
        3 & 1 & [1, 2, 3] & [3] & 0.64 & 0.1 & [0.41, 0.19, 0.13] & [0.04] \\
        3 & 1 & [1, 2, 3] & [3] & 0.70 & 0.5 & [0.39, 0.18, 0.13] & [0.03] \\ 
        3 & 1 & [1, 2, 3] & [3] & 0.79 & 1 & [0.36, 0.17, 0.13] & [0.03] \\
        4 & 1 & [1, 2, 3, 5] & [3] & 0.62 & 0.01 & [0.42, 0.19, 0.13, 0.04] & [0.04] \\ 
        4 & 1 & [1, 2, 3, 5] & [3] & 0.63 & 0.1 & [0.41, 0.19, 0.13, 0.04] & [0.04] \\ 
        4 & 1 & [1, 2, 3, 5] & [3] & 0.70 & 0.5 & [0.39, 0.18, 0.13, 0.04] & [0.03] \\ 
        4 & 1 & [1, 2, 3, 5] & [3] & 0.78 & 1 & [0.36, 0.17, 0.13, 0.04] & [0.03] \\ 
        5 & 1 & [1, 2, 3, 4, 5] & [2] & 0.60 & 0.01 & [0.42, 0.19, 0.16, 0.08, 0.04] & [0.18] \\
        5 & 1 & [1, 2, 3, 4, 5] & [2] & 0.61 & 0.1 & [0.41, 0.19, 0.16, 0.09, 0.05] & [0.17] \\ 
        5 & 1 & [1, 2, 3, 4, 5] & [2] & 0.68 & 0.5 & [0.39, 0.18, 0.14, 0.07, 0.04] & [0.16] \\ 
        5 & 1 & [1, 2, 3, 4, 5] & [2] & 0.76 & 1 & [0.36, 0.17, 0.13, 0.06, 0.03] & [0.15] \\ \hline\hline
        0 & 2 & ~ & [1, 3] & 0.79 & 0.01 & ~ & [0.43, 0.00] \\ 
        0 & 2 & ~ & [1, 3] & 0.80 & 0.1 & ~ & [0.42, 0.00] \\ 
        0 & 2 & ~ & [1, 3] & 0.85 & 0.5 & ~ & [0.39, 0.00] \\ 
        0 & 2 & ~ & [1, 3] & 0.91 & 1 & ~ & [0.35, 0.00] \\ 
        1 & 2 & [1] & [2, 3] & 0.61 & 0.01 & [0.42] & [0.05, 0.03] \\ 
        1 & 2 & [1] & [2, 3] & 0.62 & 0.1 & [0.41] & [0.05, 0.03] \\
        1 & 2 & [1] & [2, 3] & 0.69 & 0.5 & [0.39] & [0.05, 0.03] \\ 
        1 & 2 & [1] & [2, 3] & 0.77 & 1 & [0.36] & [0.04, 0.03] \\ 
        2 & 2 & [1, 2] & [2, 3] & 0.60 & 0.01 & [0.42, 0.19] & [0.18, 0.04] \\ 
        2 & 2 & [1, 2] & [2, 3] & 0.62 & 0.1 & [0.41, 0.19] & [0.17, 0.04] \\
        2 & 2 & [1, 2] & [2, 3] & 0.68 & 0.5 & [0.39, 0.18] & [0.16, 0.04] \\ 
        2 & 2 & [1, 2] & [2, 3] & 0.77 & 1 & [0.36, 0.17] & [0.15, 0.03] \\ 
        3 & 2 & [1, 2, 3] & [2, 3] & 0.60 & 0.01 & [0.42, 0.19, 0.16] & [0.18, 0.09] \\ 
        3 & 2 & [1, 2, 3] & [2, 3] & 0.61 & 0.1 & [0.41, 0.19, 0.16] & [0.17, 0.09] \\ 
        3 & 2 & [1, 2, 3] & [2, 3] & 0.68 & 0.5 & [0.39, 0.18, 0.15] & [0.16, 0.11] \\ 
        3 & 2 & [1, 2, 3] & [2, 3] & 0.77 & 1 & [0.36, 0.17, 0.13] & [0.15, 0.08] \\ 
        4 & 2 & [1, 2, 3, 6] & [2, 3] & 0.60 & 0.01 & [0.42, 0.19, 0.15, 0.04] & [0.18, 0.13] \\ 
        4 & 2 & [1, 2, 3, 6] & [2, 3] & 0.61 & 0.1 & [0.41, 0.19, 0.15, 0.04] & [0.18, 0.09] \\
        4 & 2 & [1, 2, 3, 6] & [2, 3] & 0.67 & 0.5 & [0.39, 0.18, 0.15, 0.03] & [0.16, 0.09] \\
        4 & 2 & [1, 2, 3, 6] & [2, 3] & 0.76 & 1 & [0.36, 0.17, 0.14, 0.03] & [0.15, 0.08] \\
        5 & 2 & [1, 2, 3, 4, 5] & [1, 2] & 0.55 & 0.01 & [0.42, 0.21, 0.16, 0.07, 0.04] & [0.50, 0.22] \\ 
        5 & 2 & [1, 2, 3, 4, 5] & [1, 2] & 0.56 & 0.1 & [0.41, 0.2, 0.15, 0.08, 0.05] & [0.49, 0.21] \\ 
        5 & 2 & [1, 2, 3, 4, 5] & [1, 2] & 0.63 & 0.5 & [0.39, 0.2, 0.15, 0.09, 0.04] & [0.46, 0.20] \\ 
        5 & 2 & [1, 2, 3, 4, 5] & [1, 2] & 0.71 & 1 & [0.36, 0.19, 0.14, 0.06, 0.03] & [0.42, 0.19] \\ \hline
    \end{tabular}
    }
\caption{Index case detected with weekly surveillance LFA test}
\label{TT3}
\end{table}

\nobibliography{references_covid}